\documentclass{aa}  

\usepackage{graphicx}
\usepackage{txfonts}
\usepackage[dvipsnames]{xcolor}
\usepackage{hyperref}
\newcommand{\OIII}{[\rm{O\,\textsc{iii}}]~}

\begin{document}

   \title{AGN radiative feedback as the main regulator of [O\,\textsc{iii}]~outflow activity and obscuration in X-ray AGNs
 }

 \titlerunning{AGN radiative feedback in eROSITA AGNs}

   \author{Carolina Andonie
          \inst{1}
          \and
          Andrea Merloni\inst{1}
          \and
          Catarina Aydar\inst{1,2}
          \and
          Benny Trakhtenbrot\inst{3}
          \and
          Johannes Buchner\inst{1}
          \and
          Brivael Laloux\inst{4,1}
          \and
          Mara Salvato\inst{1} 
          \and
          Peter Boorman\inst{1}
          \and
          David M. Alexander\inst{5}
          \and
          Marcella Brusa\inst{6}
          \and
          Pietro Baldini\inst{1}
          \and
          Tiago Costa\inst{7}
          \and 
          Victoria A. Fawcett\inst{8,7}
          \and
          Zsofi Igo\inst{9}
          \and
          Kirpal Nandra\inst{1}  
          }

   \institute{Max-Planck-Institut für Extraterrestrische Physik, Gießenbachstraße, D-85748 Garching, Germany\\
              \email{cpandonie [at] mpe.mpg.de}
         \and
         Excellence Cluster ORIGINS, Boltzmannstrasse 2, D-85748 Garching, Germany
         \and
             School of Physics and Astronomy, Tel Aviv University, Tel Aviv 69978, Israel
        \and 
            INAF-Osservatorio Astronomico di Capodimonte, Via Moiariello 16, 80131 Napoli, Italy
        \and 
            Centre for Extragalactic Astronomy, Department of Physics, Durham University, Durham, DH1 3LE, UK
        \and
            Dipartimento di Fisica e Astronomia “Augusto Righi”, Università di Bologna, via Gobetti 93/2, 40129 Bologna, Italy
        \and 
            School of Mathematics, Statistics and Physics, Newcastle University, Newcastle upon Tyne, NE1 7RU, UK 
        \and 
            European Southern Observatory, Karl-Schwarzschild-Strasse 2, 85748 Garching bei München, Germany 
        \and ESA, European Space Astronomy Centre (ESAC), Camino Bajo del Castillo s/n, 28692 Villanueva de la Cañada, Madrid, Spain
             }

   \date{Received September 15, 1996; accepted March 16, 1997}

  \abstract
    { Large-scale ionised outflows and nuclear obscuration are fundamental manifestations of Active Galactic Nuclei (AGNs) activity, yet direct observational evidence simultaneously linking these phenomena remains scarce.}
    {We used the eROSITA Final Equatorial Depth Survey (eFEDS), among the largest uniform optical spectroscopic datasets of X-ray–selected AGNs, to investigate how AGN accretion rate affects ionised outflow kinematics and X-ray obscuration.}
    {Our sample comprises 2,840 eROSITA AGNs at $z<0.82$ with high-quality Sloan Digital Sky Survey spectra. Through detailed optical spectral fitting, we measured Eddington ratios ($\lambda_{\rm Edd}$) and \OIII\ emission-line kinematics, tracing ionised outflows. In addition, we used a combination of archival eROSITA X-ray spectroscopy with X-ray stacking analyses to constrain the obscuration of the sample by measuring the hydrogen column density, $N_{\rm H}$.}
   { We find that (1) $\sim $35\% of the entire sample hosts \OIII outflows ($W_{80}>600 \rm \, km \, s^{-1}$), with the outflow incidence increasing with the AGN luminosity from $\approx 15\%$ at $L_{\rm AGN}<10^{44}\rm \, erg \, s^{-1}$ up to $\approx 60\%$ at $L_{\rm AGN}>10^{46}\rm \, erg \, s^{-1}$, (2) the outflow incidence increases with Eddington ratio from $\sim 29\%$ at $\log \lambda_{\rm Edd}<-2.3$ to $\sim 50\%$ at $\log \lambda_{\rm Edd}>-1.7$, and (3) AGN obscuration decreases with Eddington ratio, as X-ray stacking shows that sources with $\log\lambda_{\rm Edd}>-1.7$ are approximately five times less obscured than lower-$\lambda_{\rm Edd}$ AGNs. In addition, we find that $\sim 1\%$ of the sample populates the `forbidden region' of the $N_{\rm H}-\lambda_{\rm Edd}$ plane, where the outflow incidence peaks at $\sim 52\%$, consistent with a short-lived feedback phase. Notably, when matching the $\lambda_{\rm Edd}$ samples in AGN luminosity, these trends vanish, implying that radiation pressure drives changes in outflow activity and obscuration, while the black hole mass does not play a significant role. }
   {Our results are in agreement with AGN radiative feedback scenarios, where the Eddington ratio regulates the AGN environment by driving powerful galaxy-wide outflows and shaping the amount of circumnuclear material.}


   \maketitle
%

\section{Introduction}

Active galactic nuclei (AGNs) are powered by matter accreting onto a supermassive black hole (SMBH) and are among the most luminous objects in the Universe. During this accretion process, a fraction of the released energy may couple with matter in the host galaxy, regulating the growth of both the SMBH and the galaxy itself in a process known as AGN feedback \citep[e.g.][]{2005Springel, 2008Somerville, 2015King, 2024Harrison}. The discovery of tight scaling relations between SMBH mass and host galaxy properties, such as the bulge velocity dispersion (the $M_{\rm BH}-\sigma$ relation), provided some of the first compelling evidence of this co-evolutionary link  \citep[e.g.][]{1998Magorrian,2000Ferrarese, 2000Gebhardt, 2003Marconi,2013Kormendy}. Theoretical models propose that this regulation is achieved as the growing SMBH injects energy into its surroundings, regulating the star formation within the host galaxy  \citep[e.g.][]{2003King, 2005Springel, 2005DiMatteo}.

One of the most direct manifestations of this feedback is the presence of kiloparsec-scale outflows that reach the galaxy, which often alter the interstellar gas kinematics. In the `radiative' or `quasar' feedback mode, intense radiation from the accretion disk is believed to drive these winds \citep[e.g.][]{2014Zakamska, 2018Ishibashi, 2018RakshitWoo}. A primary diagnostic for tracing the impact of this process on the warm ionised interstellar medium (ISM) is the [\rm{O\,\textsc{iii}}]$\lambda 5007$ emission line (hereafter, \OIII). While the core of the \OIII profile typically traces the virial motion of gas in the galaxy \citep[e.g.][]{2006Bian,Le_2023}, the presence of strong, non-virial kinematic components, such as broad wings (velocity dispersion $ \gtrsim  500-600 \rm \, km\,s^{-1} $) or significant velocity shifts, is interpreted as a clear signature of an AGN-driven outflow \citep[e.g.][]{2008Komossa, 2008Nesvadba,2010Alexander, 2012Harrison}. These outflows, with velocities reaching up to 3000 km/s, are far more powerful and faster than those driven by star formation \citep{2005Rupke, 2014Cicone, 2017Concas} and are considered a signature of AGN activity \citep[see][for a review]{2024Harrison}.

Previous studies have established a fundamental link between the properties of \OIII outflows and the power of the central engine, finding positive correlations between outflow velocity and AGN luminosity \citep[e.g.][]{2013Mullaney, 2017Fiore, 2017Perna, 2020Wylezalek, 2023Musiimenta}. Several statistical studies have made significant progress in characterising \OIII outflow properties in AGN samples. For example, \citet{2013Mullaney} presented \OIII kinematics for over 20,000 optically selected AGNs from the SDSS at $z<0.4$, finding that the outflow incidence is higher in more luminous AGNs. More recently, \citet{2020Wylezalek} used MaNGA integral field spectroscopy of $\sim 2,700$ AGN host galaxies in the local Universe ($z\sim 0.05$) to investigate the spatially resolved \OIII kinematics, finding enhanced velocity dispersions in AGN hosts compared to inactive galaxies. Moreover, outflows have been found to be particularly common in dusty, reddened quasars and at high redshifts ($z \sim 2$), when galaxy and black hole growth are most intense \citep[e.g.][]{2015Brusa, 2016Kakkad, 2016Zakamska, 2020Kakkad}. 
However, our understanding of the prevalence and energetics of these outflows, and of how they depend on the fundamental properties of the accretion flow, remains incomplete. 

While large optical AGN samples have been transformative in establishing foundational statistical trends, they are inherently biased towards high-luminosity objects at low redshifts ($z \lesssim 0.4$), often missing low-luminosity AGNs that are diluted by host galaxy starlight. X-ray selection effectively mitigates this bias, enabling the identification of AGNs across a broader range of luminosities, as demonstrated by previous X-ray surveys \citep[e.g.][]{2016Menzel}. Furthermore, studies investigating the \OIII kinematics in X-ray-selected AGNs have provided key insights into feedback processes \citep[e.g.][]{2017Perna, 2023Musiimenta}; however, there remains a need for analyses based on well-controlled AGN samples matched in redshift, black hole mass, and luminosity, to robustly disentangle the interplay between outflows and accretion across a broad range of physical parameters.

Some observational and theoretical studies have also argued that AGN radiation can regulate the immediate environment of SMBHs \citep[e.g.][]{2005Springel,2005DiMatteo, 2009Fabian, 2017Ricci}. In this `radiation-regulated' model, the radiation pressure on dusty gas can blow away the circumnuclear obscuring material \citep[e.g.][]{2017Ricci}, thus successfully explaining the observed anti-correlation between the fraction of obscured AGNs and their luminosity, where low-accreting SMBHs are typically obscured, while high-accreting ones are typically unobscured \citep[e.g.][]{2007Maiolino, 2008Treister, 2014Ueda,2017Ricci}. Hence, in the AGN radiative feedback model, the accretion rate should simultaneously regulate both the SMBH nuclear obscuration and the large-scale \OIII outflow activity. However, a simultaneous investigation of both phenomena across a large, complete, and homogeneously selected sample has yet to be conducted.

In this work, we investigate how the AGN accretion rate shapes the circumnuclear environment across different physical scales, from the inner regions probed by obscuring column density ($N_{\rm H}$) to the larger-scale interstellar medium traced by the \OIII kinematics. For this, we used the eROSITA Final Equatorial Depth Survey (eFEDS; \citealp[]{2022Brunner}), a $140\,\rm deg^{2}$ extragalactic survey with homogeneous X-ray coverage from eROSITA down to  $6.5 \times 10^{-15}\rm erg \,cm^{-2}\, s^{-1}$ in the $0.5-2.0\rm \,keV$ energy band. The survey has a very rich multi-wavelength coverage, with multi-wavelength counterparts of the X-ray sources identified by \citet{2022Salvato}, and it has also extensive, high-quality optical spectroscopy from the Sloan Digital Sky Survey (SDSS; \citealp[]{2006Gunn}) and the Baryon Oscillation Spectroscopic Survey (BOSS; \citealp{2013Dawson, 2013Smee}) spanning over two decades, providing high-quality optical spectroscopy for 12,011 sources \citep{2025Aydar}. eROSITA is a soft X-ray observatory; hence, it mostly identifies unobscured, broad-line AGNs \citep{2022Liu, 2025Aydar}. A pilot study by \citet{2023Musiimenta} using a subset of eFEDS sources identified ionised outflows in $\sim 45\%$ of luminous, obscured AGNs. After extending this analysis to a larger sample ($z \sim 0.5-3.5$), they reported a weak correlation between outflow velocity and bolometric luminosity. We aim to extend and complement this work by using the entire eFEDS-SDSS AGN sample, thus enabling more robust statistical trends in the \OIII kinematic and AGN obscuration properties across a wide range of luminosities.

This paper is structured as follows. Section\,\ref{sec:sample} details the sample selection and the X-ray and optical datasets. Section\,\ref{sec:opXrayanalyses} describes our analysis methods, including optical and X-ray spectral fitting, spectral stacking, and the measurement of \OIII line kinematics. Section\,\ref{sec:results} presents our findings on the incidence of \OIII outflows in X-ray AGNs as well as the impact of the Eddington ratio on large-scale outflow activity and obscuration. In Section\,\ref{sec:discussion}, we discuss and interpret these results, and in Section\,\ref{sec:conclusions}, we summarise our main conclusions.

\section{Sample and datasets} \label{sec:sample}

In this section, we describe the datasets used in this work. In Section\,\ref{sub:data:sdss}, we present the optical spectra, and in Section\,\ref{sub:data:Xray}, we describe the X-ray data and the X-ray spectral constraints.

\subsection{Optical spectra} \label{sub:data:sdss}

The eFEDS sky area has been extensively observed by SDSS, as part of the Black Hole Mapper program\footnote{See \url{https://www.sdss.org/dr18/bhm/}.} of SDSS-V, yielding optical spectra for $13,079$ X-ray sources. The majority of the observations were obtained as part of a dedicated target program to observe eROSITA AGNs during SDSS-IV and SDSS-V. Some eFEDS sources also have serendipitous observations from earlier SDSS campaigns (SDSS-I, II, and III), and thus were obtained prior to the launch of SRG/eROSITA (see \citet{2025Aydar} for a detailed description of the SDSS observations in the eFEDS field). The optical spectra were mostly obtained with the SDSS spectrograph, which covers an observed wavelength range of $3,800 - 9,200$\AA\,  and has a spectral resolution of R $\sim 2,000$ at 6,000 \AA. At the redshifts considered in this work ($z < 0.82$), the rest-frame spectral coverage includes key diagnostic emission lines such as \OIII and  $\rm H\beta$, which are fundamental in this work (see Section\,\ref{res:sub:OIII_EddRatio}) for details). \citet{2025Aydar} performed a visual inspection of all SDSS spectra and identified 12,011 sources with reliable spectroscopic redshifts. In this context, a reliable redshift is one that has been estimated by the BOSS redshift pipeline, optimised for broad-line AGNs and passive galaxies, and subsequently confirmed through an extensive visual inspection campaign. The spectroscopic redshift completeness reaches 70\% down to $r_{\rm AB}<21.38 \rm \, mag$, meaning that spectroscopic coverage in eFEDS is limited to the optically brighter X-ray sources.

From this parent sample, we selected sources based on spectral quality and redshift range. First, following the recommendations for extragalactic sources observed with the SDSS spectrograph, we selected all the sources with a median signal-to-noise ($S/N$) such that $(S/N)^2>10$, calculated over all valid pixels in the observed wavelength range. This threshold ensures that the spectra are of sufficient quality for reliable emission-line measurements, particularly for the detection and characterisation of the  $\rm H\beta$ and \OIII profile, which are the primary focus of this work. Second, we selected all sources with a reliable spectroscopic redshift of $z < 0.82$, the redshift at which the \OIII line shifts out of the SDSS spectral range. Within this redshift range, the  $\rm H\beta$ line is always observed, which we used to estimate single-epoch black hole masses (see Section\,\ref{res:sub:OIII_EddRatio}) for details). Applying these two selection criteria yielded a sample of 2,840 sources; see Table\,\ref{t:data}. For further details on the collation of the optical spectra and their visual inspection, we refer the reader to \citet{2025Aydar}.

\begin{table}
\centering
\caption{Sample size of the X-ray AGNs in eFEDS at $z<0.82$}
\begin{tabular}{c|c|c}
 \hline
 \hline
\noalign{\smallskip}
S & Selection & N \\
\noalign{\smallskip}
 \hline
\noalign{\smallskip}

1 & X-ray AGN & 7442 \\
2 & X-ray AGN \& SDSS spectra & 3863 \\
3 & X-ray AGN \& SDSS spectra \& $(S/N)^2>10$ & 2840 \\
4 & X-ray AGN \& SDSS spectra \& $(S/N)^2>10$ \& $N_{\rm H}$ & 2722 \\
5 & X-ray AGN \& \OIII detection & 1943 \\
6 & X-ray AGN \& broad H$\beta$ detection & 2087\\
7 & X-ray AGN \& \OIII \& broad H$\beta$ detection & 1573\\
8 & X-ray AGN \& \OIII \& broad H$\beta$ detection \& $N_{\rm H}$& 1541\\
\noalign{\smallskip}

\hline
\hline

\end{tabular}
\tablefoot{ Sample (1) contains the total number of X-ray AGNs within $z<0.82$ reported in \citet{2022Salvato}, assuming that the photometric redshifts are correct. Sample (2) corresponds to the number of X-ray AGNs with SDSS spectra, and sample (3) to the number of sources that have an SDSS spectrum with $(S/N)^2>10$. Samples (4)-(8) only contain sources with $(S/N)^2>10$. Sample (4) corresponds to sources with an X-ray column density ($N_{\rm H}$) constraint. Sample (5) corresponds to sources with an \OIII emission line detection (i.e., $\rm EW(\OIII)/\sigma_{\rm EW(\OIII)} > 2$); sample (6) comprises sources with a broad H$\beta$ detection (i.e., $\rm EW(H\beta_{br})/\sigma_{\rm EW(H\beta_{br})} > 2$). Sample (7) corresponds to sources with a \OIII and broad H$\beta$ detections; and sample (8) corresponds to sources with a \OIII detection, broad H$\beta$ detections, and $N_{\rm H}$ constraints.}\label{t:data}

\end{table}

\subsection{X-ray data and spectral constraints} \label{sub:data:Xray}

The primary X-ray catalogue in eFEDS was released by \citet{2022Brunner} and contains $27,910$ sources detected in the $0.5-2 \rm \: keV$ band down to $6.5 \times 10^{-15} \rm \: erg \: cm^{-2} \: s^{-1}$. In this work, we use the X-ray measurements from \citet{2022Liu}, who performed X-ray data reduction, spectral extraction, and spectral fitting, adopting the optical counterparts presented in \citet{2022Salvato}. They presented the eFEDS AGN catalogue, containing $22,079$ sources, all of which are eROSITA point-like X-ray sources with a secure optical counterpart. These sources were classified as extragalactic by \citet{2022Salvato}, based on their multiwavelength photometric properties, thereby excluding Galactic X-ray sources and galaxy clusters. This parent sample spans the full redshift range of the eFEDS survey; of these, 7,444 sources lie at $z < 0.82$, where redshifts are taken from \citet{2025Aydar} when spectroscopic redshifts are available, and from \citet{2022Salvato} otherwise. However, the primary focus of this work is the subset of X-ray AGNs with available optical spectra, as described in Section\,\ref{sub:data:sdss} and summarised in Table\,\ref{t:data}.

Adopting a Bayesian approach, \citet{2022Liu} fitted the spectra of all the X-ray AGNs using different X-ray models. In this work, we adopted the results from the baseline model (Model 1), which fits the spectra with a single absorbed power law. Although simple, this model adequately reproduces the range of X-ray spectral shapes observed in eFEDS sources. From these results, we primarily used the intrinsic 0.5–2\,keV luminosity ($L_{\rm 0.5-2\,keV}$), the photon index ($\Gamma$), and $N_{\rm H}$. A key parameter for our analysis is the AGN bolometric luminosity ($L_{\rm AGN}$), which we estimated from the X-ray constraints. First, we converted the 0.5–2\,keV luminosity to the 2–10\,keV band, assuming the intrinsic X-ray emission can be described by a single power law with $\Gamma = 2$. This choice reflects the median of the photon-index distribution in eFEDS reported by \citet{2022Liu}, and avoids dependence on individual, highly uncertain $\Gamma$ measurements. We then applied the bolometric correction from \citet{2020Duras} to obtain $L_{\rm AGN}$. The uncertainties in $L_{\rm AGN}$ range $\sigma \log L_{\rm AGN}\approx 0.006-0.35 \rm \, dex $, with a medium value of $\sim 0.13 \rm \, dex $. Additionally, by analysing the entire $N_{\rm H}$ posterior distribution, they identified the sources where the absorption could not be constrained (\textsc{class=uninformative} in the catalogue), mostly due to low X-ray counts. We found that $\sim 95$\%  (2722 out of 2840) of our sample have an $N_{\rm H}$ measurement. The sources with class \textsc{class=uninformative} are not included in the X-ray analysis based on single measurements. We note that the initial paper used photometric redshifts from \citet{2022Salvato} for the X-ray spectral fitting; however, the authors subsequently released a new X-ray spectral catalogue using spectroscopic redshifts from \citet{2025Aydar}, which are the values used in this work. For more details on the AGN selection, X-ray data reduction and spectral fitting, see \citet{2022Liu} and \citet{2022Salvato}. Additionally, we performed X-ray stacking analyses of our sources, using the extracted spectra from \citet{2022Liu}, which we describe in Section\,\ref{subsec:Xray_stacks}.

\section{Optical and X-ray spectral analyses} \label{sec:opXrayanalyses}

In this section, we describe the approach used to model the optical spectra (Section\,\ref{sub:optfitting}) of our sample, the optical spectral stacking procedure (Section\,\ref{sub:opt_stacking}), and the calculation of the \OIII line kinematics (Section\,\ref{subsec:W80calc}). Then, in Section\,\ref{subsec:Xray_stacks}, we describe the X-ray stacking and X-ray spectral fitting of the stacks.

\subsection{Optical spectral fitting} \label{sub:optfitting}

To measure the \OIII line kinematics and the accretion rate of our sources, which we express in terms of the Eddington ratio, $\lambda_{\rm Edd}$, we first modelled the full optical spectrum to determine key parameters, including the broad H$\beta$ full width at half maximum (FWHM), the 5100\AA\, luminosity, and the \OIII line profile. The broad H$\beta$ and 5100\AA\, luminosity are then used to estimate single-epoch black hole masses, from which we subsequently derived $\lambda_{\rm Edd}$. Accurate modelling of optical spectra also requires isolating the various components, including host-galaxy and AGN emission. For this, we fit the optical spectra following the methodology outlined in \citet{aydar2026}; which has been successfully used to measure the optical spectral properties of AGNs. We first used \textsc{pPXF} \citep{2004PASPCappellari, 2017Cappellari, 2023Cappellari} to model the stellar emission (Section\,\ref{sub:ppfx}). Then, we used \textsc{PyQSOFit} \citep{2018asclGuo, 2019Shen} to fit the AGN continuum, Fe II continuum, and broad and narrow emission lines (Section\,\ref{sub:pyqsofit}).

\subsubsection{Host galaxy fitting with \textsc{pPXF}} \label{sub:ppfx}

Penalized PiXel-Fitting (\textsc{pPXF}\footnote{see \url{https://pypi.org/project/ppxf/}.}) is a general-purpose spectral fitting algorithm originally presented in \citep{2004PASPCappellari} and subsequently updated in \citet{2017Cappellari, 2023Cappellari}. It is available as a Python package but has been implemented in multiple programming languages. \textsc{pPXF} is designed to model the optical spectra of galaxies and extract the properties of their stellar populations, together with both the stellar and gas kinematics. The flexibility of \textsc{pPXF} allows the user to include additional templates to the galaxy emission model, such as the AGN continuum, Fe II continuum and emission lines. In this work, however, we used \textsc{pPXF} solely to constrain the stellar emission, as it is specifically optimised for modelling stellar populations rather than AGN spectra. In particular, \textsc{pPXF} is not well suited for fitting complex AGN emission lines that require multiple Gaussian components, such as $\rm H\beta$ and \OIII, and it does not provide built-in routines for estimating parameter uncertainties, both of which are essential for this work.  

To fit the optical spectra with \textsc{pPXF}, we followed a similar approach to \citet{2025Bernal} and \citet{aydar2026}. We first corrected for the galactic extinction using the dust maps from \citet{1998Schlegel} and the extinction curve from \citet{1989Cardelli}\footnote{Throughout this paper, we use the same approach every time we need to correct for galactic extinction.}. To model our sources, we included the stellar populations emission from the E-MILES library \citep{2016Vazdekis}, two sets of Gaussian profiles to model the narrow and broad emission lines (see Table\,\ref{t:emission_lines}), a set of templates that reproduce the Fe II pseudo-continuum \citep{1992Boroson,2001Vestergaard}, and a set of powerlaws to model the AGN continuum. The E-MILES template library used in this work considers a Salpeter initial mass function \citep{1955Salpeter} and the set of theoretical isochrones `Padova0' of \citet{2000Girardi}. We considered a wide range of ages, from 0.063 to 15.8 Gyr, and metallicities, from $\rm Z=-1.71$ to $\rm Z=0.22$. The Fe II templates cover a wide range of velocities, from $1000$ to $11800 \rm \: km \: s^{-1}$. Finally, we modelled the AGN continuum with a set of powerlaws defined as $f_{\lambda} = (\lambda/N)^{\alpha}$, where $\lambda$ corresponds to the wavelength, $N$ is the normalisation wavelength which we set to $5000\AA$, and $\alpha$ the powerlaw slope, which takes values in the range $-3 \leq \alpha \leq 0$. We note that we do not use the AGN continuum, Fe II continuum, and emission lines constraints from \textsc{pPXF}; we just included these components to properly constrain the stellar emission. 

Our routine fits the spectrum iteratively to ensure a self-consistent decomposition of all spectral components. In the first iteration, we fit the spectrum with \textsc{pPXF}, including all components except the Fe II template (i.e., the stellar continuum, broad and narrow emission lines, and the AGN power-law continuum), yielding an initial estimate of the broad-line region (BLR) velocity. Since the Fe II emission and the broad lines are both expected to arise in the BLR \citep{2013Barth}, we used this BLR velocity to select the Fe II template with the closest BLR velocity, to then perform the fit again using the selected Fe II template. We repeated this process until the Fe II template velocity and the BLR velocity converged, which, in all cases, occurred within no more than two iterations.

\subsubsection{Quasar fitting with PyQSOFit} \label{sub:pyqsofit}

\textsc{PyQSOFit}\footnote{see \url{https://github.com/legolason/PyQSOFit/tree/master}.} is a \textsc{python} package that models the optical spectra of quasars. The code includes the different components of the quasar spectrum, such as the Fe II emission, AGN continuum, and broad and narrow emission lines. \textsc{PyQSOFit} also includes a Markov Chain Monte Carlo (MCMC) routine to calculate the parameter uncertainties.

We first subtracted the stellar emission calculated with \textsc{pPXF} from the Galactic-extinction-dereddened optical spectrum. Then, we fitted the subtracted spectrum with \textsc{PyQSOFit}, only including the Fe II emission, the AGN continuum, and narrow and broad emission lines. Similar to \textsc{pPXF}, the AGN continuum is fitted with a set of powerlaws with shape $f_{\lambda} = (\lambda/3000)^{\alpha}$ with $\alpha=[-5,3]$ and the Fe II emission is modelled with a combination of templates from \citep{1992Boroson,2001Vestergaard,2006Tsuzuki,2007Salviander}. Although in this paper we used only the constraints from H$\beta$ and \OIII emission lines, we still included all emission lines in the spectra in our fit, which are listed in Table\,\ref{t:emission_lines}. The fitting of the H$\beta$+\OIII complex is explained below, while the fitting of the rest of the lines is described in Appendix\,\ref{ap:optical_fitting}. 

We jointly fitted the H$\beta$ line and the [O\,\textsc{iii}]~$\lambda\lambda4960,\,5008$ doublet. As eROSITA is a soft X-ray observatory, it mostly selects unobscured, broad-line AGNs \citep{2022Liu, 2025Aydar}; hence, we expect to have a negligible amount of Type II AGNs in our sample. H$\beta$ is modelled with a narrow Gaussian and three broad Gaussian components, each with a maximum FWHM of $17\,000\,\rm km\,s^{-1}$. Each [O\,\textsc{iii}] line is represented by a core and a wing: the core's FWHM is tied to that of the narrow H$\beta$, while the wing, capturing line asymmetry, is modelled by a Gaussian with a FWHM ranging $500-2000\,\rm km\,s^{-1}$, which are the typical outflow velocities observed at these redshifts \citep[e.g.][]{2017Perna}. The flux ratio between [O\,\textsc{iii}]~$\lambda4960$ and [O\,\textsc{iii}]~$\lambda5007$ is fixed at 1:2.99 \citep{1981Osterbrock}. Careful fitting of both the H$\beta$ and [O\,\textsc{iii}] profiles is necessary, as overlap can occur in cases of very broad H$\beta$ or a fast, blueshifted \OIII wing. Figure \ref{fig:HbOIII_ex} in Appendix\,\ref{ap:optical_fitting} shows examples of the H$\beta$+\OIII complex fitting for two sources in our sample. Our fitting approach for the H$\beta$+\OIII complex follows the recommendations of the developers of \textsc{PyQSOFit}, and has been successfully used to measure the broad H$\beta$ FWHM and the \OIII kinematics in AGNs \citep[e.g.][]{2019Shen, 2020Rakshit, 2022Wu, 2023Musiimenta, 2024Panda, aydar2026}. To further demonstrate this, we compared the performance of the two- versus three- Gaussian model to fit the broad H$\beta$ component, and we demonstrated that, for our sample, which is selected to have a median $(S/N)^2>10$, the three-Gaussian model performs better in $\sim 98\%$ of the cases, and even when the two-Gaussian model performs better, the measured broad H$\beta$ FWHMs are consistent. This analysis is described in Appendix\,\ref{ap:Hb}, and further supports that modelling the $\rm H\beta$ profile with one narrow and three broad Gaussian components provides an effective compromise between accurately fitting individual sources and maintaining a uniform methodology applicable to thousands of spectra. We note that the [O\,\textsc{iii}] kinematics is ultimately determined using the non-parametric method described in Section\,\ref{subsec:W80calc}, rather than from the Gaussian fit itself.

From our analysis, we derived key AGN spectral parameters and their associated uncertainties, including monochromatic AGN luminosities, as well as the flux, equivalent width (EW), and FWHM for each fitted emission line. To assess the consistency of our optical spectral fitting with independent X-ray measurements, we compared the optical AGN luminosity with the 2–10\,keV X-ray luminosity (see Section~\ref{sub:data:Xray}). The results of this comparison are presented in Figure~\ref{fig:LxLop} in Appendix\,\ref{ap:Lx_Lop}. We find a clear correlation between the AGN luminosity at 5100\AA\, ($L_{\rm 5100}$) and the 2–10\,keV luminosity ($L_{\rm 2-10\,keV}$), consistent with that expected from the optical and X-ray bolometric corrections from \citet{2006Richards} and \citet{2020Duras}. Figure\,\ref{fig:LxLop} also presents the luminosity of the \OIII emission line core ($L_{\rm OIII}$) against $L_{\rm 2-10\,keV}$. These results are in good agreement with the correlations reported by \citet{2015Ueda} and \citet{2006Panessa} (not shown in the figure for brevity). Overall, these comparisons validate the robustness of our optical spectral fitting methodology, confirming its consistency with established X-ray–optical relations in AGNs. 

All the analyses performed in the paper adopt the AGN bolometric luminosity estimated from $L_{\rm 2-10\,keV}$. We repeated all analyses using the AGN bolometric luminosity estimated from $L_{\rm 5100}$ and obtained fully consistent results.

\subsection{Optical spectral stacking} \label{sub:opt_stacking}

In our analysis, we use optical spectral stacking to further confirm the [O\,\textsc{iii}] line kinematics measured in single sources, as identifying and constraining a line wing can sometimes be challenging, even at high spectral $S/N$, due to difficulties in disentangling it from the continuum. Additionally, the stacked spectra provide a visual confirmation of the systematic change in [O\,\textsc{iii}] kinematics as a function of different parameters. To stack the optical spectra in our sample, we followed a similar approach to \citet{2020Comparat}. First, each spectrum is shifted to the rest frame using $\lambda_{\rm RF} = \lambda / (1+z)$ and corrected for Galactic extinction. We then interpolate both the flux and its uncertainties onto a common wavelength grid, evenly spaced in $\log_{10}(\lambda)$ from 800\,\AA{} to 11,000\,\AA{} with a spacing of $\Delta\log_{10}(\lambda)=0.0001$. Since our goal is to analyse the median \OIII profile across the sample, we next subtracted the stellar emission as estimated with \textsc{pPXF}, along with the AGN continuum, Fe II continuum, and the extended broad H$\beta$ component as calculated by \textsc{PyQSOFit}. The resulting spectrum is then normalised by the AGN flux at 5100\,\AA. Finally, we calculated the median and standard deviation for each pixel across the sample, resulting in the final stacked spectrum. Figure\,\ref{fig:velprofile} illustrates an example of a stacked \OIII profile for the sources in our sample with $\log L_{\rm AGN}/(\rm erg \, s^{-1})=45-45.5$.

\subsection{ {\rm [O\,\textsc{iii}] } line kinematics calculation } \label{subsec:W80calc}

\begin{figure}

\centering
\includegraphics[trim={0cm 0 0cm 0cm},clip,scale=0.4]{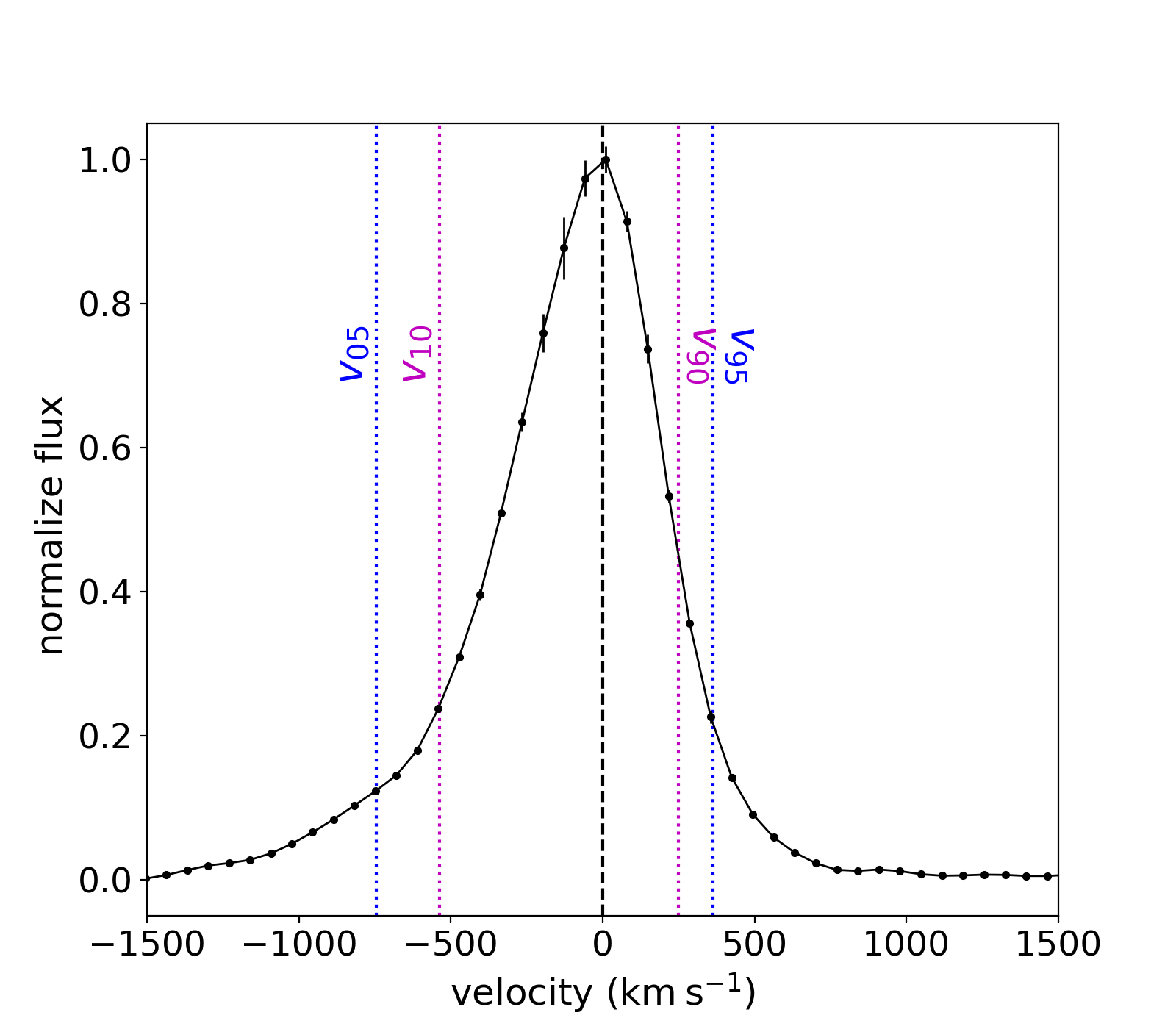}
\caption{Stacked \OIII velocity profile of the sources in our sample with $\log L_{\rm AGN}/(\rm erg \, s^{-1})=45-45.5$. The profile is normalised to the \OIII peak flux and shows a clear asymmetry. The figure also shows the velocity at which the integrated line flux reaches 5\%, 10\%, 90\%, and 95\% of the total, which are used to calculate the kinematic parameters $W_{80}$ and $\Delta v$. } 
\label{fig:velprofile}
\end{figure}

To quantify the line kinematics, we calculated the velocity (i.e. line width) and asymmetry of the \OIII $\lambda5007$ profile using the parameters $W_{80}$ and $\Delta v$, standard methods used in the literature \citep[e.g.][]{2013Liu,2014Zakamska}. These quantities require measuring the velocities at which the integrated line flux reaches 5\%, 10\%, 90\%, and 95\% of the total, denoted $v_{05}$, $v_{10}$, $v_{90}$, and $v_{95}$, respectively, which are illustrated in Fig.~\ref{fig:velprofile}. We defined the velocity: $W_{80} = v_{90} - v_{10}$, and the asymmetry: $\Delta v = (v_{05} + v_{95}) / 2$. For a perfectly symmetric line, $\Delta v = 0$, while a blueshifted profile yields $\Delta v < 0$.

We calculated the parameters $W_{80}$ and $\Delta v$ and their uncertainties using only the emission coming from the \OIII line. For this, we followed the next steps: 1) we corrected for the galactic extinction in the direction of the source; 2) we subtracted the stellar emission calculated with \textsc{pPXF}; and 3) we subtracted the AGN and Fe II continuum estimated with \textsc{PyQSOFit}, and also an extended component of the broad H$\beta$ line. After isolating the \OIII emission, we generated 100 Monte Carlo realisations of each spectrum by resampling the flux within its $1\sigma$ Gaussian errors. For each realisation, we calculate $W_{80}$ and $\Delta v$, and then determine the mean and standard deviation to provide robust measurements and uncertainties for these kinematic parameters. 

\subsection{X-ray spectral stacking and modelling} \label{subsec:Xray_stacks}

Following the same approach of Section\,\ref{sub:opt_stacking}, we performed X-ray spectral stacking to independently confirm the X-ray spectral constraints derived from individual sources. In particular, we focus on the hydrogen column density $N_{\rm H}$, which has a typical uncertainty of $\sim 0.5\,$dex when measured from single-source fits. Stacking the X-ray spectra therefore allowed us to average over these uncertainties and more robustly identify any systematic variation of $N_{\rm H}$ as a function of key physical parameters, as performed in Section\,\ref{res:sub:NH_EddRatio}. To stack the X-ray spectra of our sources, we used \textsc{Xstack}\footnote{\url{https://github.com/AstroChensj/Xstack/}} \citep{2025Chen}, a publicly available \textsc{python} package to stack eROSITA spectra. In brief, \textsc{Xstack} first sums all (rest-frame) X-ray spectra without applying any scaling, and its associated full responses, preserving the overall spectral shape and carefully correcting for Galactic extinction. \textsc{Xstack} then produces the stacked spectrum, stacked response files (ARF and RMF), and stacked background. For more details on the stacking procedure, we refer the interested reader to \citet{2025Chen}.

We fitted the X-ray spectral stacks using the Bayesian X-ray Analysis (BXA) package\footnote{See \url{https://github.com/JohannesBuchner/BXA.git}.} \citep{2014Buchner}. Our implementation links the X-ray spectral analysis environment \textsc{XSPEC} v12.14.1 \citep{1996Arnaud} with the nested sampling algorithm UltraNest \citep{2021BuchnerUl}. This approach enables an efficient exploration of the full parameter space and the incorporation of informative or non-informative priors for any parameter. BXA delivers robust parameter estimates and uncertainties, even for low-count spectra, and is particularly effective in cases where the parameter space exhibits strong degeneracies. Since the stacks follow a Poisson distribution while the backgrounds approximate a Gaussian distribution, we adopted the PG statistic\footnote{See \url{https://heasarc.gsfc.nasa.gov/docs/software/xspec/manual/XSappendixStatistics.html}.}.

To fit the stacks, we adopt an absorbed power-law model. For the absorption component, we used \textsc{disnht} \citep{2022Locatelli}, which models the absorber as a set of independent column densities following a lognormal distribution. The free parameters of this model are $\log N_{\rm H}$, representing the mean column density, and its standard deviation, $\sigma_{N_{\rm H}}$. The powerlaw model, \textsc{powdist}, has as free parameters the normalisation and the photon-index, $\Gamma$. In \textsc{XSPEC}, the model is \textsc{disnht}\,$\times$\,\textsc{powdist}.

\section{Results} \label{sec:results}

In this section, we explore the incidence of \OIII outflows in eROSITA AGNs (Section\,\ref{res:sub:OIIIin}), and the impact of the Eddington ratio in the \OIII outflows kinematics (Section\,\ref{res:sub:OIII_EddRatio}) and the AGN obscuration (Section\,\ref{res:sub:NH_EddRatio}). A discussion of the relation of \OIII outflows with obscuration and Eddington ratio follows in Section \ref{dis:sub:NH_EddRatio}.

\subsection{  {\rm [O\,\textsc{iii}]} outflow incidence in eROSITA AGNs} \label{res:sub:OIIIin}

In this section, we present the \OIII line properties of our sample. We find that $1,943$ sources have an \OIII detection, equivalent to 68\% of the spectroscopic sample; see Table\,\ref{t:data}. Following \citet{2022Wu}, we define a reliable \OIII line detection as $\rm EW([O\,III])/\sigma_{EW([O\,III])}>2$. We find that the distributions of spectral S/N for the \OIII-detected and undetected subsamples are similar, with median values of approximately 7.5 and 7.0, respectively, suggesting that the detection rate is not strongly driven by spectral quality alone. A more significant difference is found in the distribution of AGN bolometric luminosity ($L_{\rm AGN}$; Section\,\ref{sub:data:Xray} for details) derived from the X-ray luminosity. The \OIII-detected sample is more luminous with a median $\log L_{\rm AGN}/(\rm erg~s^{-1})\approx 44.6$, compared to the \OIII-undetected sample, which has a median AGN bolometric luminosity $\approx 0.4 \rm \, dex$ lower. This luminosity dependence suggests that more luminous AGNs are more likely to produce detectable \OIII emission. This is consistent with the observed correlation between the \OIII luminosity and the $2-10 \rm keV$ luminosity \citep[e.g.][]{2015Ueda}, which we present in the bottom panel of Figure\,\ref{fig:LxLop} in Appendix\,\ref{ap:Lx_Lop}.   

The \OIII line velocity width is characterised by the non-parametric quantity $W_{80}$, as described in Section\,\ref{subsec:W80calc}. We define an \OIII outflow as present when there is a significant \OIII detection and when $W_{80} > 600\,\rm km\,s^{-1}$. That threshold is based on previous studies who have found that the velocity of \OIII AGN-driven outflows is always larger than $600\,\rm km\,s^{-1}$ \citep[e.g.][]{2014Harrison,2020Kakkad}, while \OIII outflows driven by star-formation processes have velocities ranging $\approx 200-500 \rm \, km \, s^{-1}$ \citep[e.g. ][]{2005Rupke,2016Cicone, 2017Concas}. 

\begin{figure}
 
\centering
\includegraphics[trim={0cm 0 0cm 0cm},clip,scale=0.42]{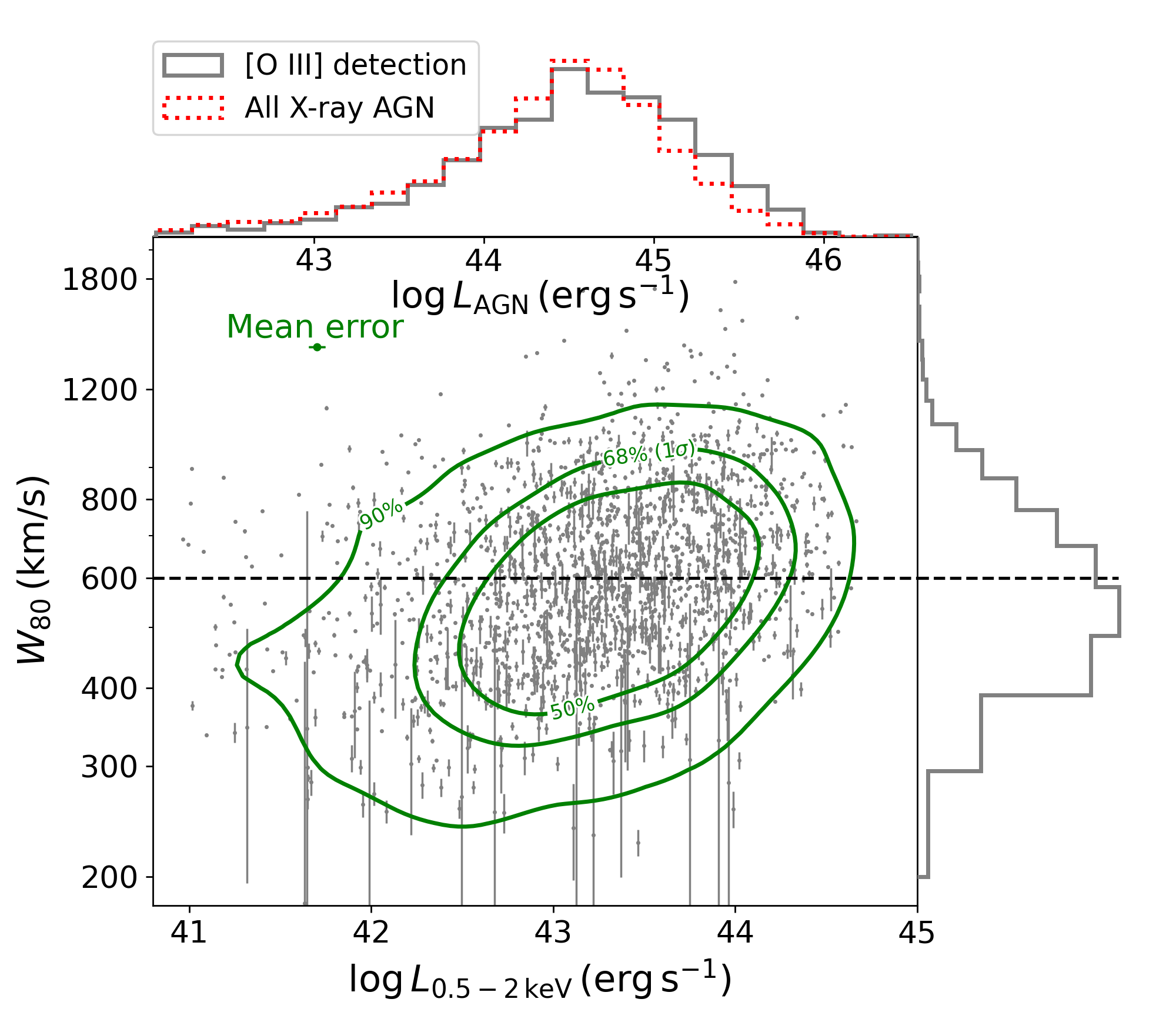}
\caption{Non-parametric \OIII velocity ($W_{80}$) plotted against intrinsic soft X-ray luminosity ($L_{0.5-2\,\mathrm{keV}}$; bottom x-axis) and AGN bolometric luminosity ($L_{\rm AGN}$, derived from $L_{0.5-2\,\mathrm{keV}}$; top x-axis). Only sources with a significant \OIII detection are shown. The $y$-axis error bars represent the $1\sigma$ uncertainties in $W_{80}$; the mean $x$-axis error is indicated by a green marker with error bars in the upper-left side of the main panel. Contours enclose 50\%, 68\%, and 90\% of the sample. The horizontal dotted black line at $W_{80} = 600\,\mathrm{km\,s}^{-1}$ denotes the threshold used to identify \OIII outflows. The top histogram shows the $L_{0.5-2\,\mathrm{keV}}$ distribution for sources with \OIII detections (grey solid line), as well as for the full sample of X-ray AGNs at $z < 0.82$, including those without optical spectra (orange dotted line). The right histogram displays the distribution of $W_{80}$ for the plotted sources. } 

\label{fig:W80_Lbol}
\end{figure}

We first look at the outflow properties of the sources with \OIII detections. Figure\,\ref{fig:W80_Lbol} shows $W_{80}$ as a function of both the intrinsic 0.5-2 keV X-ray luminosity ($L_{\rm 0.5-2\,keV}$; bottom x-axis), and the derived AGN bolometric luminosity ($L_{\rm AGN}$; top x-axis); see Section\,\ref{sub:data:Xray}. For our sample, $W_{80}$ spans approximately 200–1800 km/s, while $\log L_{\rm AGN}/(\rm erg~s^{-1})$ ranges from 42 to 46.5, probing a large range of luminosities. The majority ($68\%$) of sources cluster within $W_{80} \sim 400-900\,\rm km\,s^{-1}$ and $\log L_{\rm AGN}/(\rm erg~s^{-1})\approx 43-45.5$. While $W_{80}$ tends to increase with $L_{\rm AGN}$, the data show considerable scatter. A Spearman correlation analysis yields $r=0.33$ ($\rm p{-}value= 10^{-51}$), indicating a weak but statistically significant correlation between outflow velocity and AGN luminosity. The Spearman correlation yields the exact same results when considering the \OIII velocities on a logarithmic scale, $\log W_{80}$, most likely due to the narrow distribution of $W_{80}$ in our sample.

The horizontal histogram of Figure\,\ref{fig:W80_Lbol} compares the $L_{\rm 0.5-2\,keV}$ distributions for all eROSITA AGNs at $z<0.82$\footnote{We include sources without optical spectroscopy but robust photometric redshift measurements from \citet{2022Salvato}.} and for those with optical spectra and significant \OIII detections. The subsample with \OIII measurements has a mean X-ray luminosity only $\sim 0.2$ dex higher, suggesting that the X-ray properties of AGNs without optical spectra are not dramatically different.

\begin{figure}
 
\centering
\includegraphics[trim={0cm 0 0cm 0cm},clip,scale=0.42]{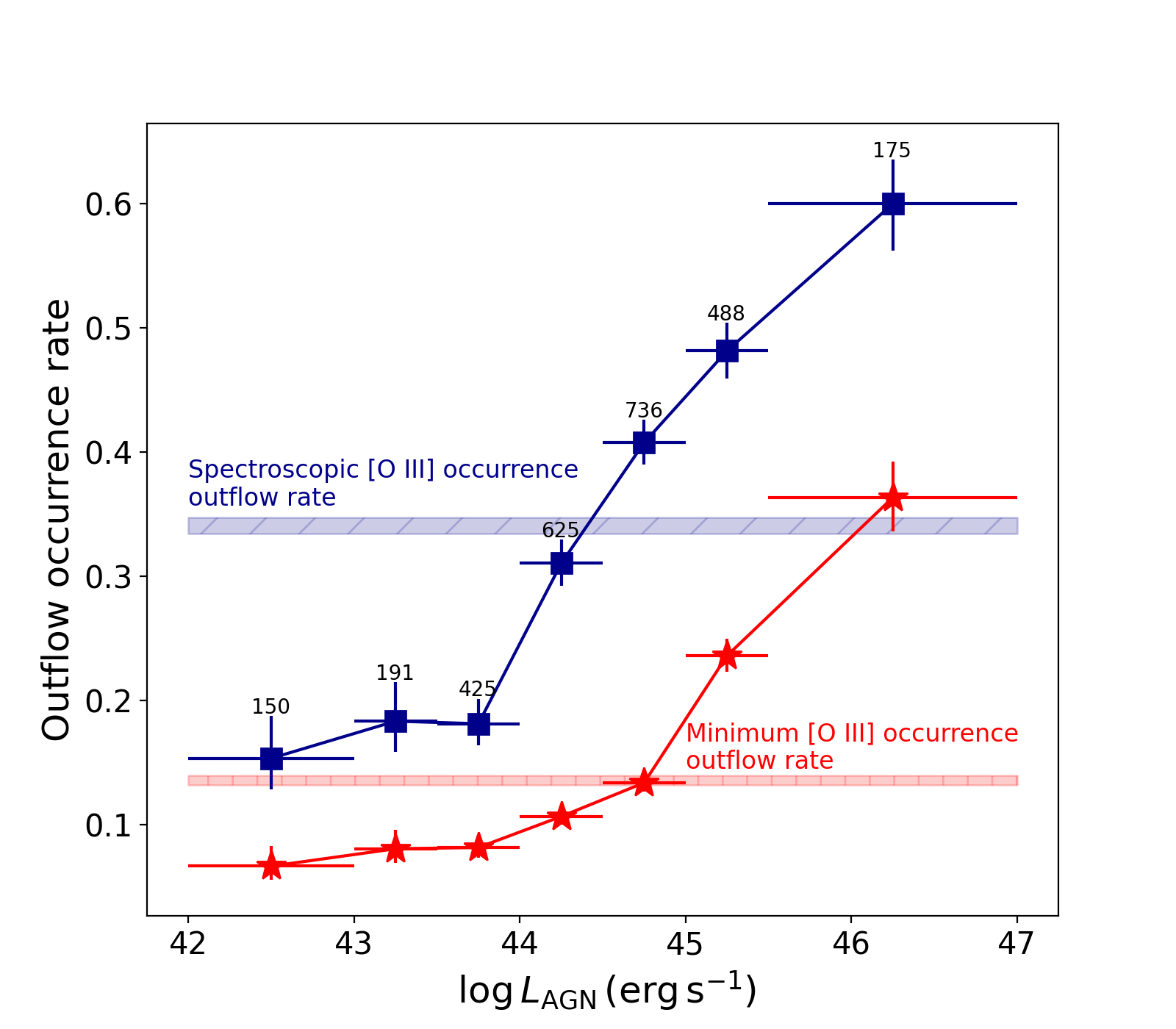}
\caption{Outflow occurrence rate as a function of bolometric AGN luminosity ($L_{\rm AGN}$) for our optical spectroscopic sample (Sample 3 in Table~\ref{t:data}; blue squares) and for the entire X-ray AGN sample (Sample 1 in Table~\ref{t:data}; red stars). For the latter, we conservatively assumed that all X-ray AGNs without optical spectra or with $(S/N)^2 < 10$ do not host outflows. An \OIII outflow is defined as one with $W_{80} > 600\,\mathrm{km\,s^{-1}}$ and a detected \OIII line by \textsc{PyQSOFit}. Horizontal error bars indicate the width of the $L_{\rm AGN}$ bin, and vertical error bars represent $1\sigma$ binomial uncertainties. The number above each data point gives the number of sources per luminosity bin. The blue-filled, diagonally hatched region marks the total \OIII outflow occurrence rate for the optical spectroscopic sample, while the red-filled, vertically hatched region shows a lower limit for the total \OIII outflow occurrence rate for the X-ray AGN sample. } 
\label{fig:OIIIrate_Lbol}
\end{figure}

We then quantify the incidence of \OIII outflows as a function of $L_{\rm AGN}$ in Figure\,\ref{fig:OIIIrate_Lbol}, assuming that sources without an \OIII detection do not host an outflow. All $L_{\rm AGN}$ bins contain at least $150$ sources. We use binomial confidence intervals to calculate the 1$\sigma$ uncertainties on the outflow occurrence rate. We find that $34 \pm 0.5\%$ of our spectroscopic sample exhibits \OIII outflows with $W_{80} > 600\,\rm km\,s^{-1}$. We also calculate a lower limit for the total outflow incidence, conservatively assuming that all X-ray AGNs without optical spectroscopy or with a $(S/N)^2<10$ lack \OIII outflows, which we find to be a $14 \pm 0.4\%$. We find that the \OIII outflow occurrence rate is relatively constant for $L_{\rm AGN}<10^{44}\rm \, erg \, s^{-1}$, ranging $\approx 15-18\%$. At $L_{\rm AGN}>10^{44}\rm \, erg \, s^{-1}$, the \OIII outflow occurrence rate rapidly increases with $L_{\rm AGN}$, reaching a maximum of $\sim 60\pm 4\%$ at $\log L_{\rm AGN}/({\rm erg \, s^{-1}}) > 45.5 $. 

To summarise, we find a mild but significant correlation between the \OIII velocity and the AGN luminosity, and a strong positive correlation between the \OIII outflow incidence and the AGN luminosity. These results indicate that AGN radiation has an important role driving \OIII outflow activity.

\subsection{Connection between  {\rm [O\,\textsc{iii}]} outflow activity and the Eddington ratio} \label{res:sub:OIII_EddRatio}

In this section, we further explore the impact of AGN radiation on the kinematic properties of the \OIII line through the Eddington ratio, $\lambda_{\rm Edd}$. This approach has two key advantages. First, it allowed us to disentangle the independent contributions of AGN luminosity and black hole mass to the observed \OIII kinematics, which is important given the observed correlation between these two quantities. Second, it enables a more direct comparison of our results with AGN radiative feedback models \citep[e.g.][]{2024Harrison}, which often make predictions in terms of the Eddington ratio rather than luminosity alone.

The Eddington ratio is a proxy for the accretion rate and corresponds to a measure of how efficiently the black hole is accreting and is defined as 

\begin{equation}
    \lambda_{\rm Edd} = \frac{L_{\rm AGN}}{L_{\rm Edd}} \propto \frac{L_{\rm AGN}}{M_{\rm BH}}, 
\end{equation}
 
\noindent where $L_{\rm AGN}$ is the AGN bolometric luminosity calculated from $L_{2-10\rm keV}$ (see Section\,\ref{sub:data:Xray}), $L_{\rm Edd}$ is the Eddington luminosity, and $M_{\rm BH}$ is the SMBH mass. The Eddington luminosity is the maximum luminosity an object can emit to maintain the equilibrium between the radiation pressure and the gravitational force, assuming spherical geometry, and is calculated as \mbox{$L_{\rm Edd}=1.36\times 10^{38} M_{\rm BH}/M_{\odot
} \rm \, erg \, s^{-1} $}. We calculated $M_{\rm BH}$ assuming virial motion of the broad H$\beta$ line and following \citet{2024Shen}

\begin{equation}
  \frac{M_{\rm BH}}{M_{\odot
}}   = 0.85 + 0.5 \log \left( \frac{L_{5100}}{10^{44} \rm erg \, s^{-1}} \right) + 2\log \left(  \frac{\rm FWHM}{\rm km \, s^{-1}} \right),
\end{equation}
  
\noindent where $L_{5100}$ is the AGN luminosity at 5100\AA\, and FWHM is the full width at half maximum of the broad component of the  $\rm H\beta$ line, calculated using \textsc{PyQSOFit} (see Section\,\ref{sub:pyqsofit}). To reliably constrain the black hole mass, we also need a robust broad H$\beta$ line measurement; hence, we select the sources with $\rm EW(H\beta_{br})/\sigma_{\rm EW(H\beta_{br})}>2$, and $\rm FWHM(H\beta_{br})>1500 \rm \, km \, s^{-1}$ a following \citet{2022Wu}. This cut excludes type II AGNs and type I AGNs where $\rm H\beta_{br}$ is not well-constrained. The broad H$\beta$ FWHMs for our sources range $\rm FWHM(H\beta_{br})\approx 1500-16000 \, km\, s^{-1}$, with a median value of $3500\rm \, km \, s^{-1}$, consistent with measurements reported in the literature for SDSS quasars at similar redshifts \citep[e.g.][]{2020Rakshit, 2022Wu}. Only $\sim 15\%$ of the sample has a $\rm FWHM(H\beta_{br})>7000\, \rm km \, s^{-1}$, indicating that extreme objects are a minority in our sample. The distributions of the black hole masses and Eddington ratios are shown in Figure\,\ref{fig:ap:deltaMBH_ER} in Appendix\,\ref{ap:distMBH_ER}. The black hole masses of the sample range \mbox{$\log M_{\rm BH}/M_{\odot}\approx 6-9.8$}, with a median uncertainty of $\sim 0.2 \rm \, dex$, and the median and standard deviation of the distribution are $\log M_{\rm BH}/M_{\odot}= 8.02\pm0.7$. The $\lambda_{\rm Edd}$ values of our sample range \mbox{$\log \lambda_{\rm Edd}\approx [-4, 1]$}, with a median uncertainty of $\sim 0.2 \rm \, dex$, and the median and standard deviation of the distribution are \mbox{$\log \lambda_{\rm Edd}\approx -1.34\pm 0.6$}.

\begin{figure}
 
\centering
\includegraphics[trim={0cm 0 0cm 0cm},clip,scale=0.42]{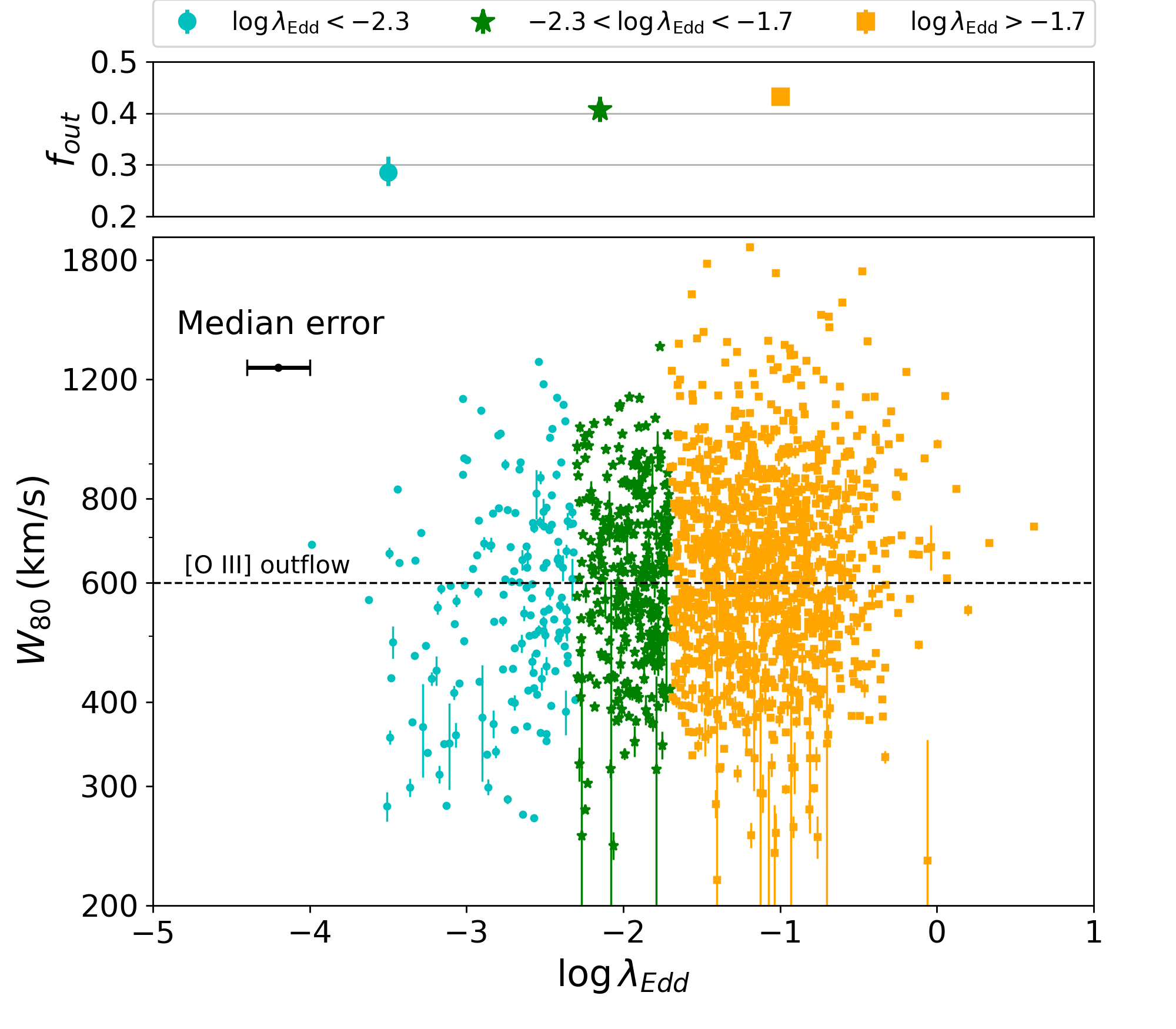}
\includegraphics[trim={0cm 0 0cm 0cm},clip,scale=0.42]{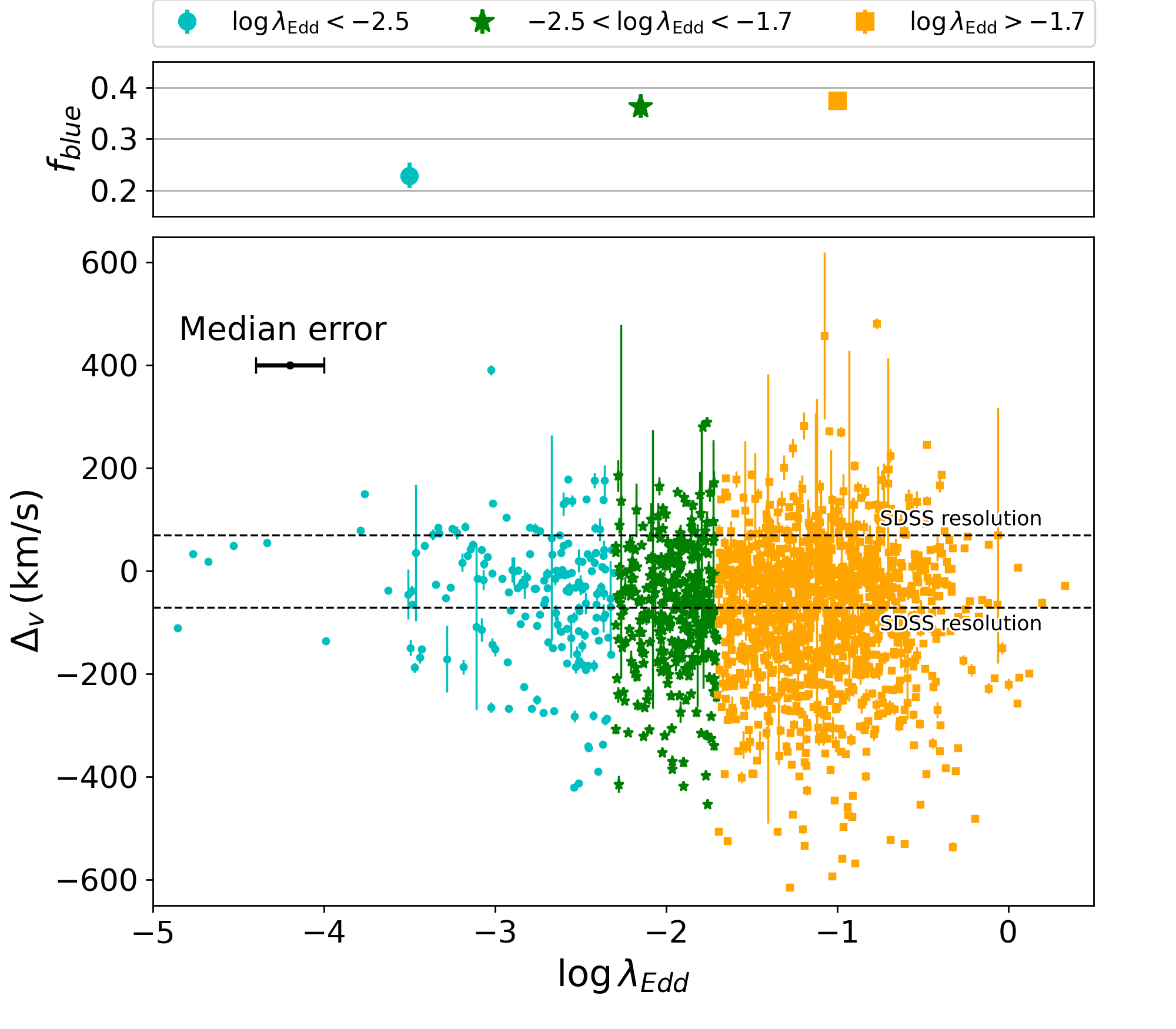}
\caption{{\it Top}: Non-parametric \OIII velocity $W_{80}$ against the logarithmic Eddington ratio $\log \lambda_{\rm Edd}$. The plot shows three samples with different Eddington ratios: $\log \lambda_{\rm Edd}<-2.3$ (cyan circles),  $-2.3<\log \lambda_{\rm Edd}<-1.7$ (green stars), and  $\log \lambda_{\rm Edd}>-1.7$ (orange squares). The horizontal dotted line at $W_{80}=600 \, \rm km \, s^{-1}$ shows the threshold typically used to identify outflows. The top panel of the figure displays the outflow occurrence rate in each Eddington ratio sample. The uncertainties, which are calculated with binomial errors, are too small to be visible. {\it Bottom}: Non-parametric \OIII asymmetry $\Delta v$ against $\log \lambda_{\rm Edd}$. The horizontal dotted lines correspond to the SDSS resolution of $\sigma\sim 70 \rm \,km\,s^{-1}$. The top panel of the figure displays the fraction of blueshifted outflows (i.e., $\Delta v<70 \, \rm km \, s^{-1}$) in each Eddington ratio sample. Similarly to $f_{out}$, the uncertainties are too small to be visible. } 
\label{fig:W80_EddRatio_nomatched}
\end{figure}

To investigate the impact of the $\lambda_{\rm Edd}$ on the \OIII kinematics, we need a reliable detection of both \OIII and the broad H$\beta$, which leads to a sample of 1573 sources (55\% of the spectroscopic sample; see Table\,\ref{t:data}). We first compare the \OIII kinematic parameters, $W_{80}$ and $\Delta v$, with the Eddington ratio. Figure\,\ref{fig:W80_EddRatio_nomatched} shows the results; the top and bottom panels show $W_{80}$ and $\Delta v$ versus $\lambda_{\rm Edd}$, respectively. We find that $W_{80}$ increases with $\lambda_{\rm Edd}$, while $\Delta v$ becomes more negative with increasing $\lambda_{\rm Edd}$, indicating that the \OIII profiles become broader and more asymmetric toward the blue at higher accretion rates. To further quantify these trends, we calculated the outflow fraction, $f_{out}$, and the blueshifted fraction, $f_{blue}$, in three $\lambda_{\rm Edd}$ bins: low ($\log \lambda_{\rm Edd}<-2.3$), medium ($-2.3<\log \lambda_{\rm Edd}<-1.7$), and high ($\log \lambda_{\rm Edd}>-1.7$). We picked these values as at $\log \lambda_{\rm Edd}\lesssim -2.3$ the accretion disk may start being radiatively inefficient \citep[e.g.][]{2009Ho,2024Cappelluti}, while at $\log \lambda_{\rm Edd}\gtrsim -1.7$ the accretion flow is radiatively efficient \citep[e.g.][]{2013Qiao, 2019Ruan}. We define \OIII profiles as blueshifted when $ \Delta v < -70\,\rm km\,s^{-1}$, which corresponds to the SDSS instrumental velocity resolution at $\lambda=5008$\,\AA\, \citep{2013Smee}. Values of $|\Delta v| < 70\,\rm km\,s^{-1}$ are therefore unresolved at this spectral resolution and cannot be confidently identified as representing a velocity shift. As in the previous section, to consider a source blueshifted or outflowing, an \OIII detection is also required. Both the outflow and blueshifted fractions, $f_{out}$ and $f_{blue}$, respectively, increase with Eddington ratio, as shown in the upper sub-panels of Figure\,\ref{fig:W80_EddRatio_nomatched}. The low-$\lambda_{\rm Edd}$ sample has $f_{out}=26\pm 3\%$ and $f_{blue}=23\pm 3 \%$, the medium-$\lambda_{\rm Edd}$ sample has $f_{out}=39\pm 2\%$ and $f_{blue}=36\pm 2 \%$, and the high-$\lambda_{\rm Edd}$ sample has $f_{out}=42\pm 1\%$ and $f_{blue}=37\pm 1 \%$. We note that the blueshifted outflow fraction changes significantly only between the low- and medium-Eddington ratio samples. We also note that the fraction of redshifted sources ($\Delta v > 70\,\rm km\,s^{-1}$) is low, around 8\%, and does not vary significantly with Eddington ratio. A summary of the properties of the three samples is reported in Table\,\ref{t:results}.

\begin{table*}
\centering
\caption{Properties of the \OIII line kinematics and obscuration of our sample for different Eddington ratio regimes: low ($\log \lambda_{\rm Edd}<-2.3$), medium ($-2.3<\log \lambda_{\rm Edd}<-1.7$), and high ($\log \lambda_{\rm Edd}>-1.7$). }
\scalebox{0.82}{

\begin{tabular}{l|l|c|c|c|c|c|c|c|c|c }
 \hline
 \hline
\noalign{\smallskip}
 & Sample & z & $\log M_{\rm BH}$& $\log L_{\rm AGN}$& $f_{out}$ & $f_{blue}$  &  $\left<W_{80} \right>$ &  $\left<\Delta v \right>$  & $\left< \log N_{\rm H} \right>$ & $f_{\rm obs,21}$ \\

 &  &  & $ (M_{\odot})$& $\rm (erg\, s^{-1})$&  &   & ($\rm km\, s^{-1}$) &  ($\rm km\, s^{-1}$) & $\rm (cm^{-2})$ & \\

(1) & (2) & (3) & (4) & (5) & (6) & (7) & (8) & (9) & (10) & (11)\\
\noalign{\smallskip}

\hline
\hline
\noalign{\smallskip}
 &  Low-$\lambda_{\rm Edd}$ & $0.29\pm0.2$& $8.51\pm 0.86$& $43.8\pm 1$ & $0.26\pm0.03$ & $0.23 \pm 0.03$ & $554\pm 211$ & $-32\pm 124$& $20.6\pm0.5$ & $0.24\pm 0.03$ \\
All &   Medium-$\lambda_{\rm Edd}$ & $0.44\pm 0.2$ &$8.27\pm 0.7$ & $44.4\pm 0.7$& $0.39 \pm 0.02$ &$0.36 \pm 0.02$ & $614\pm 195$ & $-64\pm 125$& $20.5\pm0.5$ & $0.17\pm 0.02$ \\
 &  High-$\lambda_{\rm Edd}$ & $0.47\pm 0.2$ & $7.7\pm 0.7$& $44.8\pm 0.6$& $0.42 \pm 0.01$ &$0.37 \pm 0.01$ & $627\pm 225$ & $-66\pm 136$&  $20.3\pm0.5$ & $0.13\pm 0.01$ \\
\noalign{\medskip}
\hline
\noalign{\medskip}

  &  Low-$\lambda_{\rm Edd}$ & $0.32\pm 0.02$ & $8.29\pm 0.08$ & $43.8\pm 0.09$  & $0.29 \pm 0.003$ & $0.27\pm 0.003$ & $577\pm 22$ & $-40\pm 16$& $20.8\pm 0.05$  & $0.22\pm 0.003$ \\
$M_{\rm BH}-z$  &  Medium-$\lambda_{\rm Edd}$ & $0.34\pm0.02$& $8.23\pm 0.08$& $44.4\pm0.08 $ & $0.4 \pm 0.004$ & $0.35 \pm 0.003$ & $617\pm 20$ & $-52\pm 14$ & $20.6\pm 0.02$ & $0.15\pm 0.003$ \\
matched &  High-$\lambda_{\rm Edd}$& $0.35\pm0.02$& $8.18\pm 0.08$& $45\pm 0.08$ & $0.49 \pm 0.004$ &$0.40 \pm 0.004$ & $686\pm 24$ & $-78\pm 16$& $20.4\pm 0.06$ & $0.10\pm 0.002$ \\

\noalign{\medskip}
\hline
\noalign{\medskip}

  &  Low-$\lambda_{\rm Edd}$ & $0.34\pm 0.02$& $8.7\pm 0.05$& $44.09\pm0.07$ & $0.35 \pm 0.004$ &$0.18 \pm 0.003$ & $611\pm 21$ & $-48\pm 15$& $20.8\pm 0.05$ & $0.25\pm 0.003$\\
$L_{\rm AGN}-z$  &  Medium-$\lambda_{\rm Edd}$ & $0.34\pm 0.02$& $8\pm 0.07$& $44.12\pm0.06$ & $0.34 \pm 0.004$ & $0.18 \pm 0.003$ & $595\pm 19$ & $-49\pm 13$ & $20.7\pm 0.05$ & $0.19\pm 0.003$\\
 matched &  High-$\lambda_{\rm Edd}$& $0.33\pm 0.02$& $7.2\pm 0.5$& $44.13\pm 0.07$ & $0.31 \pm 0.003$ &$0.18 \pm 0.002$ & $608\pm 22$ & $-66\pm 15$ & $20.7\pm 0.05$ & $0.2\pm 0.002$\\
\noalign{\smallskip}

\hline
\hline

\end{tabular}}
\tablefoot{Column description: (1) Sample description: entire sample or the three Eddington ratio bins samples matched by $M_{\rm BH}-z$ or $L_{\rm AG}-z$; (2) Eddington ratio regime; (3) mean and standard deviation of the redshift of the sample; (4) mean and standard deviation of the black hole mass of the sample; (5) mean and standard deviation of the AGN bolometric luminosity of the sample; (6)  fraction of the sample with \OIII outflows with $W_{80}>600 \,\rm km \, s^{-1}$; (7) fraction of the sample with blueshifted \OIII outflows with $\Delta v<-70 \,\rm km \, s^{-1}$; (8) mean and standard deviation of the \OIII velocity $W_{80}$; (9) mean and standard deviation of the \OIII asymmetry $\Delta v$; (10) mean and standard deviation of the column density of the sample using single measurements. }\label{t:results}

\end{table*}

\begin{figure}
 
\centering
\includegraphics[trim={0cm 0 0cm 0cm},clip,scale=0.42]{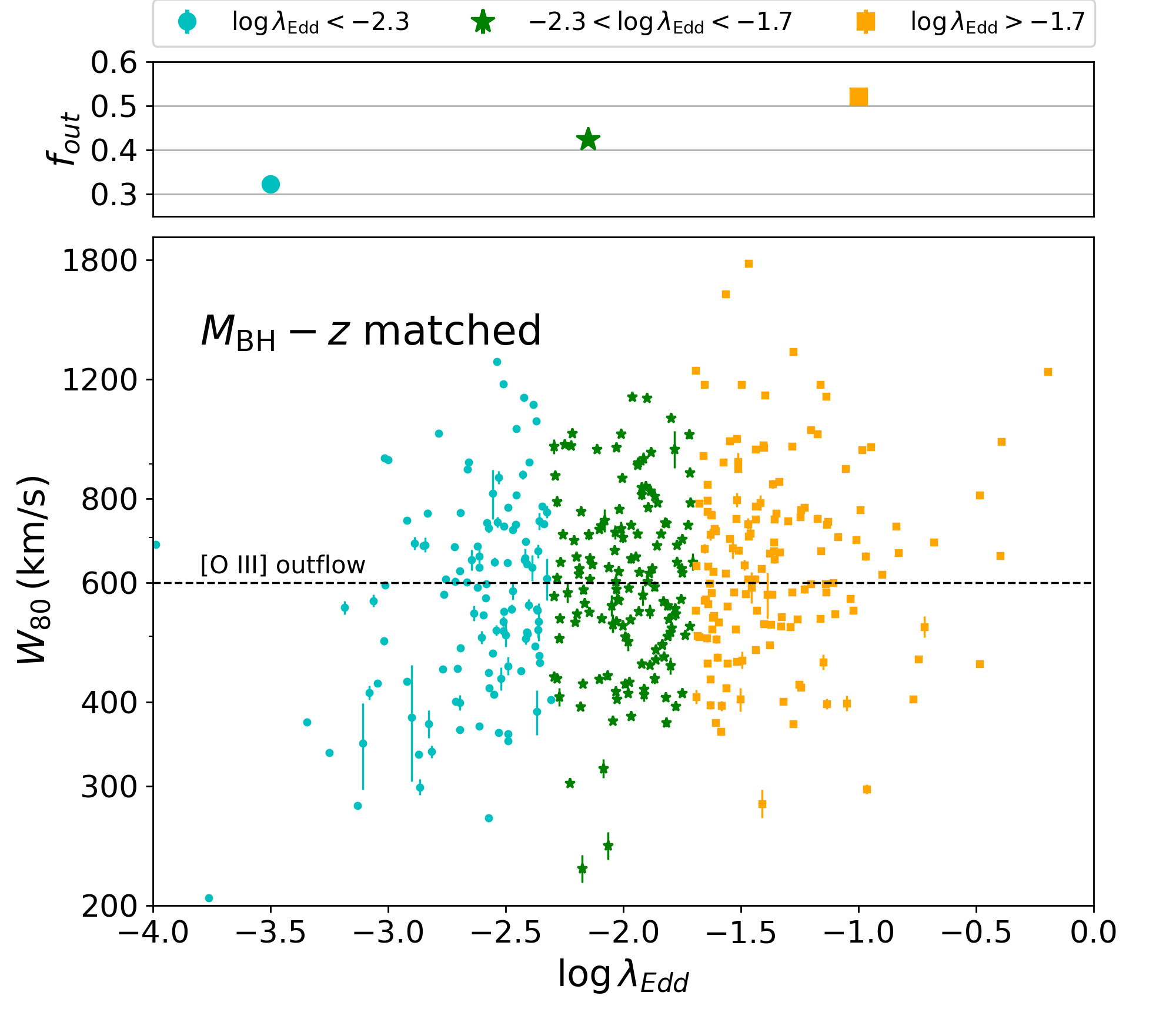}
\includegraphics[trim={0cm 0 0cm 0cm},clip,scale=0.42]{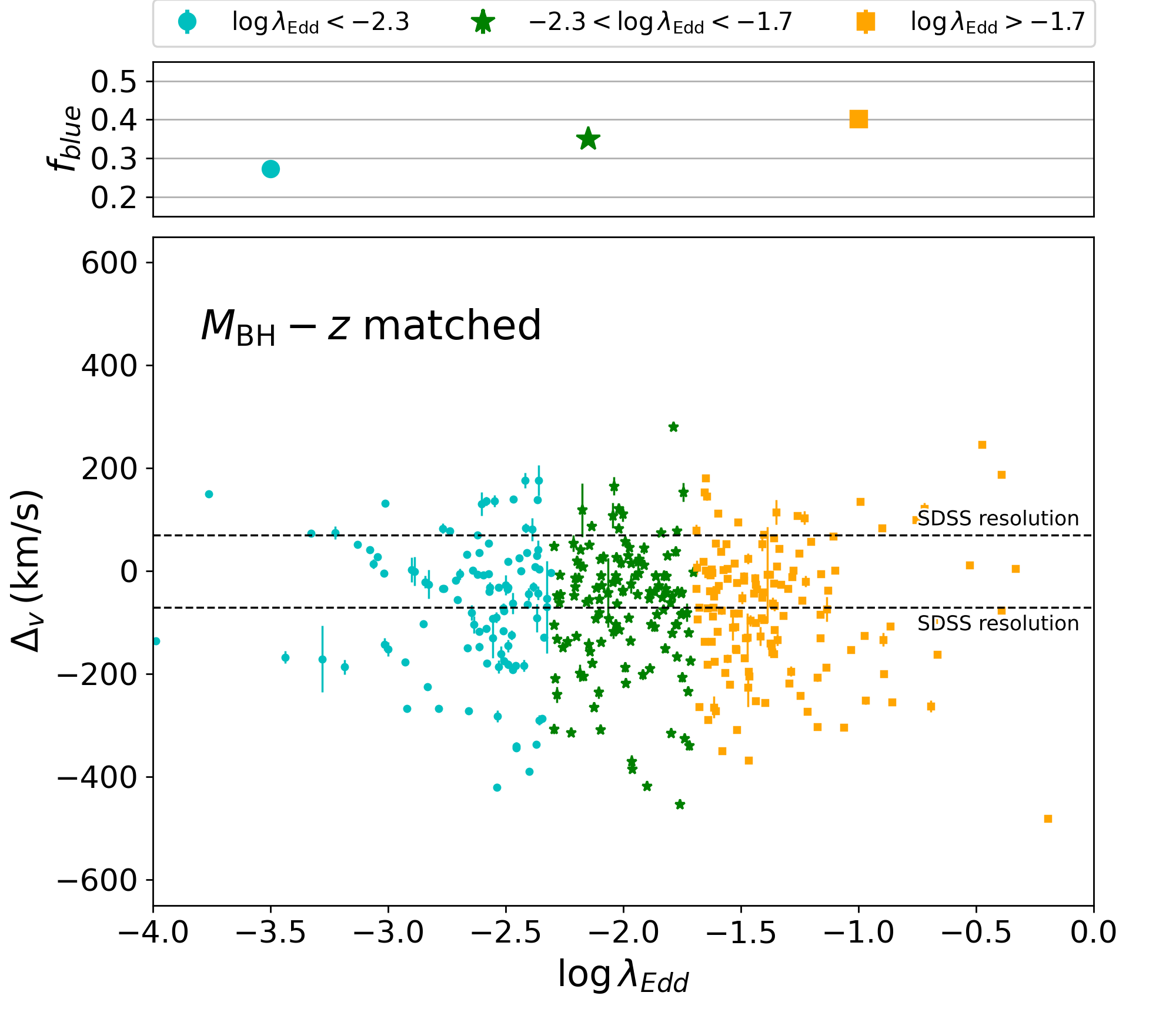}
\caption{Same as Figure\,\ref{fig:W80_EddRatio_nomatched}, but the three Eddington ratios samples are matched in $M_{\rm BH}$ and redshift. } 
\label{fig:W80_EddRatio_MBHmatched}
\end{figure}

As the $\lambda_{\rm Edd}$ is a parameter that depends on the AGN luminosity and black hole mass, the observed correlation between $\lambda_{\rm Edd}$ and the \OIII line kinematics could be primarily driven by $L_{\rm AGN}$ or $M_{\rm BH}$ (or by a combination of the two). To address this problem, we match the three $\lambda_{\rm Edd}$ samples in $M_{\rm BH}$ and redshift, thereby comparing samples with similar physical properties. For this, we randomly pick sources in the three samples that lie within $\Delta \log M_{\rm BH}=0.3 \rm \, dex$ and $\Delta z=0.06\times (1+z)$, resulting in three equally populated samples containing $\sim 170$ sources each. As the matching process has a random factor, we repeat it 100 times. In each iteration, we calculate the outflow fraction $f_{out}$, the blueshifted outflow fraction $f_{blue}$, the mean $W_{80}$ and $\Delta v$ parameters, and we perform Anderson-Darling (AD) tests between the kinematic properties of the samples to determine whether they are statistically different. The results are shown in Figure\,\ref{fig:W80_EddRatio_MBHmatched}. As before, we find that the outflow occurrence rate systematically increases with the Eddington ratio, rising from $f_{out}=29\pm 0.3$ for the low-Eddington-ratio sample to $f_{out}=49\pm 0.4$ for the high-Eddington-ratio sample. The blueshifted fraction has the same behaviour as in the unmatched sample: it increases from $f_{blue}=27\pm 0.3$ for the low-$\lambda_{\rm Edd}$ sample to $f_{blue}=40\pm 0.4$ for the high-$\lambda_{\rm Edd}$ sample.

\begin{figure}
 
\centering
\includegraphics[trim={1.5cm 1.5cm 1.5cm 0cm},clip,scale=0.32]{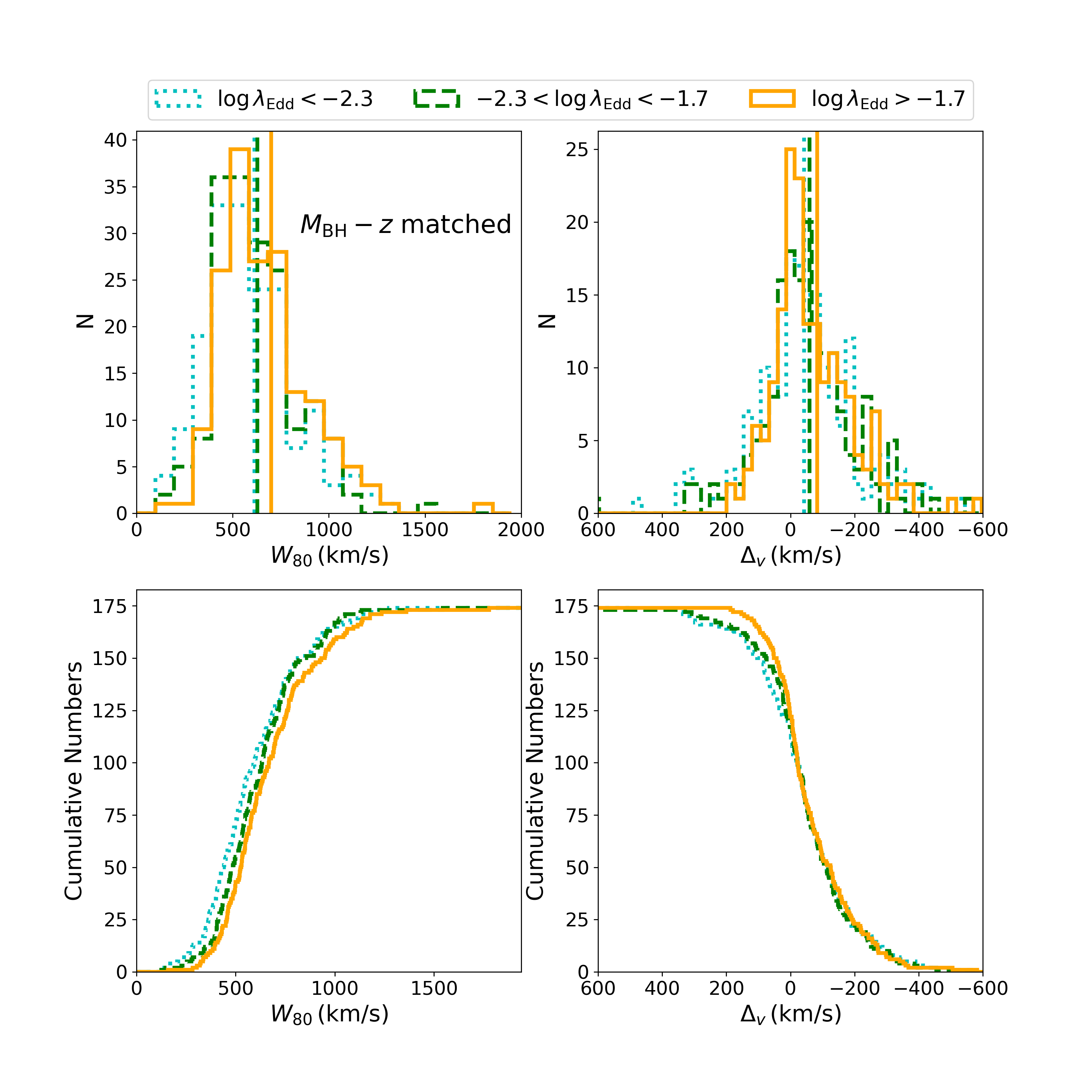}
\caption{{\it Top panels}: Distributions of the non-parametric velocity ($W_{80}$) and asymmetry ($\Delta v$) of the same sample $\lambda_{\rm Edd}$ $M_{\rm BH}-z$ matched samples plotted in Figure\,\ref{fig:W80_EddRatio_MBHmatched}. {\it Bottom panels}: Cumulative distributions of $W_{80}$ and $\Delta v$. } 
\label{fig:W80dVdist_EddRatio_MBHmatched}
\end{figure}

We show the distributions of $W_{80}$ and $\Delta v$ for the $M_{\rm BH}-z$ matched samples in Figure\,\ref{fig:W80dVdist_EddRatio_MBHmatched}. It is clear from the histograms (top panels) and cumulative distributions (bottom panels) that the $W_{80}$ distributions systematically shift towards higher velocities with increasing Eddington ratio. The median \OIII velocity values increase from \mbox{$\left<W_{80} \right>=577\pm22 \, \rm km\, s^{-1}$} for the low-$\lambda_{\rm Edd}$ sample to \mbox{$\left<W_{80} \right>=686\pm24 \, \rm km\, s^{-1}$} for the higher $\lambda_{\rm Edd}$ sample. The AD test among the samples indicates that the $W_{80}$ distribution of the low-$\lambda_{\rm Edd}$ sample is statistically different from the medium- and high-$\lambda_{\rm Edd}$ samples, with $\rm p{-}value=0.04$ and $\rm p{-}value=0.001$, respectively.

We then analyse the $\Delta v$ distributions, which look much more similar, with a very mild shift towards negative velocities with increasing Eddington ratio. The median $\Delta v$ values do not represent the true asymmetry of each sample, as there are positive and negative values. However, the comparison of the median $\Delta v$ between samples is insightful.  We observe a larger difference between the properties of the low- and high-$\lambda_{\rm Edd}$ samples, with median \mbox{$\left<\Delta v \right> = -40 \pm 16 \, \rm km\, s^{-1}$} and \mbox{$\left<\Delta v \right> = -78 \pm 16 \, \rm km\, s^{-1}$}, respectively, and a $\rm p{-}value=0.05$, suggesting that the two distributions are different. However, these differences are less significant than the ones observed in the $W_{80}$ distributions. The $\Delta v$ distributions of the low- and medium-$\lambda_{\rm Edd}$ samples do not show major differences, with a $\rm p{-}value=0.2$. A summary of the median $W_{80}$ and $\Delta v$ of each distribution is reported in Table\,\ref{t:results}.

\begin{figure}
 
\centering
\includegraphics[trim={0cm 0 0cm 0cm},clip,scale=0.4]{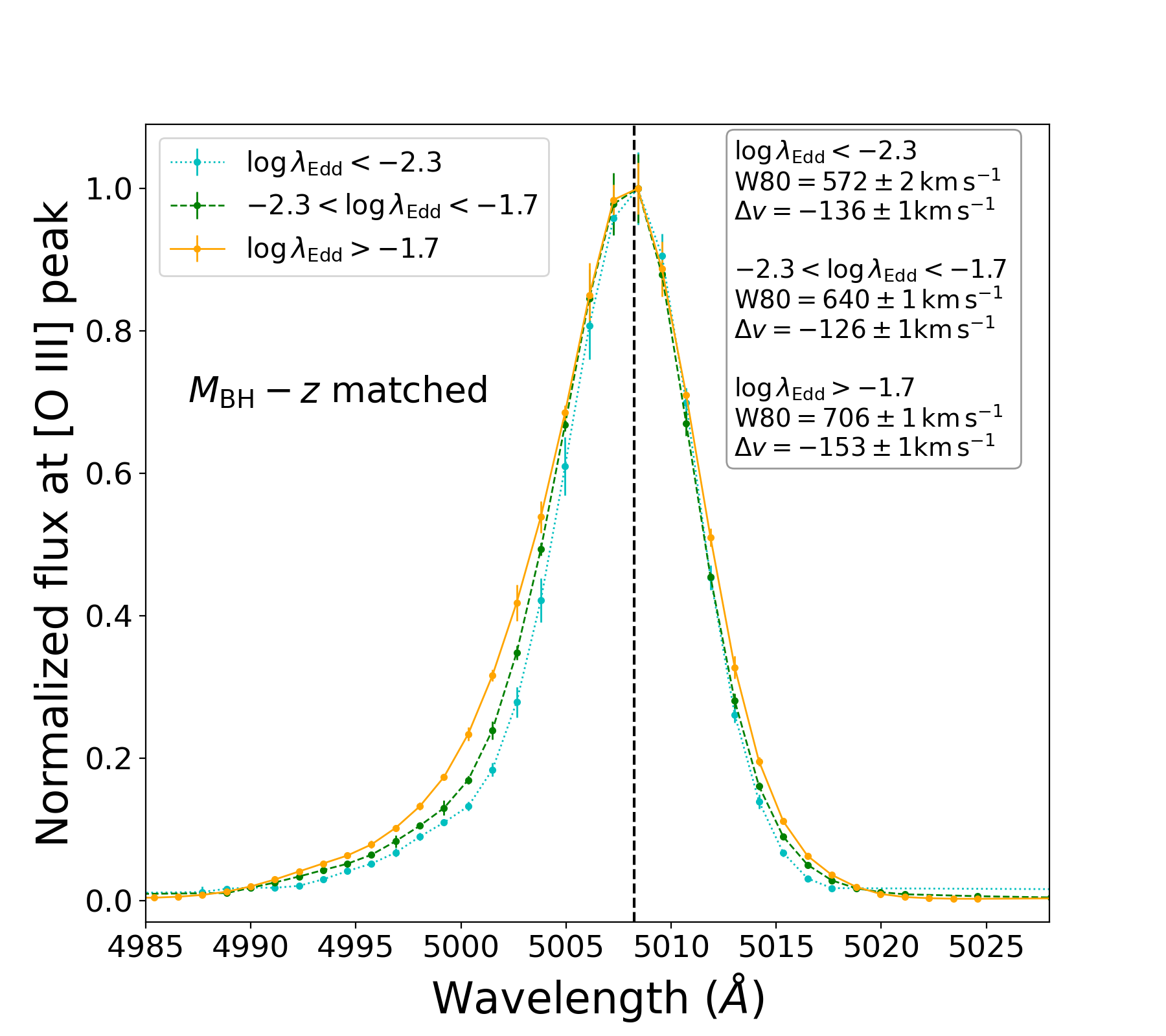}
\caption{Stacked \OIII spectra for the same Eddington ratio matched samples as in Figure\,\ref{fig:W80_EddRatio_MBHmatched}. The stacks have been normalised to their maximum flux to highlight differences in the \OIII line profiles across the Eddington ratio bins. The central dotted black line marks the \OIII line peak at $5008.24$\,\AA.} 
 
\label{fig:OIIIstacks_EddRatio_MBHmatched}
\end{figure}

To further confirm the differences in the \OIII profiles among the $\lambda_{\rm Edd}$ samples, we stacked the \OIII optical spectra for each $M_{\rm BH}-z$ matched sample following the procedure described in Section\,\ref{sub:opt_stacking}. Spectral stacking significantly enhances the signal-to-noise ratio, enabling the detection of extended components in the \OIII profile that may be challenging to identify in individual spectra \citep[e.g.][]{2017Perna}. As shown in Figure~\ref{fig:OIIIstacks_EddRatio_MBHmatched}, the stacked \OIII profiles become progressively broader with increasing $\lambda_{\rm Edd}$. The measured kinematic parameters for the stacks are $W_{80} = 572 \pm 2\,\rm km\,s^{-1}$ and $\Delta v = -136 \pm 1\,\rm km\,s^{-1}$ for the low-$\lambda_{\rm Edd}$ sample; $W_{80} = 640 \pm 1\,\rm km\,s^{-1}$ and $\Delta v = -126 \pm 1\,\rm km\,s^{-1}$ for the medium $\lambda_{\rm Edd}$ sample; and $W_{80} = 706 \pm 1\,\rm km\,s^{-1}$ and $\Delta v = -153 \pm 1\,\rm km\,s^{-1}$ for the high-$\lambda_{\rm Edd}$ sample. The stacked $W_{80}$ values are consistent within $1\sigma$ with the medians obtained from individual sources. Regarding the asymmetry, we find that the highest $\lambda_{\rm Edd}$ stack shows the largest blueshift. Overall, the stacks confirm the single-object results: the outflows traced by the \OIII emission line become stronger with Eddington ratio.

Finally, we repeated the analysis, this time matching the samples by AGN bolometric luminosity and redshift. In this case, the black hole mass distributions differ across the three $\lambda_{\rm Edd}$ bins. The corresponding results are presented in Appendix~\ref{ap:oIII_ER} (Figures\,\ref{fig:W80dV_EddRatio_Lbolmatched}, \ref{fig:W80dVdist_EddRatio_Lbolmatched} and \ref{fig:OIIIstacks_EddRatio_LAGNmatched}) and in Table~\ref{t:results}. Unlike our previous findings when matching by $M_{\rm BH}$, we detect no variation in the \OIII outflow properties with Eddington ratio. The outflow fraction, blueshifted fraction, and mean $W_{80}$ and $\Delta v$ remain essentially constant across all $\lambda_{\rm Edd}$ bins. The stacked \OIII profiles for the different bins (see Figure\,\ref{fig:OIIIstacks_EddRatio_LAGNmatched}) likewise show no visible differences, reinforcing this result. This analysis implies that large-scale outflows are not solely determined by the Eddington ratio; a high accretion rate must be accompanied by high AGN luminosity for radiation pressure to play a dominant role in driving ionised outflows. Nevertheless, we still measure a significant outflow fraction ($f_{\rm out} \approx 10$–20\%) even at very low $\lambda_{\rm Edd}$ and $L_{\rm AGN}$ (see Figures\,\ref{fig:OIIIrate_Lbol}, \ref{fig:W80_EddRatio_nomatched}, and \ref{fig:W80_EddRatio_MBHmatched}). This suggests that at low luminosities and accretion rates, other physical mechanisms are dominant to power large-scale outflows, a point explored further in Section~\ref{sec:discussion}.

The results of this section indicate that the \OIII outflow activity increases with $\lambda_{\rm Edd}$ only when the AGN luminosity increases, indicating that AGN radiation is a key driver of large-scale outflows in AGNs.

\subsection{Connection between $N_{\rm H}$ and Eddington ratio}\label{res:sub:NH_EddRatio}

In this section, we investigate the impact of the $\lambda_{\rm Edd}$ in the AGN obscuration as traced by the hydrogen column density, $N_{\rm H}$, measured from the X-ray spectra. In the AGN radiative feedback scenario, at very high Eddington ratios, the radiation pressure acting on the circumnuclear dusty gas is sufficient to blow it away \citep[e.g.][]{2009Fabian, 2018Ishibashi}. Therefore, we expect to observe a reduction of the $N_{\rm H}$ with the Eddington ratio, as shown in previous works \citep[e.g.][]{2014Merloni,2017Ricci,2024Laloux}. Here, we test this scenario using the eFEDS AGN sample that has a broad H$\beta$ detection (2303 sources; see Table\,\ref{t:data}), allowing us to estimate the black hole mass, and consequently, the Eddington ratio. As described in Section\,\ref{sub:data:Xray}, we adopted the $N_{\rm H}$ measurements derived for the eFEDS sample in \citet{2022Liu}, to which we refer the reader for a comprehensive characterisation of the X-ray spectral properties of these sources.

\begin{figure}
 
\centering

\includegraphics[trim={0cm 0cm 0cm 0cm},clip,scale=0.4]{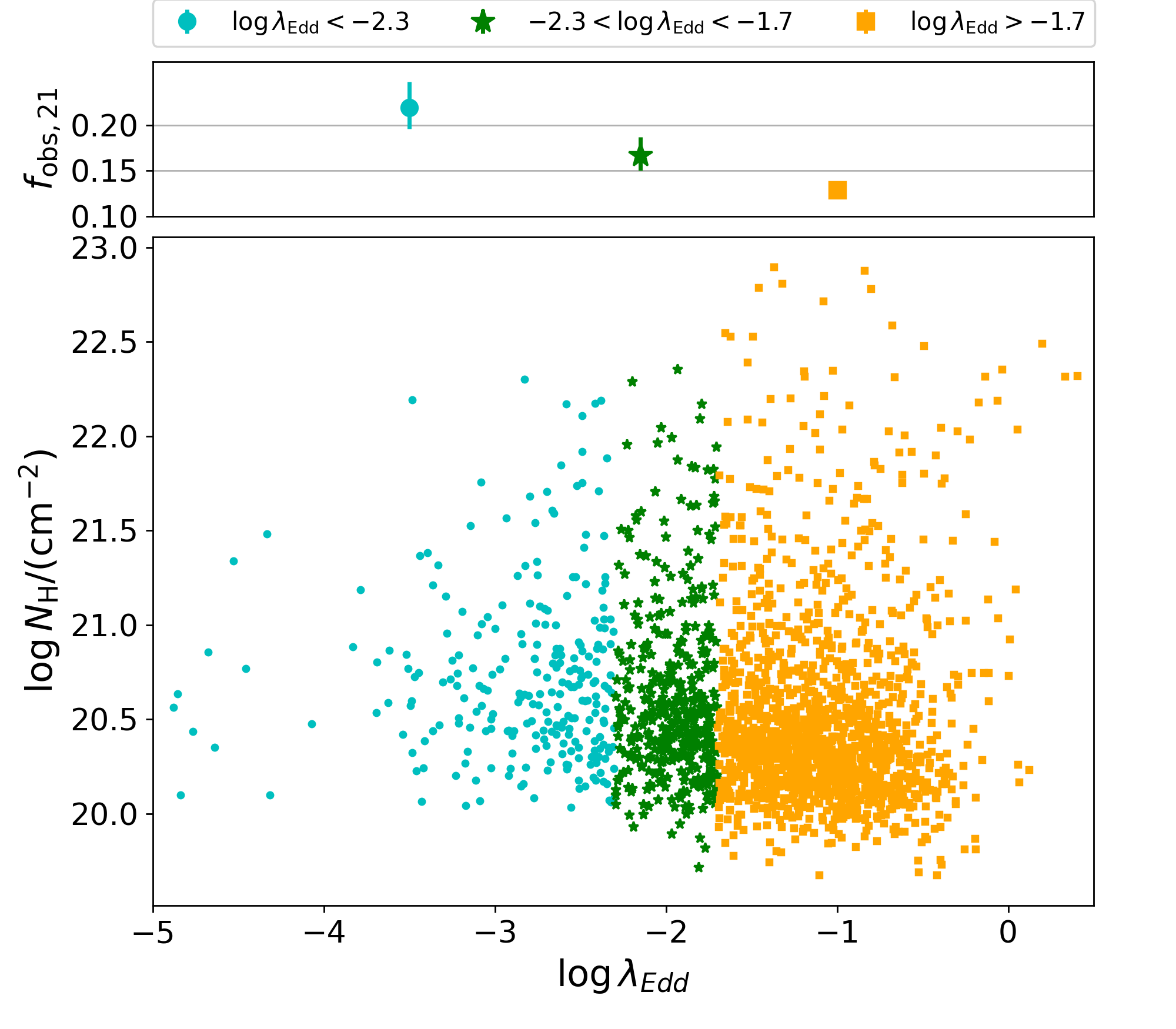}

\caption{Line of sight column density $N_{H}$ against Eddington ratio  $\log \lambda_{\rm Edd}$. The plot shows three samples with different Eddington ratios: $\log \lambda_{\rm Edd}<-2.3$ (cyan circles),  $-2.3<\log \lambda_{\rm Edd}<-1.7$ (green stars), and  $\log \lambda_{\rm Edd}>-1.7$ (orange squares). The top panel of the figure displays the fraction of sources with $\log N_{\rm H}/{\rm cm^{-2}}>21$, $f_{\rm obs,\,21}$, in each Eddington ratio sample. The error bars correspond to 1$\sigma$ uncertainties calculated with binomial errors.}
\label{fig:NH_ER_3ERbins}
\end{figure}

\begin{figure}
 
\centering

\includegraphics[trim={2cm 20 2cm 2cm},clip,scale=0.32]{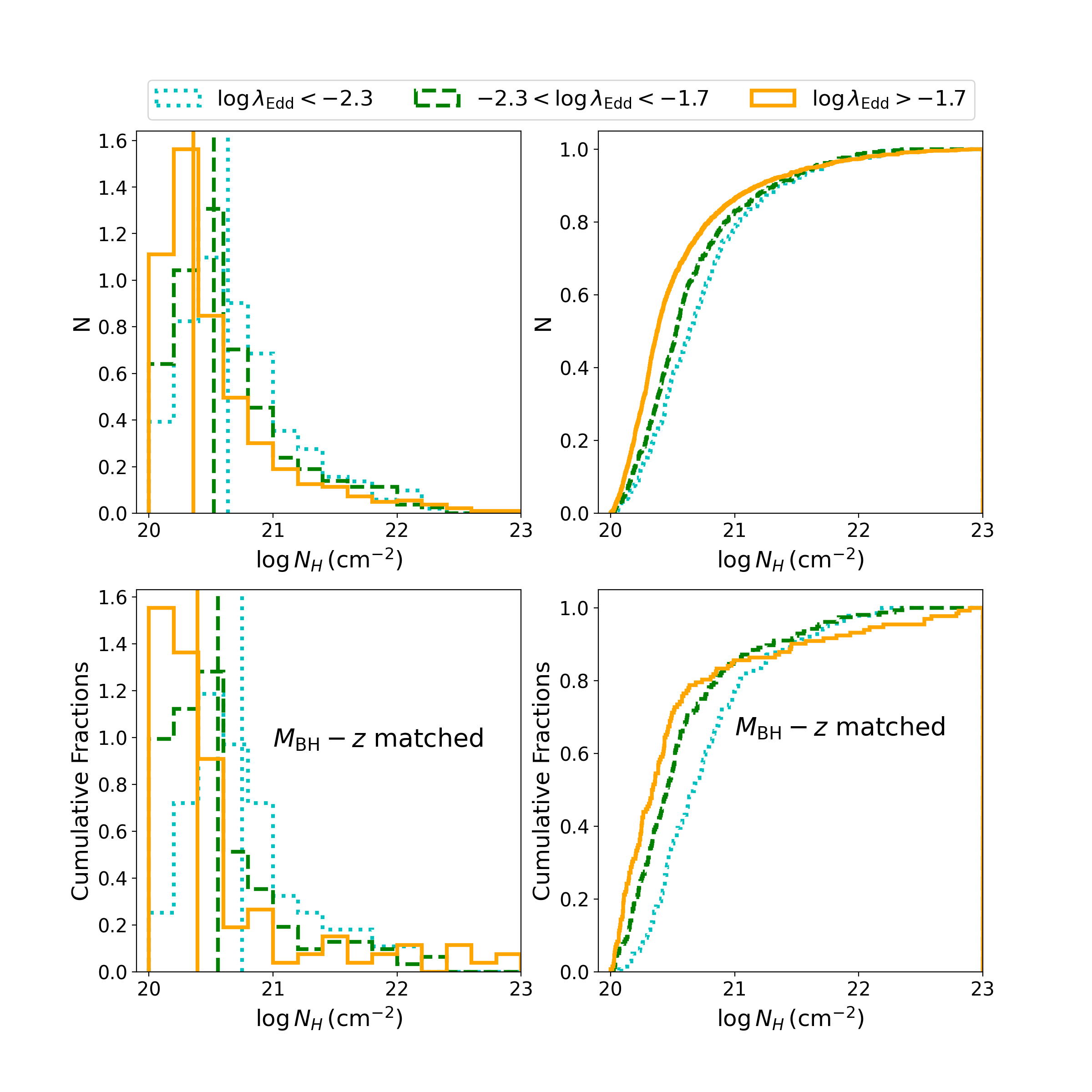}

\caption{{\it Top panels:} Distributions (left) and cumulative distributions (right) of the line of sight column densities $N_{H}$ of sources with $\log \lambda_{\rm Edd}<-2.3$ (cyan dotted),  $-2.3<\log \lambda_{\rm Edd}<-1.7$ (green dashed), and  $\log \lambda_{\rm Edd}>-1.7$ (orange solid). The vertical lines indicate the mean $N_{\rm H}$ of each $\lambda_{\rm Edd}$ sample. {\it Bottom panels:} Same as the top panels, but the Eddington ratio samples are matched in $M_{\rm BH}$ and redshift.}
\label{fig:NH_dist_3ERbins}
\end{figure}

Figure\,\ref{fig:NH_ER_3ERbins} displays $N_{\rm H}$ as a function of $\log \lambda_{\rm Edd}$, dividing the sample into the same Eddington ratio bins as in the previous section: low ($\log \lambda_{\rm Edd} < -2.3$), medium ($-2.3 < \log \lambda_{\rm Edd} < -1.7$), and high ($\log \lambda_{\rm Edd} > -1.7$). Only sources with measured $N_{\rm H}$ are included (see Section\,\ref{sub:data:Xray}). We observe a clear trend of decreasing $N_{\rm H}$ with increasing $\lambda_{\rm Edd}$, which is even more evident in the $N_{\rm H}$ distributions and cumulative distributions shown in the top panels of Figure\,\ref{fig:NH_dist_3ERbins}. Both the mean column density, $\left< N_{\rm H} \right>$, and the fraction of sources with \mbox{$\log (N_{\rm H}/{\rm cm^{-2}}) > 21$} ($f_{\rm obs,21}$), decline as $\lambda_{\rm Edd}$ increases: for $\log \lambda_{\rm Edd} < -2.3$, we find \mbox{$\left<\log  (N_{\rm H}/{\rm cm^{-2}}) \right> = 20.6 \pm 0.5$} and \mbox{$f_{\rm obs,21}=24\pm3\%$}; for \mbox{$-2.3 < \log \lambda_{\rm Edd} < -1.7$}, \mbox{$\left<\log  (N_{\rm H}/{\rm cm^{-2}}) \right> = 20.5 \pm 0.5$} and $f_{\rm obs,21}=17\pm2\%$; and for $\log \lambda_{\rm Edd} > -1.7$, \mbox{$\left<\log  (N_{\rm H}/{\rm cm^{-2}}) \right>=20.3 \pm 0.5$} and \mbox{$f_{\rm obs,21}=13\pm1\%$}. The uncertainties on $f_{\rm obs}$ correspond to 1$\sigma$ binomial errors. To ensure a fair comparison, we match the three samples in $M_{\rm BH}$ and redshift, repeating the matching 100 times. In each iteration, we compute $\left< N_{\rm H} \right>$ and $f_{\rm obs,21}$; the bottom panels of Figure\,\ref{fig:NH_dist_3ERbins} show the distributions for one representative realisation. Vertical lines indicate the mean $N_{\rm H}$ from the 100 matched samples. These results are consistent with the unmatched sample but show even larger differences in $N_{\rm H}$ distributions between $\lambda_{\rm Edd}$ regimes, providing evidence that $N_{\rm H}$ decreases as the $\lambda_{\rm Edd}$ increases. The results are summarised in Table\,\ref{t:results}.

\begin{figure}
 
\centering

\includegraphics[trim={0cm 0 0cm 0cm},clip,scale=0.45]{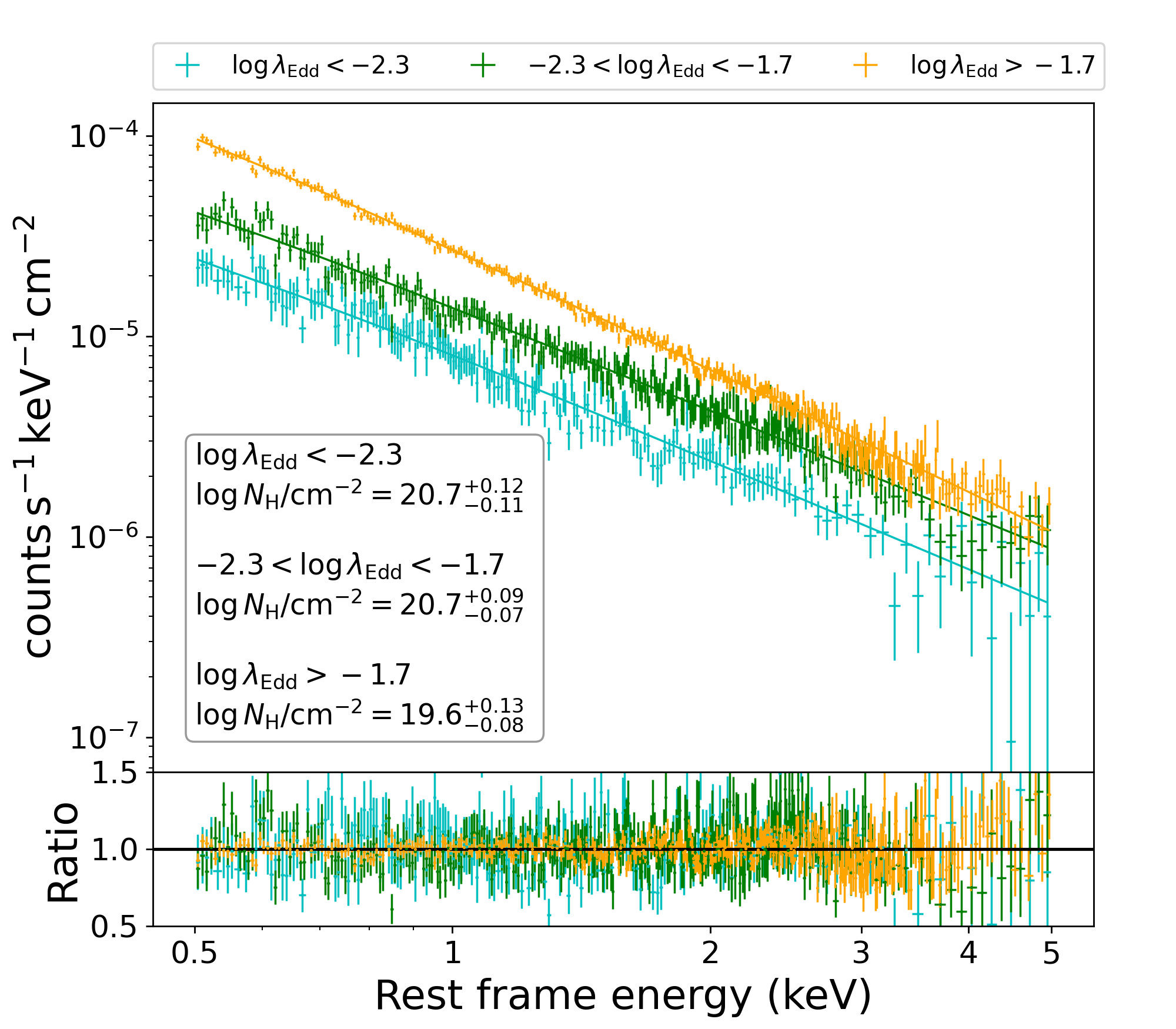}

\caption{Rest-frame X-ray spectral stacks and their best-fitting model for the low (cyan), medium (green), and high (orange) Eddington ratio regimes. The points with error bars correspond to the data, and the solid line is the best-fitting model. The bottom panel shows the ratio between the model and the data. The text in the figure also shows the best-fitting $\left< \log N_{\rm H} \right>$ of the three stacks.}

\label{fig:xray_stacks}
\end{figure}

Although there is a clear trend of decreasing $N_{\rm H}$ with the Eddington ratio, the $N_{\rm H}$ measurements have large uncertainties, with a median error of $\sigma_{\log N_{\rm H, err}}\approx 0.5 \rm \, dex$; hence, the differences in the median $N_{\rm H}$ for the different $\lambda_{\rm Edd}$ samples are less reliable. To further investigate whether there is a change in the $N_{\rm H}$ with $\lambda_{\rm Edd}$, we stacked the eROSITA X-ray spectra in the three $M_{\rm BH}-z$ matched $\lambda_{\rm Edd}$ bins following the procedure described in Section\,\ref{subsec:Xray_stacks}. The rest-frame 0.5-5\,keV spectra stacks are shown in Figure\,\ref{fig:xray_stacks}. We first fitted the rest-frame 2-5\,keV stacks using an absorbed powerlaw model (see Section\,\ref{subsec:Xray_stacks}) to constrain the power-law of the intrinsic spectrum of each stack. The spectrum below 2\,keV is more sensitive to absorption and can also be affected by other X-ray features, such as the soft excess \citep[e.g.][]{2000Reeves,2004Gierlinski, 2024Waddell}, which is particularly important in bright, broad-line AGNs \citep[e.g.][]{2016Boissay}. We find that the three stacks have a consistent photon index of $\Gamma \approx 2-2.1$, in agreement with the mean $\Gamma$ distribution reported in \citet{2022Liu}. As expected, we find that the $N_{\rm H}$ posterior distributions are unconstrained, as the mild level of absorptions observed in our sample is not expected to affect the X-ray spectrum above 2\,keV. 

We then fitted the full 0.5–5\,keV X-ray spectral stacks using a Gaussian prior on the photon index, centred at $\Gamma=2$ with a standard deviation of $\sigma_{\Gamma}=0.05$. The best-fitting models and parameters are shown in Figure~\ref{fig:xray_stacks}. The stacks for \mbox{$\log \lambda_{\rm Edd}< -2.3$} and \mbox{$-2.3 < \log \lambda_{\rm Edd} < -1.7$} yield similar obscuration levels, with \mbox{$\log (N_{\rm H}/\mathrm{cm}^{-2}) \approx 20.7 \pm 0.1$}. In contrast, the stack for $\log \lambda_{\rm Edd} > -1.7$ shows the lowest obscuration, with \mbox{$\log (N_{\rm H}/\mathrm{cm}^{-2}) = 19.6^{+0.13}_{-0.08}$}. The posterior distributions for $N_{\rm H}$ and $\Gamma$ are presented in Figure~\ref{fig:posteriorNHGamma} (Appendix~\ref{ap:Xray_modellign}). The $N_{\rm H}$ posteriors for the low- and medium-$\lambda_{\rm Edd}$ stacks strongly overlap, while the posterior for $\log \lambda_{\rm Edd} > -1.7$ is clearly shifted toward much lower obscuration levels. This provides strong evidence of a decline in obscuration at high Eddington ratios. As noted earlier, a soft X-ray excess could affect the spectra of some sources, particularly those at $\log \lambda_{\rm Edd} > -1.7$. To investigate this, we refitted the X-ray stacks, including an additional component to model the soft excess. This analysis is described in Appendix\,\ref{ap:Xray_modellign}. The results remain consistent: the $N_{\rm H}$ for the high-$\lambda_{\rm Edd}$ stack is at least $\sim$0.7\,dex lower than that of the lower $\lambda_{\rm Edd}$ stacks.

As in the previous section, we repeated the analysis, this time matching the samples by $L_{\rm AGN}$ and redshift. The obscuration constraints derived from individual sources (Table~\ref{t:results}, right panel of Figure\,\ref{fig:NH_dist_3ERbins_LAGNmatch}) show no clear change in $N_{\rm H}$ or $f_{\rm obs,21}$ with Eddington ratio. We then stacked the spectra for the three $\lambda_{\rm Edd}$ bins and performed X-ray spectral fitting. The resulting $N_{\rm H}$ posterior distributions (see right panel of Figure\,\ref{fig:posteriorNHGamma}) largely overlap across the bins. Although the high-$\lambda_{\rm Edd}$ sample shows a posterior distribution shifted toward lower obscuration compared to the low and medium-$\lambda_{\rm Edd}$ samples, the substantial overlap between the three distributions prevents us from drawing firm conclusions about differences in $N_{\rm H}$ among the $L_{\rm AGN}$–$z$ matched samples.  

In summary, our results indicate that AGN obscuration decreases with increasing Eddington ratio, results that become clearer when controlling for $M_{\rm BH}$ and redshift, consistent with a scenario in which radiation pressure on the circumnuclear dusty gas expels the material from the nuclear region.

\section{Discussion} \label{sec:discussion}

In this work, we have examined the prevalence of large-scale outflows in X-ray-selected AGNs and explored the relationship between the Eddington Ratio, AGN obscuration, and outflow activity. Our analysis used the eROSITA/eFEDS sample, which is among the largest uniform optical spectroscopic follow-up of X-ray AGNs to date. The final sample consists of 2,840 AGNs at $z<0.82$ with high-quality SDSS optical spectra (median continuum $(S/N)^2>10$). Through parallel optical spectral and X-ray analysis, we identified three key results: (1) The \OIII outflow incidence systematically increases with the AGN luminosity; (2) after controlling for SMBH mass and redshift, both the outflow rate and the fraction of blueshifted \OIII lines increase with Eddington ratio; and (3) similarly, AGN obscuration ($N_{\rm H}$) decreases with increasing Eddington ratio, result which is more significant when controlling by SMBH mass and redshift. These results suggest that AGN radiation plays a key role in regulating obscuration and driving outflows. In this section, we discuss the implications of these findings in the context of AGN feedback.

\subsection{AGN radiation as the main driver of  {\rm [O\,\textsc{iii}]} outflow activity}

The \OIII outflow rate in our spectroscopic sample is approximately 34\%, with a conservative lower limit of 14\% (see Figure\,\ref{fig:OIIIrate_Lbol}), indicating that large-scale outflows are common among X-ray-selected AGNs. This prevalence is consistent with results from AGNs selected using different methods \citep[e.g.][]{2013Mullaney, 2014Harrison, 2016HarrisonC, 2016WooJ, 2017Perna,2023Musiimenta, 2018RakshitWoo}. For example, \citet{2017Perna} analysed the SDSS spectra of a sample of $\sim 600$ X-ray AGNs identified by different X-ray facilities and found that 41\% of the sample shows outflow signatures, which is consistent with our findings. 

It is possible that previous works and our own results are biased againts obscuration, as most X-ray AGNs with optical spectroscopy are optically bright. Based on the photometric redshifts from \citet{2022Salvato}, we estimate that 52\% of all eROSITA AGNs at $z<0.82$ have corresponding optical spectra from SDSS. At $z<0.82$, the spectroscopic completeness reaches about 70\% for sources brighter than $r_{\rm AB}<21.3 \rm \, AB$ mag, indicating that the majority of missing spectra correspond to optically faint sources. By examining the intrinsic $2$–$10$ keV X-ray luminosity (see top histogram of Figure\,\ref{fig:W80_Lbol}) and column density ($N_{\rm H}$) of sources with and without optical spectra, we find that those without spectra have a median $L_{2-10\rm \, keV}$ about $0.2$ dex lower and a median $N_{\rm H}$ about $0.4$ dex higher. This suggests that these optically fainter sources are primarily affected by higher levels of absorption, rather than by lower intrinsic AGN luminosities. Hence, we do not expect the \OIII rate among X-ray AGNs without optical spectroscopy to differ substantially. In fact, previous works have found that obscured AGNs have a higher outflow incidence than unobscured AGNs \citep[e.g.][]{2015Brusa, 2016Zakamska,2019Perrotta,2023Musiimenta,2024Tozzi}; therefore, we may even expect the population without optical spectra to host a higher incidence of outflows.

Our analysis further reveals that the \OIII outflow occurrence rate rises with $L_{\rm AGN}$, from $\approx 15\%$ at \mbox{$\log L_{\rm AGN}/(\rm erg \, s^{-1})=42-43$} to $\approx 60\%$ at \mbox{$L_{\rm AGN} > 10^{45.5}$\,erg\,s$^{-1}$}, suggesting that radiation pressure plays an important role driving ionised outflow activity. This is further confirmed in Section\,\ref{res:sub:OIII_EddRatio}, where we find that the \OIII outflow velocity and occurrence rate increase with the Eddington ratio only when controlling for black hole mass; i.e., the change in the Eddington ratio is solely driven by the AGN luminosity. When we instead match the Eddington ratio sample by AGN luminosity, we find no differences in the \OIII outflow incidence as a function of $\lambda_{\rm Edd}$, indicating the black hole mass is not a driver of outflow activity. So far, there was no clear consensus on whether \OIII outflow activity and $\lambda_{\rm Edd}$ are correlated. Some works have found a positive correlation \citep[e.g.][]{2014Bae,2023Ayubinia,2025Vivek}, while others have found no significant correlation \citep[e.g.][]{2020Rojas, 2022Kakkad}. However, this is the first time that a well-controlled study, matching by $L_{\rm AGN}$, $M_{\rm BH}$, and redshift, between the \OIII line kinematics and $\lambda_{\rm Edd}$ is performed. 

Overall, our results provide compelling evidence that the observed \OIII outflows are predominantly driven by radiation pressure from the central AGN \citep[e.g.][]{2005King,2005Murray,2014Costa,2018Harrison}. The positive correlations between AGN luminosity, outflow incidence, and outflow velocity strongly support the conclusion that the AGN is the primary power source behind the ionised outflow activity. Moreover, the finding that a strong positive relationship between the Eddington ratio and outflow incidence emerges only when controlling for SMBH mass further indicates that radiation pressure is the key physical mechanism that launches and accelerates the surrounding gas \citep{2016WooJ,2018RakshitWoo}. These observational results are consistent with predictions from hydrodynamical simulations, which likewise identify AGN radiation pressure as the dominant driver of large-scale outflows \citep{2014Costa,2020Costa,2024Ward}. We therefore conclude that AGN radiative feedback is the principal mechanism powering outflow activity in our sample.

We note that at $\log \lambda_{\rm Edd}<-2.3$ or $L_{\rm AGN}<10^{44}\rm \, erg \,s^{-1}$, where the accretion flow is likely radiatively inefficient, we still find an \OIII outflow occurrence rate of $\approx 10-30 \%$, suggesting that in this regime, the physical mechanisms driving outflow activity are different. Previous works have shown that at low accretion states, the AGNs are in `jet mode', where relativistic jets can efficiently propagate through the galaxy, creating shocks in the ISM \citep{2007Sijacki}, and eventually driving ionised outflows \citep[e.g.][]{2020Hardcastle, 2023Kukreti, 2024Harrison}.  Investigating the origin of outflows in this low-accretion regime, perhaps using LOFAR \citep[]{2017Shimwell} radio data in eFEDS \citep{2024Igo}, is an interesting avenue for future work.

\subsection{Outflow incidence in the $N_{\rm H} - \lambda_{\rm Edd}$ plane } \label{dis:sub:NH_EddRatio}

\begin{figure*}
\centering

\includegraphics[trim={0cm 0 0cm 0cm},clip,scale=0.45]{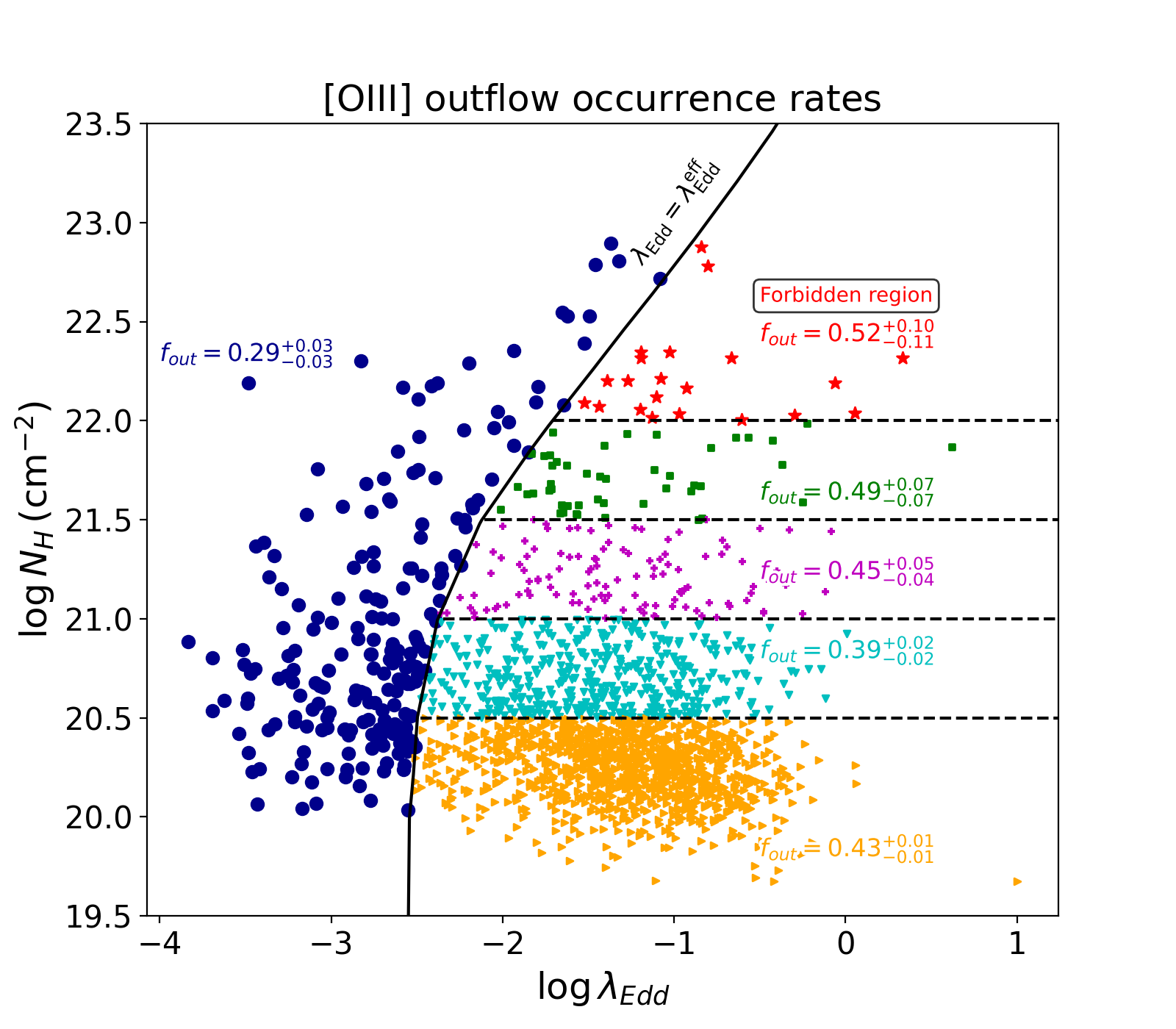}
\includegraphics[trim={0cm 0 0cm 0cm},clip,scale=0.45]{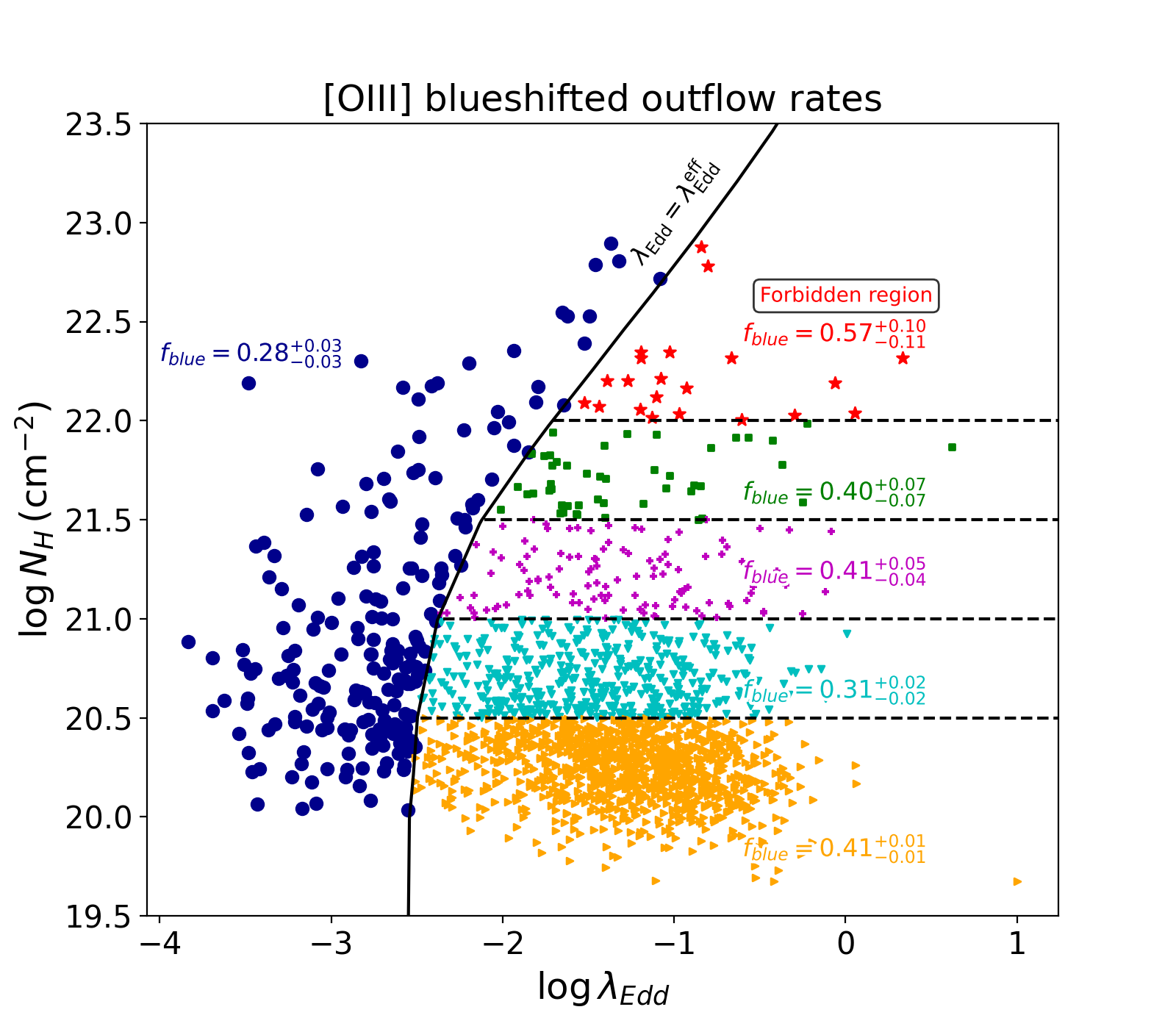}

\caption{{\it Left panel}: Line of sight column density, $N_{H}$, against the Eddington ratio, $\log \lambda_{\rm Edd}$ for our sample with broad $\rm H\beta$ detections. The black curve corresponds to the effective Eddington ratio ($\lambda_{\rm Edd}^{\rm Eff}$) for the dust composition of the Milky Way ISM, taken from \citet{2009Fabian}. The $N_{H}-\lambda_{\rm Edd}$ plane is subdivided into six different sections, and for each one, we calculated the outflow occurrence rate, $f_{out}$. The text in the figure reports the $f_{out}$ of each region. {\it Right panel}: Same as the left panel, but we report the \OIII blueshifted rate ($f_{blue}$) instead.} 
\label{fig:NH_EddRatio}
\end{figure*}

The $N_{\rm H} - \lambda_{\rm Edd}$ plane provides important information on the physical processes that shape the circumnuclear environment of the central SMBH in AGNs, and it has been studied in several works \citep[e.g.][]{2008Fabian, 2018Ishibashi, 2017Ricci, 2024Laloux}. In the radiative feedback scenario, intense radiation from the AGN exerts pressure on circumnuclear dusty gas, potentially driving it outwards. As a result, AGNs with high accretion rates are expected to be predominantly unobscured, or if obscured, the obscuring material must reside at large distances where radiation pressure is insufficient to disrupt it. Conversely, low-accretion sources are more likely to remain obscured, as radiation pressure is insufficient to clear the surrounding medium. The critical threshold at which radiation pressure becomes strong enough to expel the gas is known as the effective Eddington limit ($\lambda_{\rm Edd}^{\rm eff}$), which has been derived in various theoretical models \citep[e.g.][]{2007Honig, 2008Fabian, 2018Ishibashi}. In this work, we adopt the $\lambda_{\rm Edd}^{\rm eff}$ curve from \citet{2008Fabian}, based on a Milky Way-like interstellar dust composition (solid line in Figure~\ref{fig:NH_EddRatio}).

Given that at the redshift of our sources the intrinsic obscuration from host galaxies exceed $\log N_{\rm H} / (\mathrm{cm}^{-2}) \gtrsim 22$ in a minority of AGNs \citep[e.g.][]{2017Buchner}, models predict a `forbidden region' in the $N_{\rm H}$–$\lambda_{\rm Edd}$ plane where nuclear obscuring clouds are short-lived due to rapid destruction by radiation pressure. In wind-driven AGN feedback scenarios, the timescale for clearing circumnuclear material is typically $1$–$10\,\mathrm{Myr}$ \citep{2010King,2014Costa,2016Hopkins}, comparable to the period over which SMBHs can sustain high accretion rates \citep[e.g.][]{2016Hopkins}. For sources in our sample within the forbidden region, which have a median $\log L_{\rm AGN} / (\mathrm{erg\,s^{-1}}) \approx 44.9$, the quasar lifetime is estimated to be $\sim100$–$300\,\mathrm{Myr}$ \citep{2006Hopkins}, implying an expected fraction of $\sim1$–$3\%$ in this region. We measure that $\sim 1.1\%$ of our sample lies in this region, fully consistent with theoretical expectations and previous observational results \citep[e.g.][]{2017Ricci,2022Toba,2024Laloux}. This agreement supports the view that the immediate vicinity of the SMBH is dynamically unstable under strong radiation fields, reinforcing the role of AGN feedback in regulating the circumnuclear environment.

This forbidden region is often referred to as the `feedback phase', during which AGNs are expected to drive powerful outflows as they expel surrounding material \citep[e.g.][]{2008Fabian, 2022Brusa, 2023Musiimenta}. To test this scenario, we explore the incidence of outflows and blueshifted \OIII kinematics across different regions of the $N_{\rm H} - \lambda_{\rm Edd}$ plane; the results are illustrated in Figure~\ref{fig:NH_EddRatio}. We find that the forbidden region exhibits the highest outflow occurrence rate of $f_{\rm out} = 0.52^{+0.1}_{-0.11}$, although the uncertainties are substantial given that there are only 21 objects in this region. In contrast, sources with $\lambda_{\rm Edd} < \lambda_{\rm Edd}^{\rm eff}$ show the lowest outflow fraction ($f_{\rm out} = 0.29^{+0.03}_{-0.03}$), as expected from the lack of sufficient radiation pressure to drive outflows. At $\lambda_{\rm Edd} > \lambda_{\rm Edd}^{\rm eff}$ and $N_{\rm H} < 10^{21.5}\,\rm cm^{-2}$, the outflow rate does not substantally vary ($f_{\rm out} \approx 0.4-0.49$), indicating that once the effective Eddington limit is exceeded, outflow activity becomes robust and largely independent of moderate obscuration. Similarly, the fraction of sources with blueshifted \OIII emission peaks in the forbidden region ($f_{\rm blue} = 0.57^{+0.1}_{-0.11}$), while the lowest blueshifted fraction is again found at $\lambda_{\rm Edd} < \lambda_{\rm Edd}^{\rm eff}$ ($f_{\rm blue} = 0.28^{+0.03}_{-0.03}$), further supporting the link between high accretion rates and enhanced outflow activity.

To corroborate these results, we stack the \OIII spectra across the different regions of the $N_{\rm H} - \lambda_{\rm Edd}$ plane. Figure~\ref{fig:OIIIstacks_EffEddRatio} and Table~\ref{t:W80_stacks_EffER} show the stacking results. The stacked \OIII line kinematics are in qualitative agreement with the outflow and blueshifted fractions. At $\lambda_{\rm Edd} < \lambda_{\rm Edd}^{\rm eff}$, the \OIII profile shows weak outflow signatures: $W_{\rm 80} = 531 \pm 4\,\rm km\,s^{-1}$ and $\Delta v = -142 \pm 2\,\rm km\,s^{-1}$, indicating minimal outflow activity. In contrast, the forbidden region exhibits the broadest profile: $W_{\rm 80} = 747 \pm 13\,\rm km\,s^{-1}$ and $\Delta v=-180 \pm 3\rm \, km\, s^{-1}$. However, this stack is based on only 21 sources, resulting in a lower signal-to-noise ratio than those of other regions. Notably, despite the strong blueshift, the \OIII line peak is shifted to the red side, suggesting a more complex outflow dynamics.

We have also directly investigated the relationship between \OIII outflow velocity and AGN obscuration; see Figure\,\ref{fig:W80_NH}. We find no significant trend between these two quantities, as confirmed by a Spearman correlation test ($r=-0.08$ and $\rm p-value=0.0008$). This result suggests that the enhanced \OIII outflow incidence observed in the forbidden region is driven by the connection between the \OIII line kinematics and the Eddington ratio, rather than by an intrinsic physical link between nuclear obscuration and large-scale ionised outflows.

Section\,\ref{res:sub:NH_EddRatio} shows evidence of a decrease in the obscuration with the Eddington ratio, mostly coming from objects in the $N_{\rm H}=10^{20}-10^{21} \rm \, cm^{-2}$ range. The difference in $N_{\rm H}$ between $\lambda_{\rm Edd}$ bins becomes more pronounced when the samples are matched in SMBH mass and redshift, highlighting the role of AGN radiation pressure in shaping the distribution of obscuring material. Given that eROSITA is a soft X-ray observatory, and the wide majority ($\sim 90 \% $) of the sources are unobscured with $N_{\rm H}<10^{21.5} \rm \, cm^{-2}$ \citep{2022Liu}, it is difficult to probe higher levels of obscuration. Nevertheless, it is still impressive that we can observe a change in the column density given the observational limitations. Our results are in qualitative agreement with \citet{2017Ricci}, who used much deeper hard X-ray observations in the local Universe to demonstrate a decrease in the fraction of obscured sources at $N_{\rm H}=10^{22}-10^{24}  \rm \, cm^{-2}$ with the Eddington ratio. 

Several studies have reported enhanced outflow activity in obscured and type-II AGN \citep[e.g.][]{2015Brusa, 2016Zakamska, 2019Perrotta, 2023Musiimenta, 2024Tozzi, 2026Donnan, 2026RamosAlmeida}, which might appear to contradict our findings. However, these studies focus on highly luminous obscured quasars ($\log L_{\rm AGN}/(\rm erg\,s^{-1}) \gtrsim 45-46$), whose luminosities also imply large Eddington ratios. Hence, many of these objects inherently occupy the forbidden region; in fact, some of these works explicitly target these objects \citep[e.g.][]{2015Brusa,2023Musiimenta}. Consequently, these samples are directly comparable to the AGNs in our own forbidden region (see Figure\,\ref{fig:NH_EddRatio}), which exhibit the highest outflow occurrence rates and blueshifted fractions across the $N_{\rm H} - \lambda_{\rm Edd}$ plane. This reinforces the hypothesis that the forbidden region represents a short-lived evolutionary phase: a transition from obscured, slow accretion to rapid accretion, where powerful outflows are driven into the host galaxy while the SMBH is still embedded in gas and dust. Subsequently, radiation pressure clears this material, leading to a decrease in column density \citep[e.g.][]{2017Ricci}. 

Overall, our results align with a scenario in which the SMBH regulates its immediate surroundings and the host galaxy, with AGN radiation regulating the amount of circumnuclear material and driving powerful outflows into the ISM \citep[e.g.][]{2005DiMatteo, 2016Hopkins}.

\section{Conclusions} \label{sec:conclusions}

We have investigated the prevalence of large-scale outflows in X-ray-selected AGNs and explored how the Eddington ratio shapes both ionised outflow activity and AGN obscuration. Our analyses are performed in eROSITA/eFEDS, a 140\,deg$^2$ field which has been extensively observed by SDSS, offering one of the largest uniform optical spectroscopic follow-up of X-ray AGNs to date. Specifically, we study 2,840 AGN detected by eROSITA at $z<0.82$ with good quality ($(S/N)^2>10$) SDSS optical spectra. Through optical spectral fitting, we extract key properties including the FWHM of the broad  $\rm H\beta$ emission line (used to estimate $M_{\rm BH}$ under the assumption of virialised motion) and AGN luminosity at 5100\,\AA. We further determine the velocity and asymmetry of the \OIII emission using the non-parametric parameters $W_{80}$ and $\Delta v$. Finally, we constrain the obscuration properties of the sample by combining archival X-ray $N_{\rm H}$ measurements with results from X-ray spectral stacking. Our main results are as follows:

\begin{enumerate}
    \item Approximately 34\% of the sample exhibit \OIII outflows with velocities $W_{80}>600~\rm km~s^{-1}$ (see Figures\,\ref{fig:W80_Lbol} and\,\ref{fig:OIIIrate_Lbol}). We find a weak but statistically significant correlation between the \OIII outflow velocity and AGN luminosity. Additionally, we find that the outflow occurrence rate is $\approx 15\%$ for $\log L_{\rm AGN}/(\rm erg~s^{-1})<44$, and then systematically increases with the AGN bolometric luminosity, reaching a maximum of $\sim$60\% at $\log L_{\rm AGN}/(\rm erg~s^{-1})> 45.5$ (see Section\,\ref{res:sub:OIIIin}).

    \item When controlling for $M_{\rm BH}$ and redshift, we find that the outflow and blueshifted fractions ($f_{out}$, $f_{blue}$) progressively increase with $\lambda_{\rm Edd}$ (see Fig.~\ref{fig:W80_EddRatio_MBHmatched},\,\ref{fig:W80dVdist_EddRatio_MBHmatched}, and\,\ref{fig:OIIIstacks_EddRatio_MBHmatched}), going from $f_{out}=0.29$ and $f_{blue}=0.27$ at $\log \lambda_{\rm Edd}<-2.3$ to $f_{out}=0.49$ and $f_{blue}=0.4$ at $\log \lambda_{\rm Edd}>-1.7$. This trend is also seen in the stacked \OIII spectra (see Section\,\ref{res:sub:OIII_EddRatio}). Notably, when controlled by $L_{\rm AGN}$ instead (i.e., when the black hole mass drives the changes in Eddington ratio, we do not observe any change in the average \OIII outflow kinematic properties with Eddington ratio, indicating that the increase in the outflow occurrence rate is radiatively driven, while $M_{\rm BH}$ does not impact the outflow activity.

    \item We find evidence of a decline in the AGN obscuration ($N_{\rm H}$) with increasing $\lambda_{\rm Edd}$ (see Figure\,\ref{fig:NH_dist_3ERbins} and Table\,\ref{t:results}). The fraction of sources with $N_{\rm H}>10^{21} \rm \, cm^{-2}$ and the median obscuration decrease from $f_{\rm obs,21}=0.21$ and $\left< \log N_{\rm H} \right>/(\rm cm^{-2})=20.7$ at $\log \lambda_{\rm Edd}<-2.3$ to $f_{\rm obs,21}=0.1$ and $\left< \log N_{\rm H}\right>/(\rm cm^{-2})=20.4$ at $\log \lambda_{\rm Edd}>-1.7$. We qualitatively confirm these results using detailed X-ray stacking analyses, and find that the high-$\lambda_{\rm Edd}$ stack has an average column density $\sim 1$\,dex lower than the low- and medium-$\lambda_{\rm Edd}$ stacks, when controlling by $M_{\rm BH}$ and redshift. Similarly to our \OIII outflows results, the $N_{\rm H}$ dependency on $\lambda_{\rm Edd}$ becomes much less significant when matching the $\lambda_{\rm Edd}$ samples in $L_{\rm AGN}$ (see Fig.~\ref{fig:xray_stacks} and \ref{fig:posteriorNHGamma}), indicating that the decrease in the AGN obscuration is driven by increasing AGN luminosities, while the black hole mass does not play a significant role.

    \item We analyse the source density and outflow incidence across the $N_{\rm H}-\lambda_{\rm Edd}$ plane (see Figure\,\ref{fig:NH_EddRatio}). We find that 1.1\% of our sample lies in the `forbidden region' ($\lambda_{\rm Edd}>\lambda_{\rm Edd}^{\rm Eff}$ and $N_{\rm}>10^{22}\rm \, cm^{-2}$), which also corresponds to the region with the largest \OIII outflow incidence ($f_{out}= 0.52^{+0.1}_{-0.11}$ and $f_{blue}=0.57_{-0.1}^{+0.11}$). This picture is confirmed by \OIII spectral stacks (see Figure\,\ref{fig:OIIIstacks_EffEddRatio} and Table\,\ref{t:W80_stacks_EffER}), but with large uncertainties due to the small sample size. At $\lambda_{\rm Edd}>\lambda_{\rm Edd}^{\rm Eff}$ and $N_{\rm H}<10^{22}\,\rm cm^{-2}$, we find no strong change between $N_{\rm H}$ and the \OIII outflow properties. The lowest outflow occurrence rate is observe at $\lambda_{\rm Edd}<\lambda_{\rm Edd}^{\rm Eff}$ with $f_{out}= 0.29^{+0.03}_{-0.03}$ and $f_{blue}=0.28_{-0.03}^{+0.03}$.
    
\end{enumerate}

Overall, our results show that highly accreting X-ray-selected AGNs host faster, more blueshifted \OIII outflows and are less obscured than AGN with lower accretion rates. We find that these transition occurs progressively with $\lambda_{\rm Edd}$. These findings are consistent with models of AGN radiative feedback and SMBH self-regulation, in which rapid accretion enables black holes to clear their immediate environment, drive powerful winds, and exert significant influence on the host galaxy.

In the future, the 4MOST AGN survey will deliver deeper optical spectroscopy for the remaining X-ray AGNs in eFEDS, which are too optically faint (and potentially obscured) to be detected by SDSS, as well as for all eROSITA AGNs in the southern hemisphere. This will extend such studies to lower Eddington ratios, higher obscuration levels, and significantly larger samples.

\begin{acknowledgements}

CA acknowledges support and resources from Alexander von Humboldt Foundation.

This work is based on data from eROSITA, the primary instrument aboard SRG, a joint Russian-German science mission supported by the Russian Space Agency (Roskosmos), in the interests of the Russian Academy of Sciences represented by its Space Research Institute (IKI), and the Deutsches Zentrum für Luft- und Raumfahrt (DLR). The SRG spacecraft was built by Lavochkin Association (NPOL) and its subcontractors, and is operated by NPOL with support from the Max Planck Institute for Extraterrestrial Physics (MPE). The development and construction of the eROSITA X-ray instrument was led by MPE, with contributions from the Dr. Karl Remeis Observatory Bamberg \& ECAP (FAU Erlangen-Nürnberg), the University of Hamburg Observatory, the Leibniz Institute for Astrophysics Potsdam (AIP), and the Institute for Astronomy and Astrophysics of the University of Tübingen, with the support of DLR and the Max Planck Society. The Argelander Institute for Astronomy of the University of Bonn and the Ludwig Maximilians Universität Munich also participated in the science preparation for eROSITA. The eROSITA data shown here were processed using the eSASS software system developed by the German eROSITA consortium. 

Funding for the Sloan Digital Sky Survey V has been provided by the Alfred P. Sloan Foundation, the Heising-Simons Foundation, the National Science Foundation, and the Participating Institutions. SDSS acknowledges support and resources from the Center for High-Performance Computing at the University of Utah. The SDSS website is \url{www.sdss.org}.

SDSS is managed by the Astrophysical Research Consortium for the Participating Institutions of the SDSS Collaboration, including the Carnegie Institution for Science, Chilean National Time Allocation Committee (CNTAC) ratified researchers, Caltech, the Gotham Participation Group, Harvard University, Heidelberg University, The Flatiron Institute, The Johns Hopkins University, L'Ecole polytechnique f\'{e}d\'{e}rale de Lausanne (EPFL), Leibniz-Institut f\"{u}r Astrophysik Potsdam (AIP), Max-Planck-Institut f\"{u}r Astronomie (MPIA Heidelberg), Max-Planck-Institut f\"{u}r Extraterrestrische Physik (MPE), Nanjing University, National Astronomical Observatories of China (NAOC), New Mexico State University, The Ohio State University, Pennsylvania State University, Smithsonian Astrophysical Observatory, Space Telescope Science Institute (STScI), the Stellar Astrophysics Participation Group, Universidad Nacional Aut\'{o}noma de M\'{e}xico, University of Arizona, University of Colorado Boulder, University of Illinois at Urbana-Champaign, University of Toronto, University of Utah, University of Virginia, Yale University, and Yunnan University.

This research was supported by the Munich Institute for Astro-, Particle and BioPhysics (MIAPbP), which is funded by the Deutsche Forschungsgemeinschaft (DFG, German Research Foundation) under Germany´s Excellence Strategy – EXC-2094 – 390783311.

CAy, AM and MS acknowledge the support of the Excellence Cluster ORIGINS, which is funded by the Deutsche Forschungsgemeinschaft (DFG, German Research Foundation) under Germany’s Excellence Strategy– EXC-2094– 390783311.

VAF acknowledges funding from a United Kingdom Research and Innovation grant (code: MR/V022830/1).

DMA acknowledges support from the Science and Technology Facilities Council (grant code: ST/X001075/1)

\end{acknowledgements}

\bibliographystyle{aa} 
\bibliography{references} 

%
%

\begin{appendix}

\section{Optical spectral fitting} \label{ap:optical_fitting}

\subsection{Further details on emission line fitting}

Here, we described in detail the fitting process for the emission lines not used in this work, but that were still included in the optical fitting described in Section\,\ref{sub:pyqsofit}.
All the emission lines included in the optical spectral fitting procedure are listed in Table\,\ref{t:emission_lines}.

\begin{figure}
 
\centering

\includegraphics[trim={0cm 0 0 1cm},clip,scale=0.4]{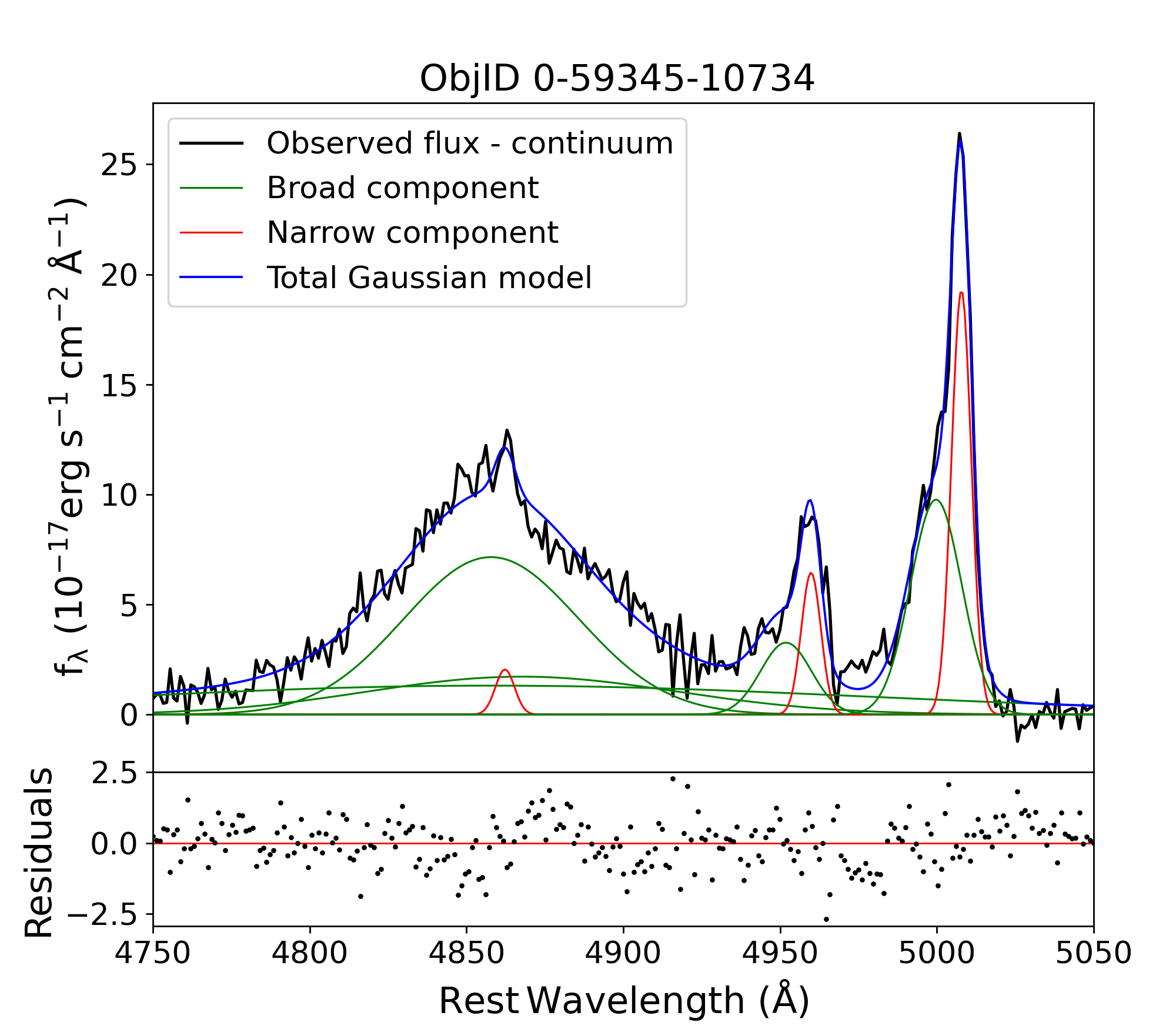}
\includegraphics[trim={0cm 0 0 1cm},clip,scale=0.4]{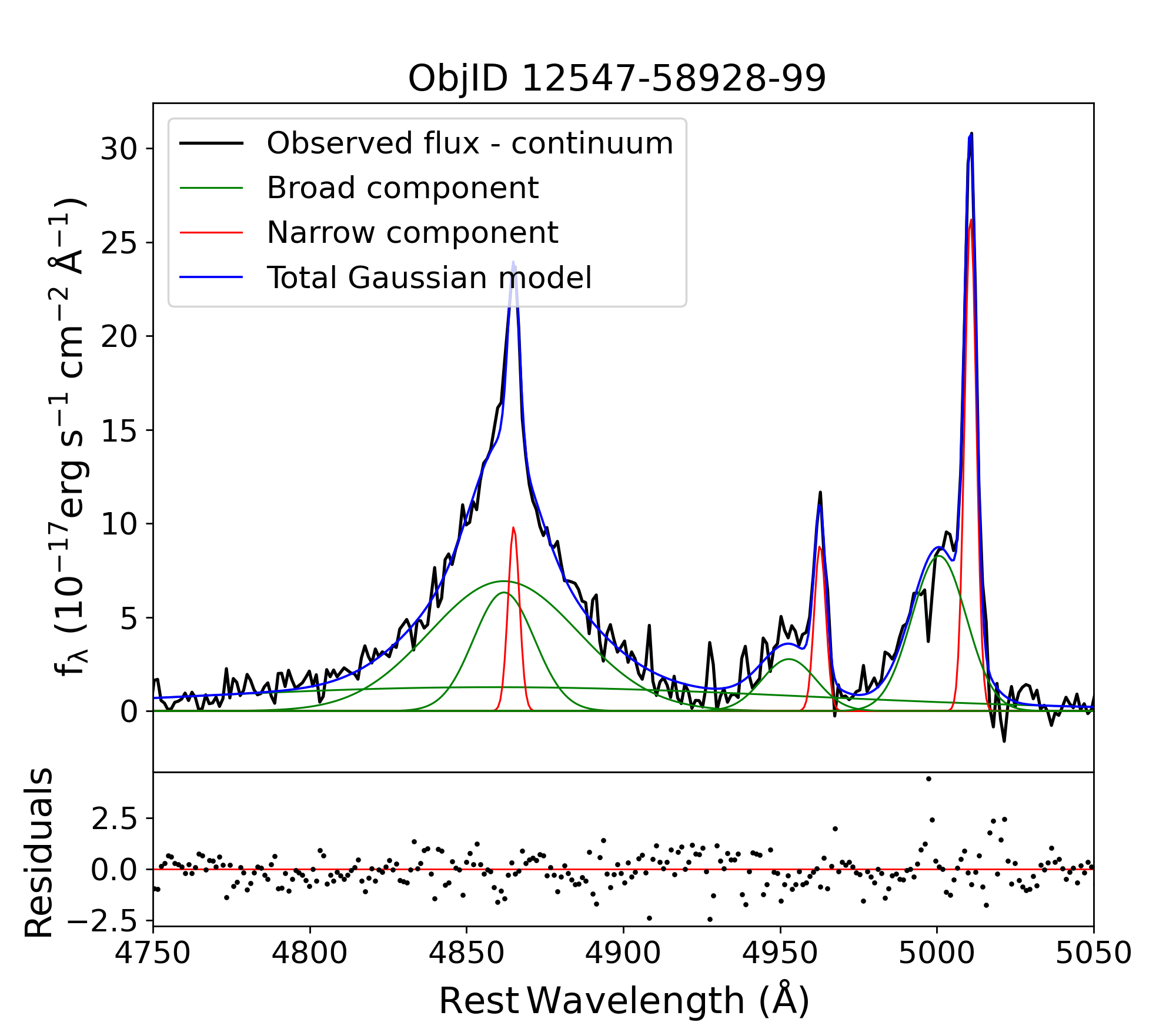}

\caption{ Example of the fitting of the $\rm H\beta$+\OIII complex for two sources in our sample with a median $S/N\approx 20$. The black spectrum corresponds to the continuum-subtracted spectrum (i.e., with the stellar emission, Fe II continuum, and AGN continuum removed). The red Gaussians represent the narrow  $\rm H\beta$ and the core of the [O\,\textsc{iii}]~$\lambda\lambda4960,\,5008$ lines, while green Gaussians show the modelling of broad  $\rm H\beta$ and the wing of the [O\,\textsc{iii}]~$\lambda\lambda4960,\,5008$ lines. The blue line corresponds to the total Gaussian model. The bottom sub-panels show the residuals between the model and the data.} 
\label{fig:HbOIII_ex}
\end{figure}

In addition, Figure\,\ref{fig:HbOIII_ex} shows two exampled of the fitting of the  $\rm H\beta$ plus \OIII complex.

\begin{enumerate}
    \item \textbf{H$\alpha$ Complex}:  In this complex, we fit the H$\alpha$, [N\,\textsc{ii}]~$\lambda\lambda6549,\,6585$, and [S\,\textsc{ii}]~$\lambda\lambda6718,\,6732$ lines simultaneously. The H$\alpha$ emission is modelled with one narrow Gaussian component (maximum FWHM of $1000\,\rm km\,s^{-1}$) and three broad Gaussian components (each with a maximum FWHM of $17\,000\,\rm km\,s^{-1}$); these FWHM limits are applied similarly in other line complexes. Each [N\,\textsc{ii}] and [S\,\textsc{ii}] line is fit with a single narrow Gaussian whose width is tied to that of the narrow H$\alpha$ component. The flux ratio between [N\,\textsc{ii}]~$\lambda6549$ and [N\,\textsc{ii}]~$\lambda6585$ is fixed at 1:2.99, following \citet{1981Osterbrock}.
    \smallskip

    \item \textbf{[O\,\textsc{ii}] Complex}:  this complex includes H$\gamma$, H$\delta$, H$\varepsilon$, [O\,\textsc{iii}]~$\lambda4363$, [Ne\,\textsc{iii}]~$\lambda3869$, and the [O\,\textsc{ii}]~$\lambda\lambda3726,3728$ doublet. As with previous complexes, the H$\gamma$ line is fitted with one narrow Gaussian and three broad Gaussian components. The fainter Balmer lines, H$\delta$ and H$\varepsilon$, are each modelled with a narrow Gaussian plus a single broad Gaussian, as they do not require multiple broad components. The [O\,\textsc{iii}]~$\lambda4363$ and [Ne\,\textsc{iii}]~$\lambda3869$ lines are each fitted with a single narrow Gaussian. The [O\,\textsc{ii}] doublet is modelled with two narrow Gaussians centred at 3728\,\AA. The line width of all the narrow lines in this complex are tied to the [O\,\textsc{ii}] line width. Additionally, the velocity offsets of all broad lines are tied to that of [O\,\textsc{ii}] to improve the robustness of the fit, as broad components in this spectral region are challenging to constrain.
    \smallskip

    \item \textbf{[Ne\,\textsc{v}] Lines}:  we fit the [Ne\,\textsc{v}]~$\lambda3346$ and [Ne\,\textsc{v}]~$\lambda3426$ lines independently. [Ne\,\textsc{v}]~$\lambda3346$ is modelled with a single narrow Gaussian, while [Ne\,\textsc{v}]~$\lambda3426$ is fitted with both a narrow and a broad Gaussian component.
    \smallskip
    
    \item \textbf{[Mg\,\textsc{ii}] Profile}:  the [Mg\,\textsc{ii}] profile is fitted using one narrow Gaussian and three broad Gaussian components. Here, the velocity of the broad components can be as high as $30,000\,\rm km\,s^{-1}$, as [Mg\,\textsc{ii}] emission often exhibits greater velocity widths than the Balmer lines.
\end{enumerate}

\begin{table}
\caption{Emission lines included in the spectral fitting}             
\label{t:emission_lines}      
\centering                          
\begin{tabular}{c c c c }        
\hline\hline   
\noalign{\smallskip}
Complex & Line & N & Central $\lambda$ (\AA) \\ 
\noalign{\smallskip}
\hline           
\noalign{\smallskip} 
$\rm H\alpha$ & Narrow $H\rm \alpha$    & 1 & 6564.61  \\      
\noalign{\smallskip} 
              & Broad  $H\rm \alpha$    & 3 & 6564.61 \\
\noalign{\smallskip} 
              & $\rm [NII]\lambda 6549$ & 1 & 6549.85 \\
\noalign{\smallskip} 
              & $\rm [NII]\lambda 6585$ & 1 & 6585.28 \\
\noalign{\smallskip} 
              & $\rm [SII]\lambda 6718$ & 1 & 6718.29 \\ 
\noalign{\smallskip} 
              & $\rm [SII]\lambda 6732$ & 1 & 6732.67 \\ 
\noalign{\medskip} 
$\rm H\beta$  & Narrow $H\rm \beta$     & 1 & 4862.68  \\      
\noalign{\smallskip} 
              & Broad $H\rm \beta$      & 3 & 4862.68  \\      
\noalign{\smallskip} 
              & $\rm \OIII\lambda4959$ core & 1 & 4960.30  \\      
\noalign{\smallskip} 
              & $\rm \OIII\lambda4959$ wing & 1 & 4960.30  \\      
\noalign{\smallskip} 
              & $\rm \OIII\lambda5007$ core & 1 & 5008.24  \\      
\noalign{\smallskip} 
              & $\rm \OIII\lambda5007$ wing & 1 & 5008.24  \\      
\noalign{\smallskip} 
 $\rm OII$  & Narrow $H\rm \gamma$     & 1 & 4341  \\      
\noalign{\smallskip} 
                 & Broad $H\rm \gamma$      & 1 & 4341  \\      
\noalign{\smallskip} 
                 & $\rm \OIII\lambda 4363$ & 1 & 4363  \\      
\noalign{\smallskip} 
    & Narrow $H\rm \delta$     & 1 & 4101  \\      
\noalign{\smallskip} 
                 & Broad $H\rm \delta$      & 1 & 4101  \\      
\noalign{\smallskip} 
      &  $\rm [OII]\lambda3726+3728\AA$ & 2 & 3728.48  \\      
\noalign{\smallskip} 
    $\rm [NeIII]$      &  $\rm [NeIII]\lambda3870\AA$  & 1 & 3869  \\      
\noalign{\smallskip} 
            
  $\rm Ne\:V$    &  $\rm [Ne \: V]\lambda 3346$     & 1 & 3346  \\      
\noalign{\smallskip} 
                 & Narrow $\rm [Ne \: V]\lambda 3426$     & 1 & 3426.84  \\      
\noalign{\smallskip} 
                 & Broad $\rm [Ne \: V]\lambda 3426$     & 1 & 3426.84  \\      
\noalign{\smallskip} 
  $\rm Mg II$    &  Narrow $\rm Mg \: II\lambda 2796+2803\AA$ & 1 & 2798.75 \\      
\noalign{\smallskip} 
  $\rm Mg II$    &  Broad $\rm Mg \: II\lambda 2796+2803\AA$ & 2 & 2798.75 \\      
\noalign{\smallskip}

\hline                                   
\end{tabular}
\end{table}

\subsection{Modelling of the broad H$\beta$ emission line} \label{ap:Hb}

To model the emission lines in the optical spectra of AGNs, PyQSOFit allows users to specify a fixed number of Gaussian components, which are then used to reconstruct the overall shape of the emission line profile. The key principle is not to determine the "optimal" number of components via model selection, but rather to use an empirically validated configuration that reliably captures the complexity of the line profile. In our analysis, to model the H$\beta$ emission line, we adopted three broad Gaussian components plus one narrow component, following the recommendations of the PyQSOFit developers \citep{2019Shen}. This approach has been successfully applied in multiple studies to measure the broad H$\beta$ FWHM and derive single-epoch black hole masses for SDSS AGN \citep[e.g.][]{2019Shen, 2020Rakshit, 2022Wu, 2024Panda, aydar2026}. \citet{2019Shen} demonstrated that three broad Gaussians are necessary to accurately model asymmetric line profiles, which are common in AGNs. Furthermore, they found that the 3 broad plus 1 narrow Gaussian fitting scheme remains robust even at moderate $S/N$, with only minor degradation expected for sources with $S/N \lesssim 3$. Given that our sample includes only sources with median $S/N > 3.162$, we are confident that the broad  $\rm H\beta$ profiles are well-reproduced in the vast majority of cases.

Nevertheless, in this Appendix, we demonstrate that a three-broad-Gaussian model can accurately reproduce the broad $\rm H\beta$ profile for our sample. To this end, we conducted a systematic test to assess whether three Gaussians are indeed necessary for our sample. Following \citet{2022Wu}, we identify all sources in which one of the three broad Gaussian components contributed less than 5\% of the total broad H$\beta$ flux. This criterion helps identify cases where a component may be effectively redundant. We found that 144 out of 2087 sources ($\sim 7\%$ of the H$\beta$-detected sample) met this condition. For these 144 sources, we refitted the H$\beta$ complex using a two-broad-Gaussian + 1-narrow-Gaussian model. Then, we use the Bayesian Information Criterion (BIC; \citealp{1978Schwarz}) to evaluate whether additional Gaussian components are justified in spectral line fitting \citep[e.g.][]{2020Vietri}. Following standard practice, we considered a BIC difference greater than 10 as decisive; thus, we select the two-Gaussian model as preferred if $|BIC_{2G}-BIC_{3G}|>10$ and $BIC_{2G}<BIC_{3G}$. 

\begin{figure*}

\centering

\includegraphics[trim={0cm 0 2cm 0cm},clip,scale=0.48]{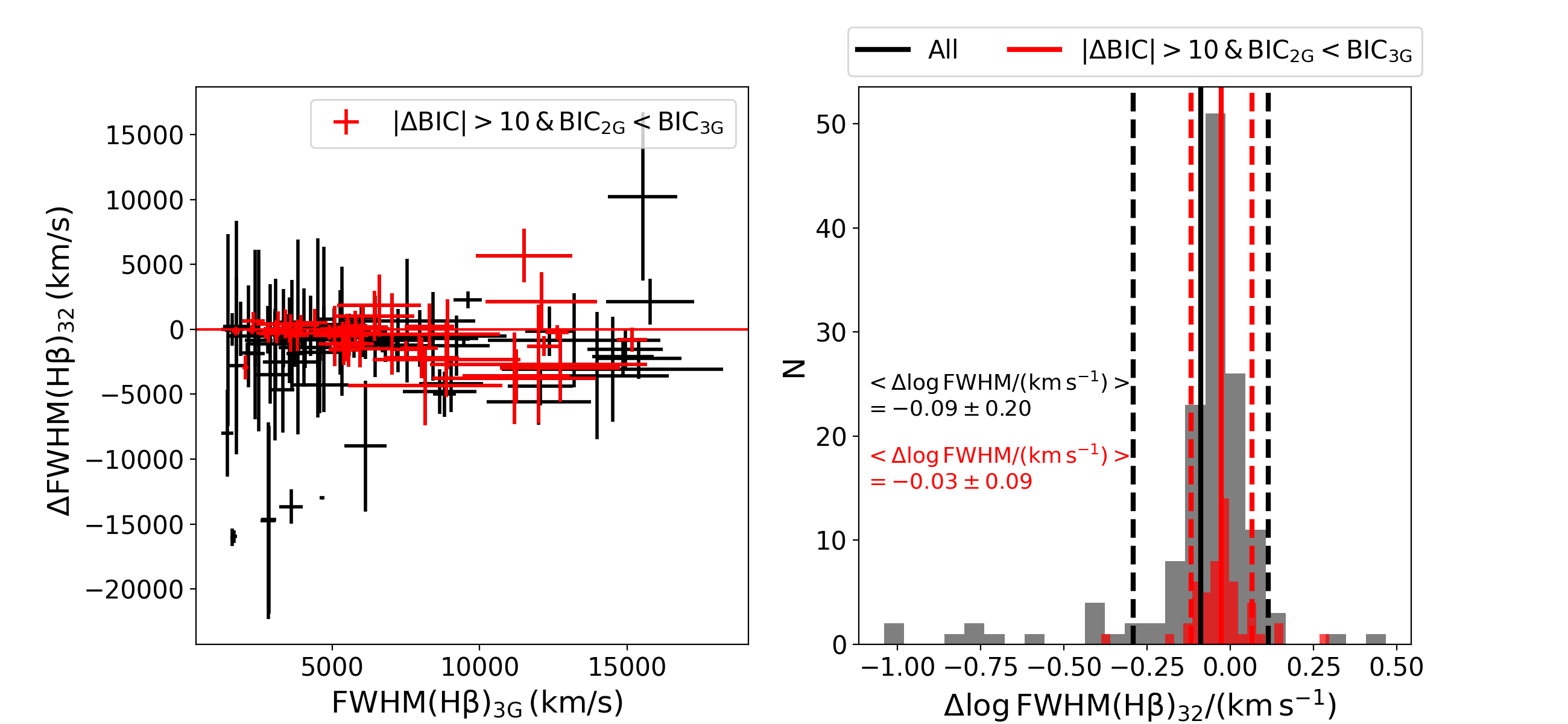}

\caption{{\it Left:} Difference in broad HFWHM between the three-Gaussian and 2-Gaussian fits ($\Delta \rm FWHM(H\beta)_{32}= FWHM(H\beta)_{3G}-FWHM(H\beta)_{2G}$), plotted against the broad  $\rm H\beta$ FWHM derived from the three-Gaussian fit, $FWHM(H\beta)_{3G}$. Red points indicate sources where the 2-Gaussian fit is statistically worse than the three-Gaussian fit (i.e., $\chi_{r,3G}^2-\chi_{r,2G}^2<-0.1$), meaning the three-Gaussian model reproduces the data better. Green points represent sources where the 2-Gaussian model performs better (i.e., $\chi_{r,2G}^2-\chi_{r,3G}^2<-0.1$). {\it Right panel}: Distribution of  $\Delta \log \rm FWHM(H\beta)_{32}$ across the sample. The orange vertical lines show the mean and standard deviation of the full sample, while the magenta lines indicate the mean and standard deviation for sources where the 2- and three-Gaussian models perform equally well, statistically  ($|\chi_{r,2G}^2-\chi_{r,3G}^2|<0.1$). } 
\label{fig:ap:deltaFWHM}
\end{figure*}

The results are illustrated in Figure\,\ref{fig:ap:deltaFWHM}. The left panel of Figure\,\ref{fig:ap:deltaFWHM} shows the differences between the 3- and two-Gaussian fit ($\Delta \rm FWHM(H\beta)_{32}= FWHM(H\beta)_{3G}-FWHM(H\beta)_{2G}$) against the value measured from the 3-Gaussian fit, and the right panel displays the full distribution of FWHM differences in logarithmic scale ($\Delta \log \rm FWHM(H\beta)_{32}$), with the mean and standard deviation shown on the Figure. We found that in 52 out of 144 sources ($\sim 2.5\%$ of the total broad  $\rm H\beta$-detected sample), the broad H$\beta$ profile is statistically better reproduced by two Gaussians; these cases are highlighted in red in Figure\,\ref{fig:ap:deltaFWHM}. For those sources, the mean and standard deviation of the FWHM differences between the two- and three-Gaussian fits are only  $\left< \Delta \log \rm FWHM(H\beta)_{32} \right>=-0.03\pm0.09\rm \, dex$, which is consistent with the expected systematic uncertainty of the parameter \citep[e.g.][]{2011Shen, 2020Rakshit}.

\begin{figure*}
\centering

\includegraphics[trim={0cm 0 2cm 0cm},clip,scale=0.48]{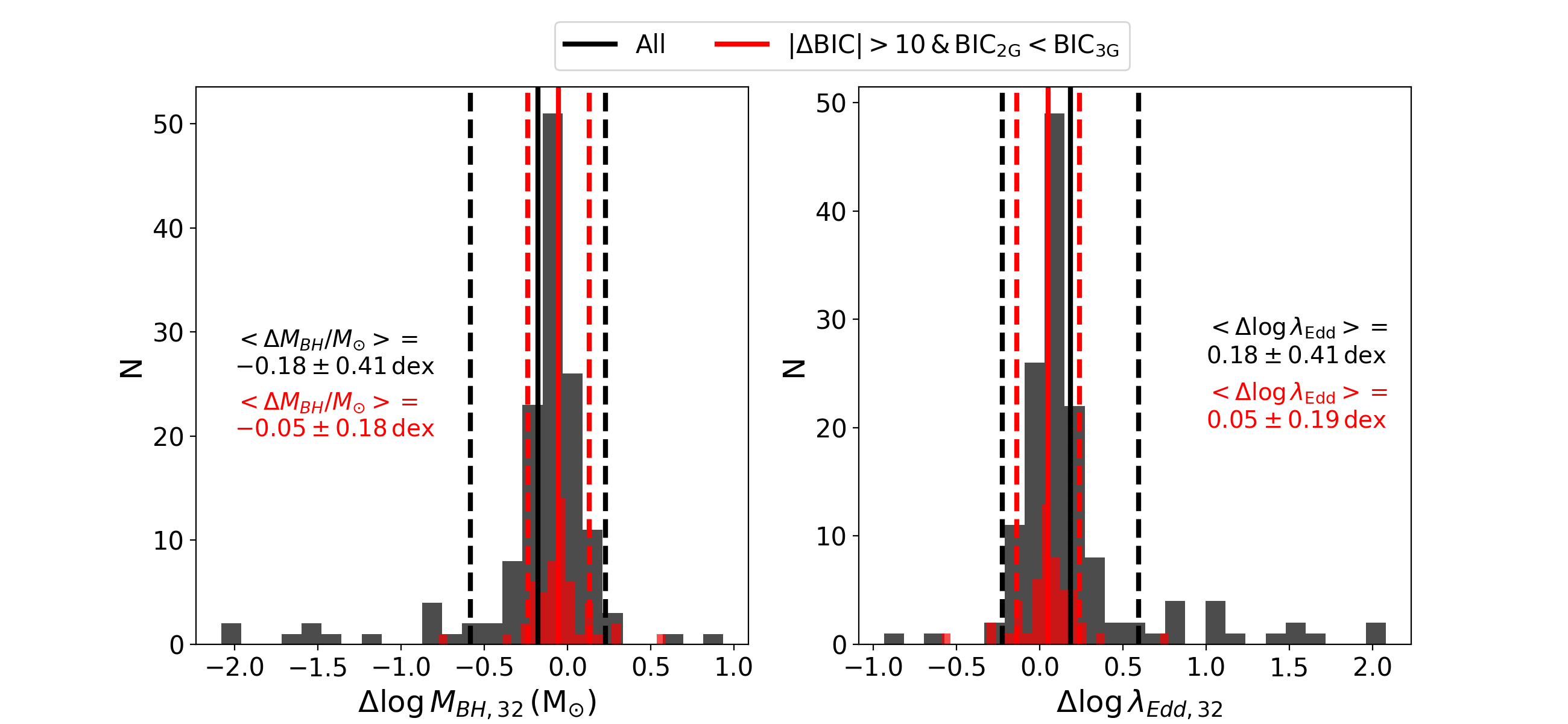}

\caption{Same as the right panel of Figure\,\ref{fig:ap:deltaFWHM}, but for the black hole mass (left) and Eddington ratio (right).  } 
\label{fig:ap:deltaMBH_ER}
\end{figure*}

Similarly, Figure\,\ref{fig:ap:deltaMBH_ER} shows the black hole masses (left panel) and Eddington ratio (right panel) differences between the three- and two-Gaussian fits. When the 2-Gaussian model is preferred, the mean differences in both parameters are less than $\lesssim 0.2\rm \,dex$, again well within the systematic uncertainties. Given that our analysis groups sources into broad Eddington ratio bins (typically spanning 0.6$-$1 dex), this level of uncertainty has no meaningful impact on our conclusions.

In summary, the three-Gaussian model is validated both empirically and by its robustness over a wide S/N range. Our test confirms that fitting all sources with three broad Gaussians yields accurate and consistent H$\beta$ FWHM measurements across our large sample. With the continued growth of AGN optical spectroscopic surveys such as SDSS, DESI, and the forthcoming 4MOST, it becomes increasingly important to adopt a common, empirically validated fitting approach. Such consistency enables reliable characterisation of the complex H$\beta$ emission line profiles in individual spectra, while ensuring uniformity across datasets containing tens to hundreds of thousands of objects.

\subsection{Comparison between X-ray and optical luminosities.} \label{ap:Lx_Lop}

\begin{figure}
\centering

\includegraphics[trim={0.5cm 0 1.5cm 0cm},clip,scale=0.42]{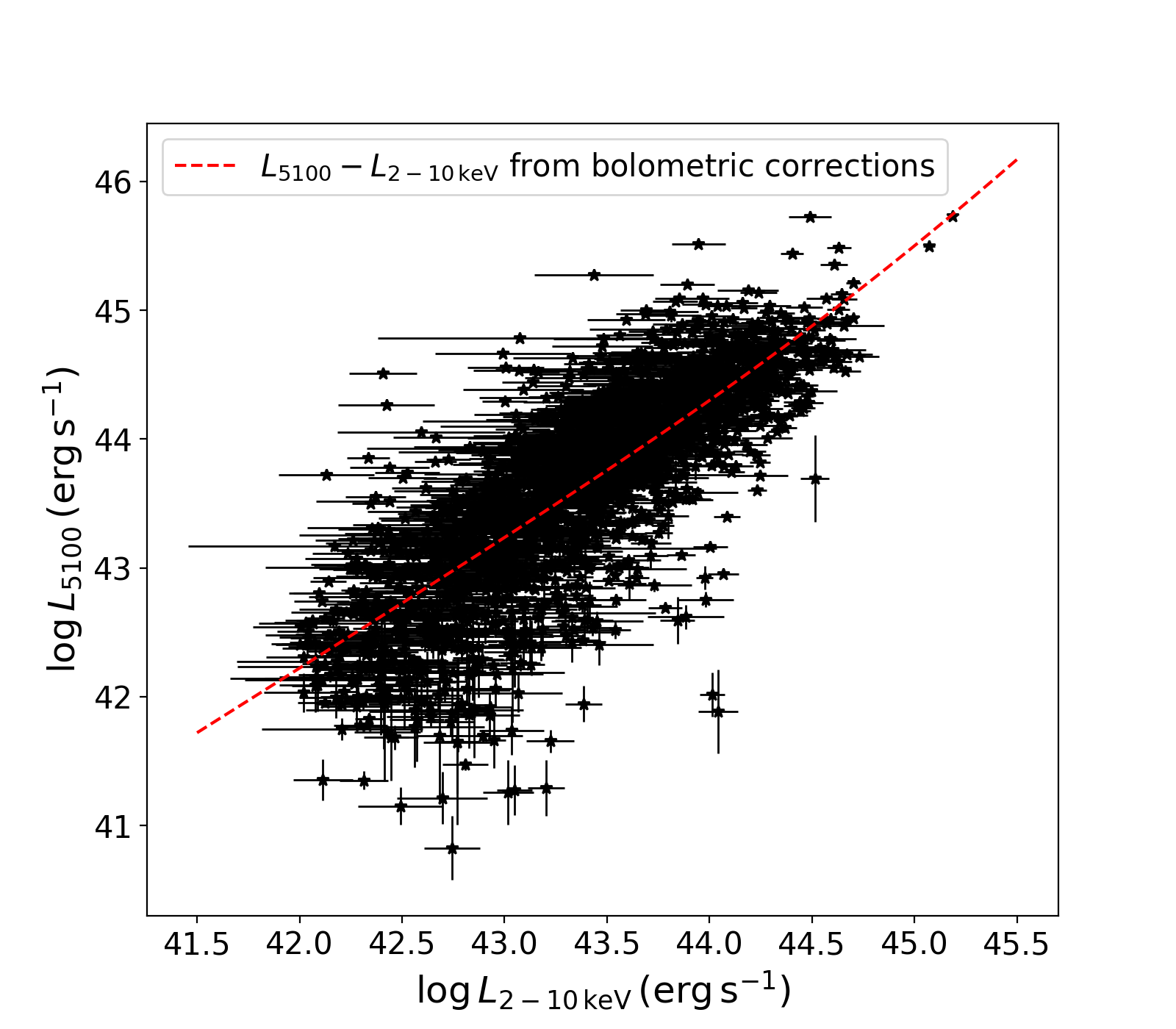}
\includegraphics[trim={0.5cm 0 1.5cm 0cm},clip,scale=0.42]{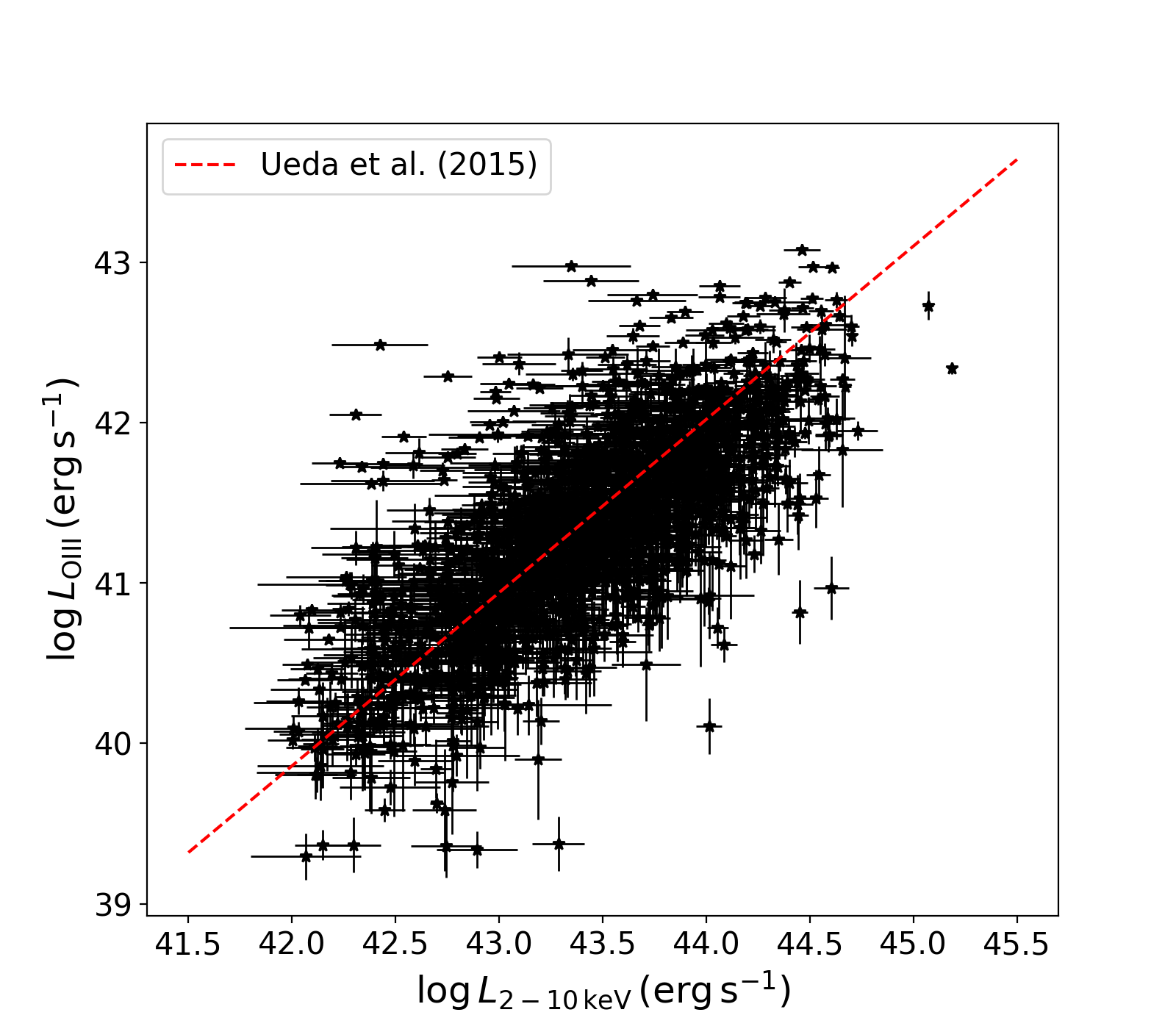}

\caption{{\it Top:} 5100\AA\, AGN luminosity against the intrinsic 2-10\,keV AGN luminosity of our sample. The error bars represent 1$\sigma$ uncertainties. The red dotted line shows the $L_{5100}-L_{2-10\rm keV}$ relationship expected when assuming the optical bolometric correction from \citet{2006Richards} ($K_{5100}=9.26$) and the X-ray bolometric correction from \citet{2020Duras}. {\it Bottom}: Luminosity of the \OIII line core against the intrinsic 2-10\,keV AGN luminosity. The red line corresponds to the $L_{\rm OIII}-L_{2-10\rm keV}$ relationship derived by \citet{2015Ueda}.}

\label{fig:LxLop}
\end{figure}

In this subsection, we compare two optical AGN luminosities derived from our spectral fitting analysis (see Section~\ref{sub:optfitting}) with independent measurements of the X-ray luminosity (see Section~\ref{sub:data:Xray}). Figure~\ref{fig:LxLop} presents these comparisons: the AGN continuum luminosity at 5100\AA, versus the 2–10\,keV luminosity, and the luminosity of the \OIII emission line core versus the 2–10\,keV luminosity.

\subsection{Black hole mass and Eddington ratio distributions} \label{ap:distMBH_ER}

\begin{figure}
 
\centering

\includegraphics[trim={0.5cm 0 1.5cm 0cm},clip,scale=0.42]{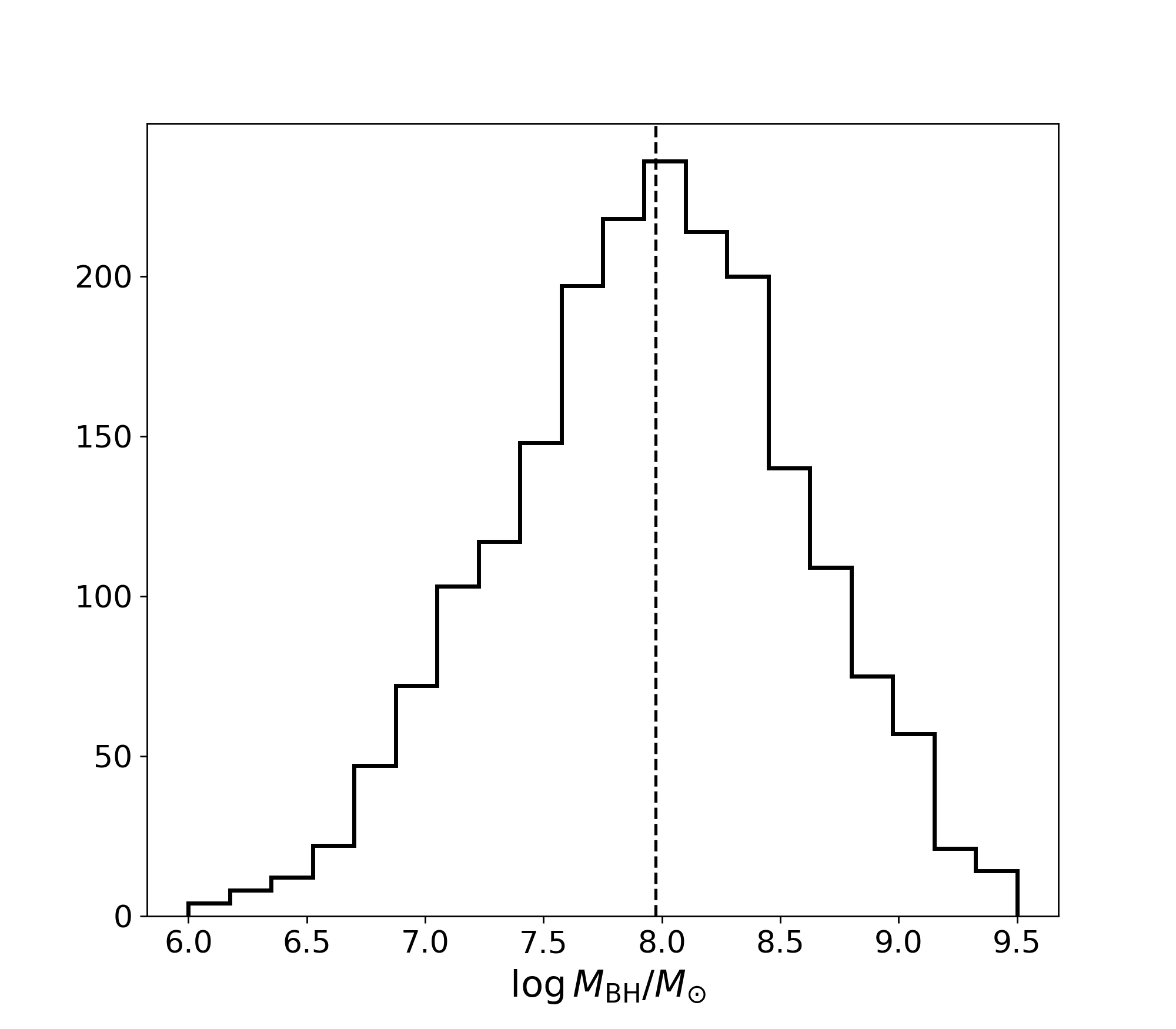}
\includegraphics[trim={0.5cm 0 1.5cm 0cm},clip,scale=0.42]{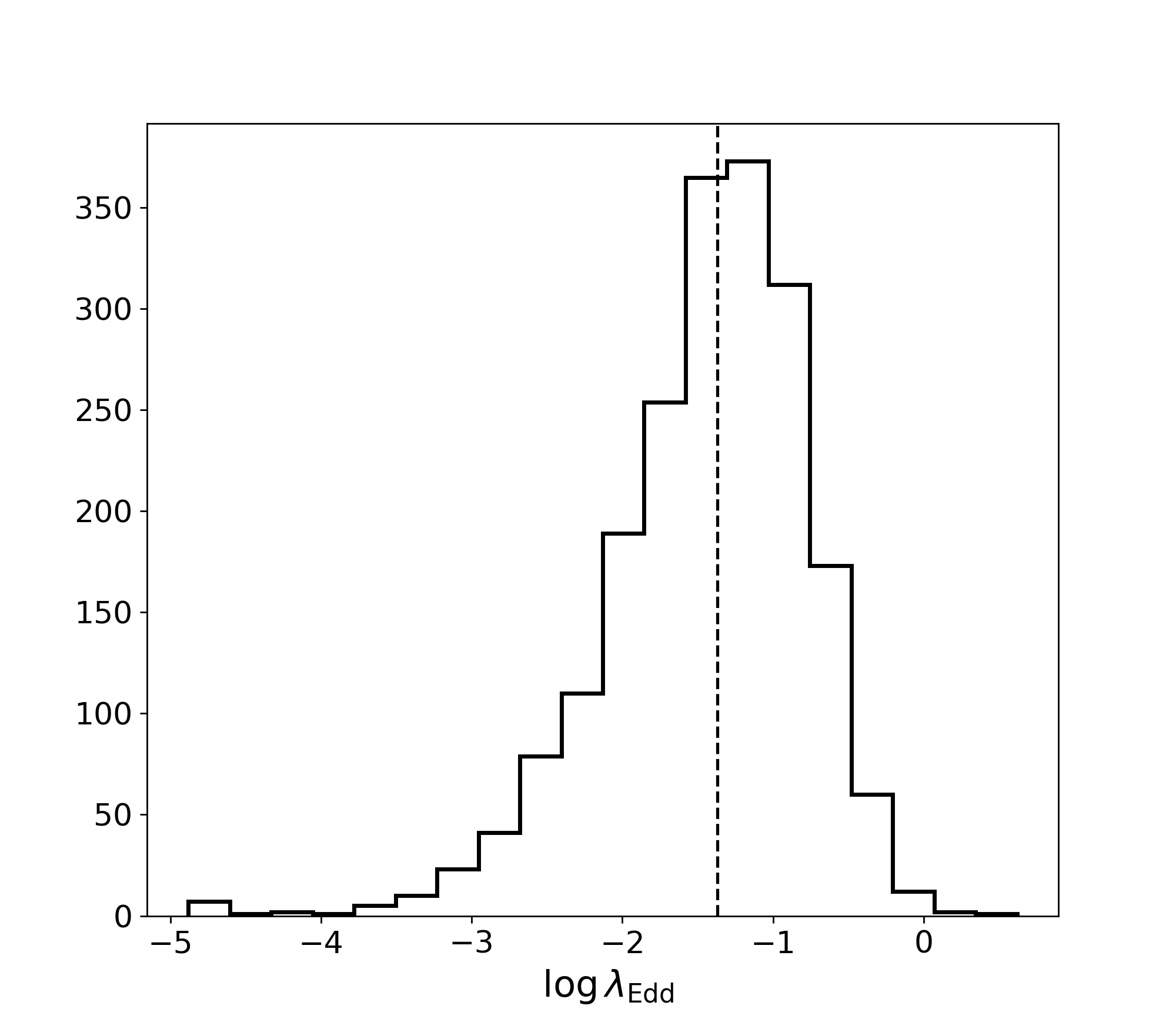}

\caption{ Distribution of the logarithmic black hole masses ($M_{\rm BH}$; top panel), and Eddington ratio ($\lambda_{\rm Edd}$; bottom panel) for the sources in our sample with a broad  $\rm H\beta$ detection.  } 
\label{fig:MBH_ER_dist}
\end{figure}

Figure\,\ref{fig:MBH_ER_dist} shows the distributions for the black hole masses and Eddington ratio of our broad  $\rm H\beta$ detected sample.

\section{ {\rm [O\,\textsc{iii}]} kinematics  against Eddington ratio controlled by $L_{
\rm bol}$} \label{ap:oIII_ER}

\begin{figure}
 
\centering
\includegraphics[trim={0cm 0 0cm 0cm},clip,scale=0.4]{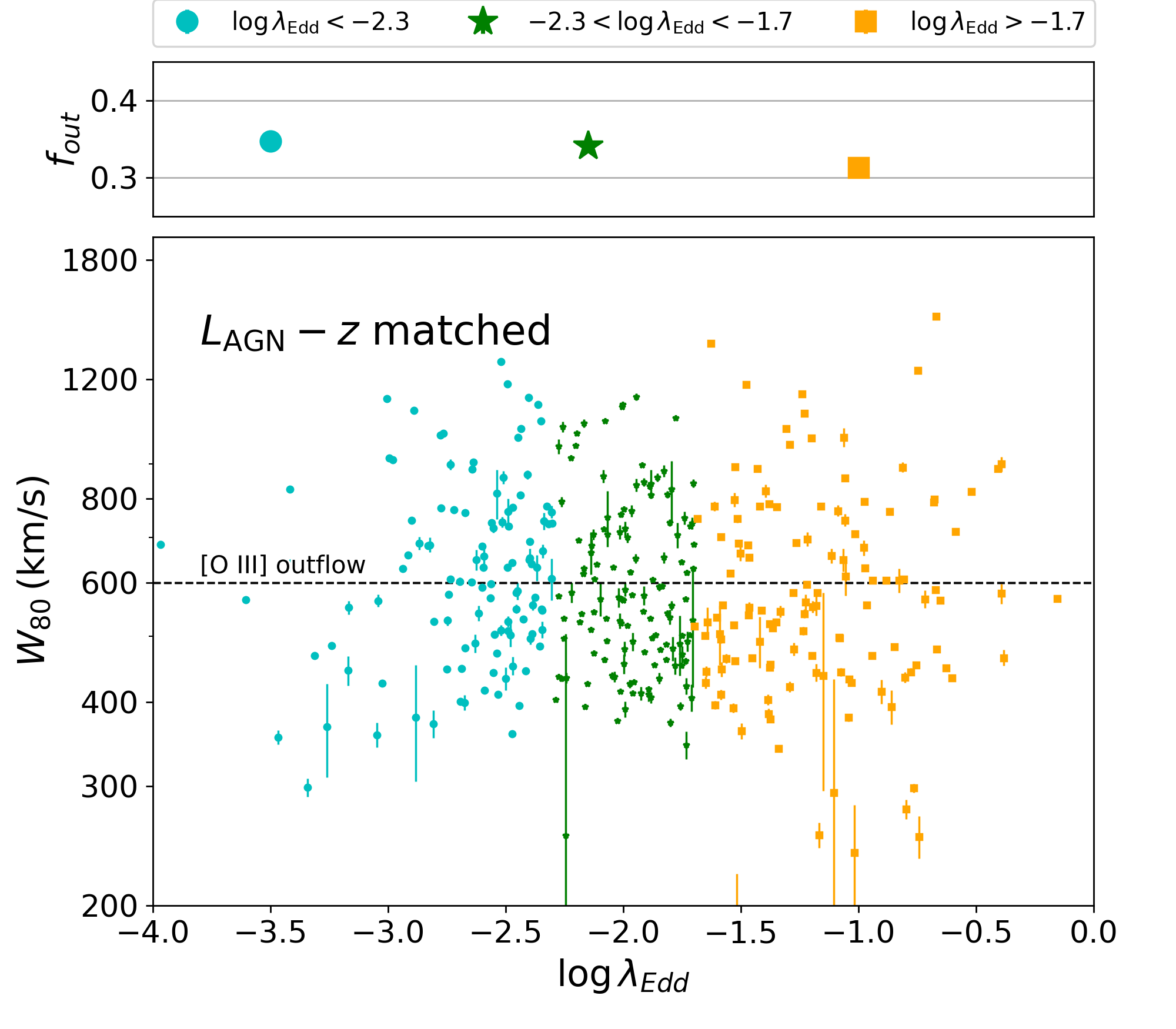}
\includegraphics[trim={0cm 0 0cm 0cm},clip,scale=0.4]{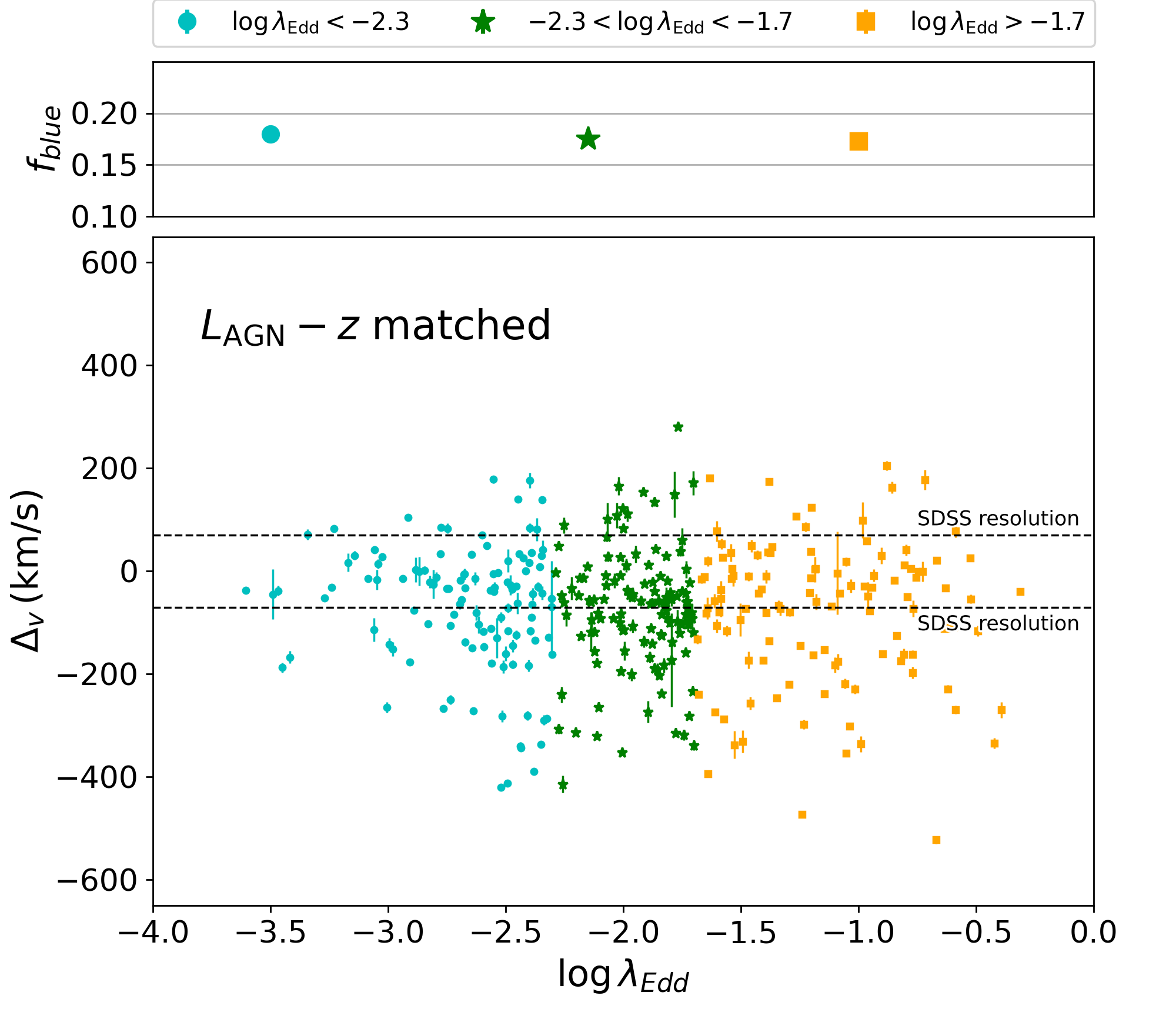}
\caption{Same as Figure\,\ref{fig:W80_EddRatio_MBHmatched} but the three $\lambda_{\rm Edd}$ samples are matched in AGN bolometric luminosity and redshift.} 
\label{fig:W80dV_EddRatio_Lbolmatched}
\end{figure}

\begin{figure}
 
\centering
\includegraphics[trim={1cm 1.5cm 2cm 0cm},clip,scale=0.32]{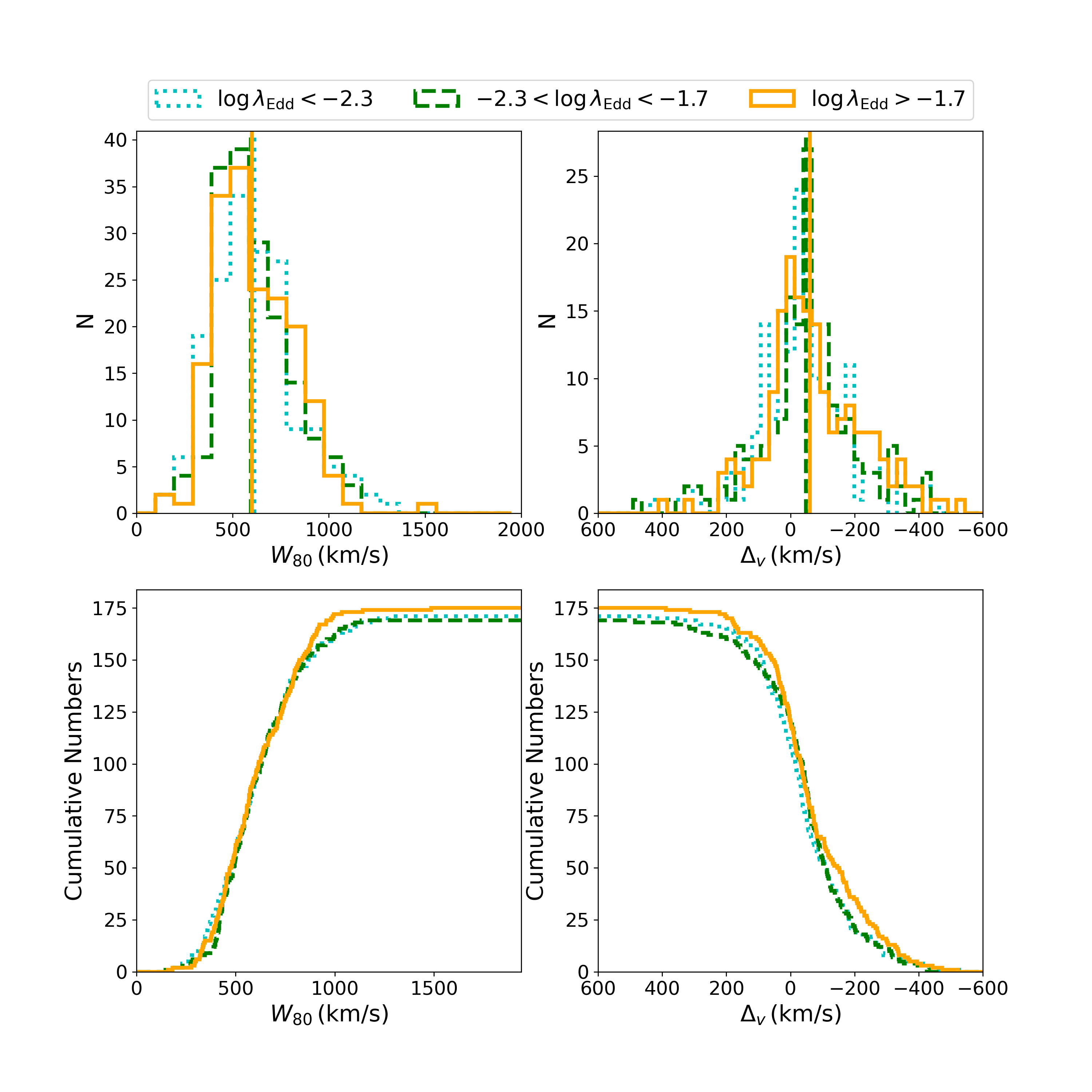}
\caption{Same as Figure\,\ref{fig:W80dVdist_EddRatio_MBHmatched} but the three $\lambda_{\rm Edd}$ samples are matched in AGN bolometric luminosity and redshift} 
\label{fig:W80dVdist_EddRatio_Lbolmatched}
\end{figure}

\begin{figure}
 
\centering
\includegraphics[trim={0cm 0 0cm 0cm},clip,scale=0.35]{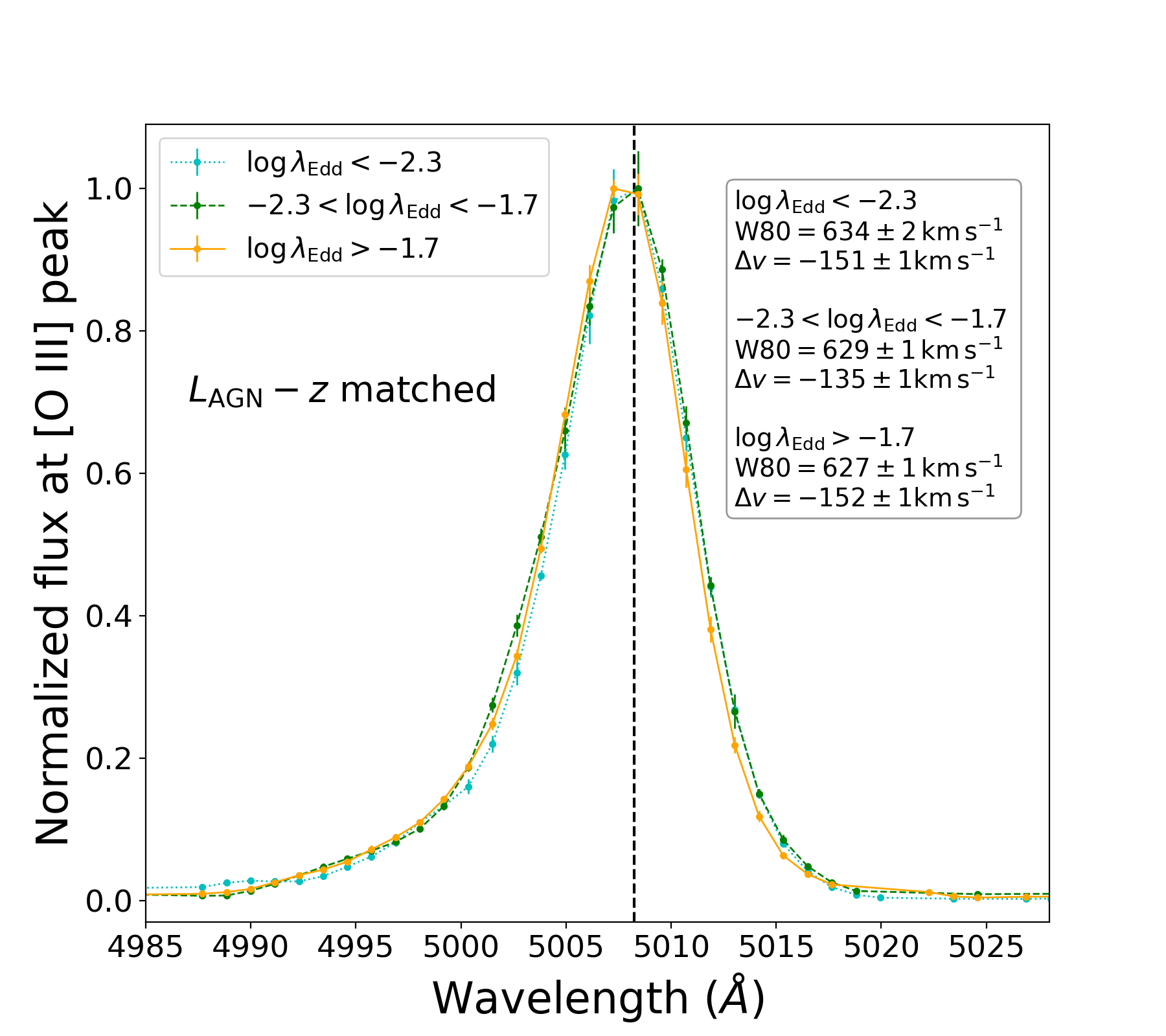}
\caption{ Same as Figure\,\ref{fig:OIIIstacks_EddRatio_MBHmatched}, but the Eddington ratio samples are matched in AGN bolometric luminosity ($L_{\rm AGN}$) and redshift. } 
\label{fig:OIIIstacks_EddRatio_LAGNmatched}
\end{figure}

In this section, we redo the analyses performed in Sections\,\ref{res:sub:OIII_EddRatio} and\,\ref{res:sub:NH_EddRatio}, where we explore the impact of the Eddington ratio on the \OIII outflow kinematics and the AGN obscuration, but matching the three Eddington ratio samples (i.e., $\log \lambda_{\rm Edd}<-2.3 $, $-2.3<\log \lambda_{\rm Edd}<-1.7$, and $\log \lambda_{\rm Edd}>-1.7$) in AGN luminosity and redshift. Figure\,\ref{fig:W80dV_EddRatio_Lbolmatched} shows the \OIII outflow velocity ($W_{\rm 80}$) and asymmetry ($\Delta v$) as a function of $\lambda_{\rm Edd}$, and Figure\,\ref{fig:W80dVdist_EddRatio_Lbolmatched} shows the distributions of $W_{\rm 80}$ and $\Delta v$ for the three $L_{\rm AGN}-z$ matched Eddington ratio samples. We see no change in the \OIII outflow kinematics with the Eddington ratio; see also Table\,\ref{t:results}. Figure\,\ref{fig:OIIIstacks_EddRatio_LAGNmatched} shows \OIII stacked profiles of the three Eddington ratio samples, which show no differences.

\begin{figure}
 
\centering

\includegraphics[trim={2cm 0 2cm 0cm},clip,scale=0.35]{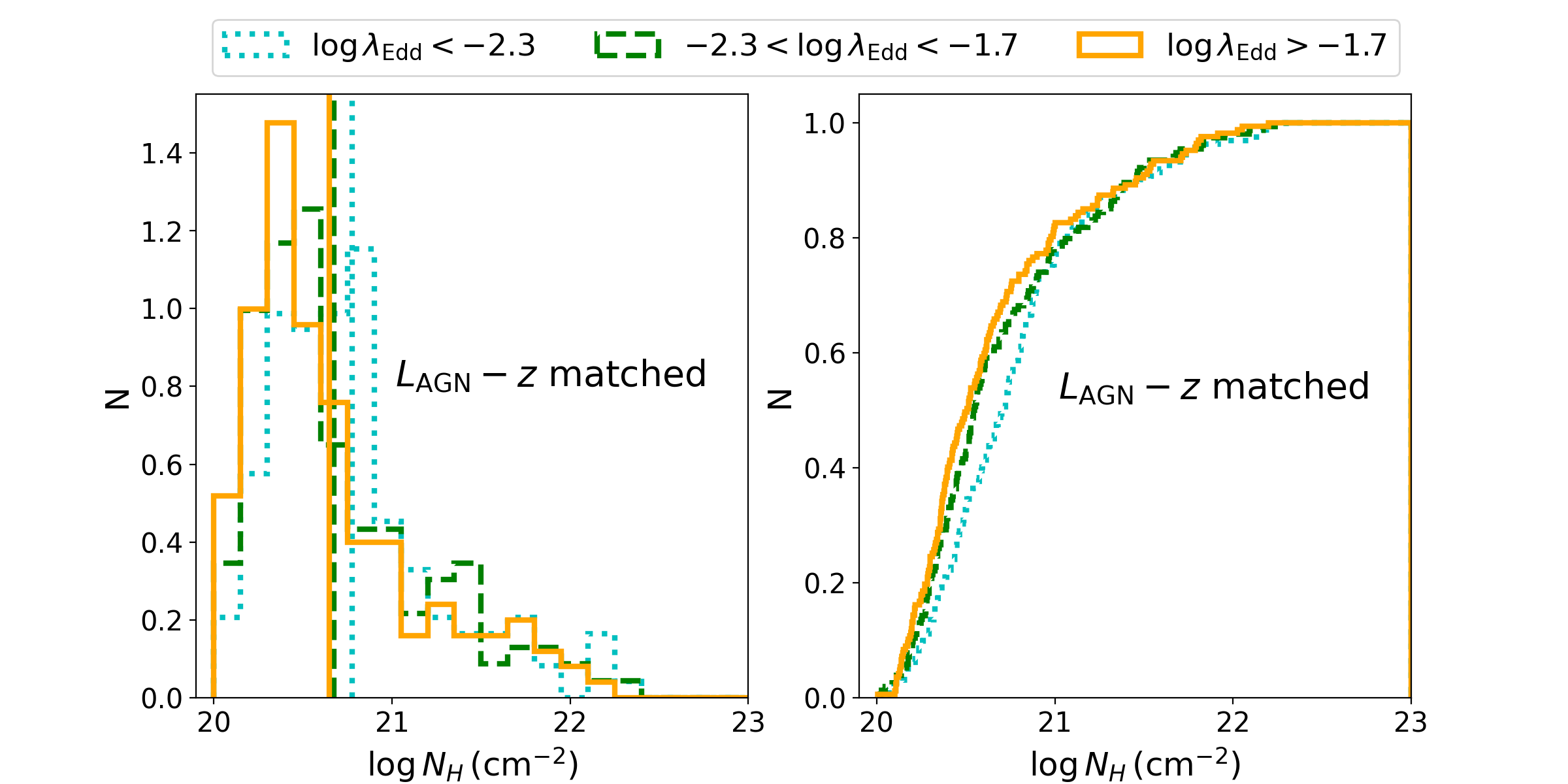}

\caption{Distributions (left) and cumulative distributions (right) of the line of sight column densities $N_{H}$ of sources with $\log \lambda_{\rm Edd}<-2.3$ (cyan dotted),  $-2.3<\log \lambda_{\rm Edd}<-1.7$ (green dashed), and  $\log \lambda_{\rm Edd}>-1.7$ (orange solid). The three Eddington ratio samples are matched in $L_{\rm AGM}$ and redshift.}
\label{fig:NH_dist_3ERbins_LAGNmatch}
\end{figure}

Finally, Figure\,\ref{fig:NH_dist_3ERbins_LAGNmatch} shows the $N_{\rm H}$ distributions for the $L_{\rm AGN}-z$ matched Eddington ratio samples. There are not differences in the $N_{\rm H}$ distributions of the medium- and high-$\lambda_{\rm Edd}$ samples, while the low-$\lambda_{\rm Edd}$ sample has a mean $\log N_{\rm H}$ just $\sim 0.1 \rm dex$ higher; see Table\,\ref{t:results}. 

\section{X-ray spectral stacks modelling} \label{ap:Xray_modellign}

In this section, we present and analyse the posterior distributions of $N_{\rm H}$ and $\Gamma$, which are key parameters in our X-ray spectral analyses.  

\begin{figure}
 
\centering
\includegraphics[trim={1cm 0 0cm 1cm},clip,scale=0.4]{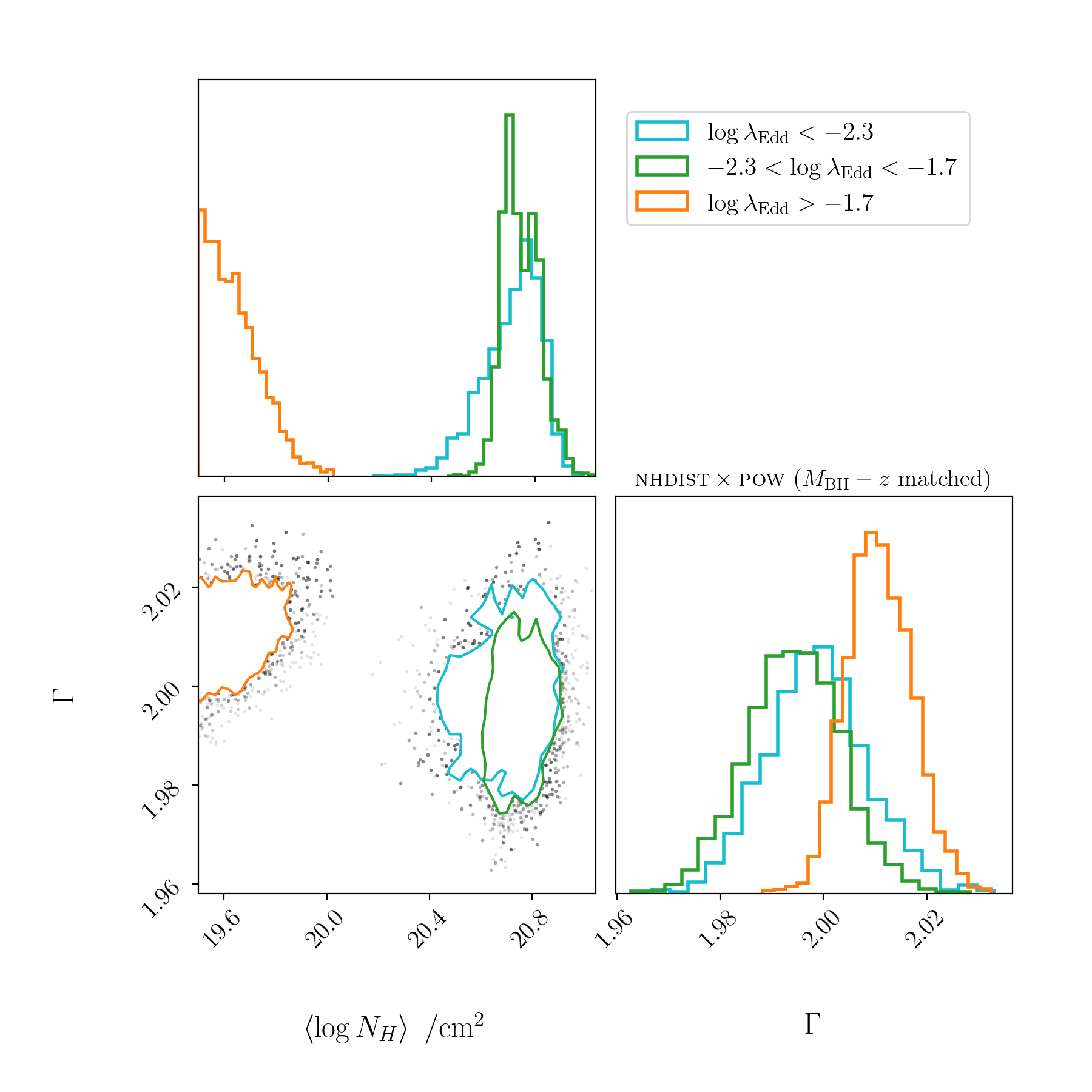}
\includegraphics[trim={1cm 0 0cm 1cm},clip,scale=0.4]{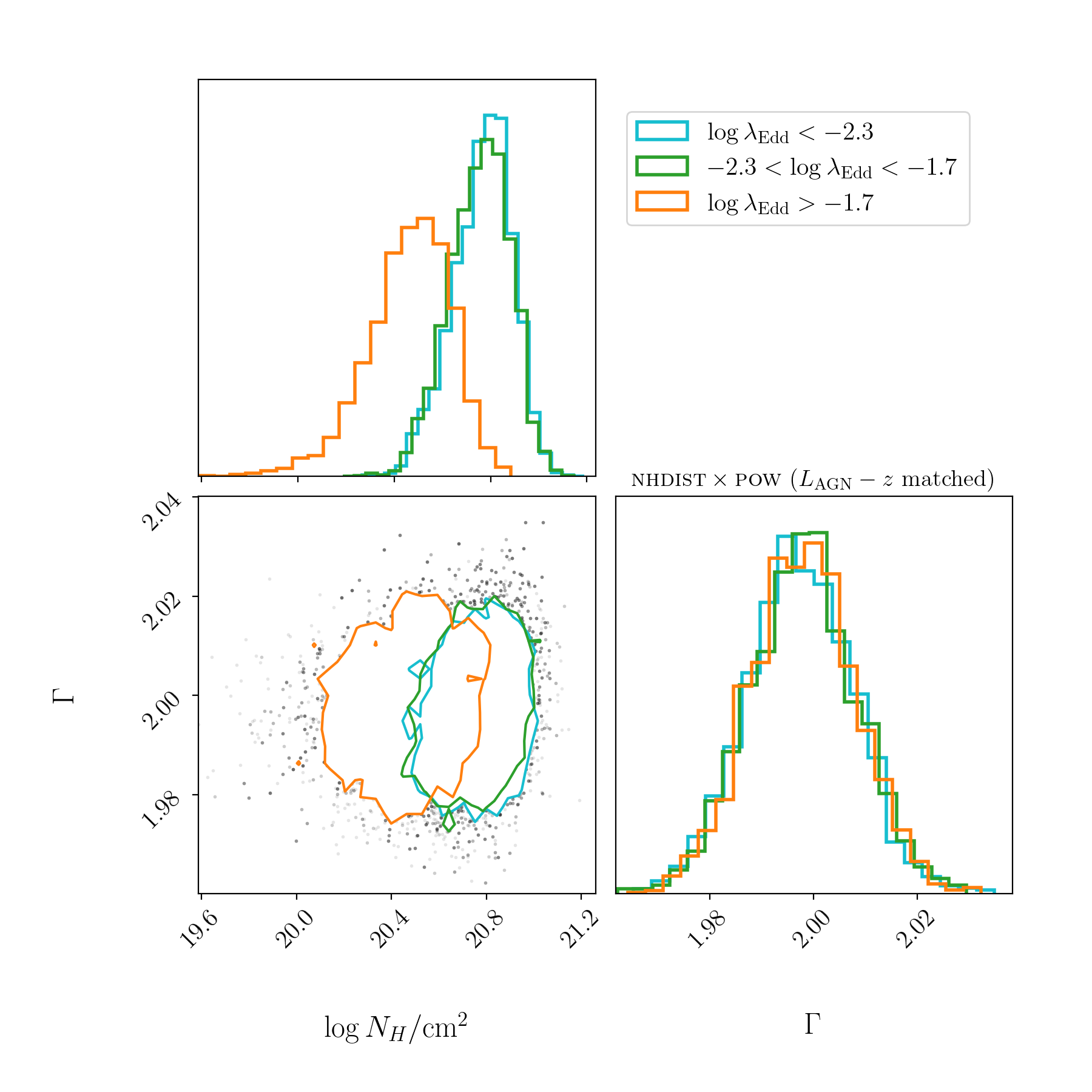}
\caption{Contour plot showing the posterior distributions of $N_{\rm H}$ and $\Gamma$ for the absorbed powerlaw model (\textsc{nhdist}$\times$\textsc{powerlaw}). The left plot shows the results for the $M_{\rm BH}-z$ matched Eddington ratio samples, while the right plot shows the results for the $L_{\rm AGN}-z$ samples.} 
\label{fig:posteriorNHGamma}
\end{figure}

Figure\,\ref{fig:posteriorNHGamma} shows the joint posterior distribution of $N_{\rm H}$ and $\Gamma$ for the absorbed power-law model (\textsc{nhdist}$\times$\textsc{powerlaw}) for the three Eddington ratio samples, when matching by SMBH mass and redshift (left panel) and by AGN luminosity and redshift (right panel). The $\Gamma$ posteriors are very narrow because we adopt a Gaussian prior centred at $\Gamma = 2$ with a standard deviation $\sigma_{\Gamma} = 0.05$ (see Section~\ref{res:sub:NH_EddRatio}).  
 
We also investigate the impact of including a soft-excess component. The soft excess refers to an enhancement of flux in the soft X-ray regime ($E < 2$\,keV) in addition to the hard X-ray power-law emission ($E > 2$\,keV) produced by the hot corona \citep[e.g.][]{1991Haardt}. Its physical origin remains debated, but previous studies have found it to be more prominent in unobscured, high–Eddington ratio AGN \citep{2016Boissay,2020Waddell,2025Chen}. In addition, eROSITA-selected AGNs often show a strong soft-excess component \citep{2024Waddell,2025Chen}.  

To test whether the soft excess influences the $N_{\rm H}$ values obtained with the absorbed power-law model, we re-fit the $\lambda_{\rm Edd}$ stacks including a soft-excess component, following the modelling approach of \citet{2025Chen}. They modelled the soft excess with a power law including a high-energy exponential cutoff, and find average parameters of $\Gamma_{\rm SE} \approx 2.5$ and $E_{\rm cut} \approx 0.5$\,keV. We therefore adopt the model, which in \textsc{XSPEC} is
\[
\textsc{nhdist} \times (\textsc{powerlaw} + \textsc{const} \times \textsc{cutoffpl}),
\]  
where the primary power law has a fixed Gaussian prior on $\Gamma$ ($\Gamma = 2$, $\sigma_{\Gamma} = 0.05$), the soft-excess cutoff power law has fixed parameters $\Gamma_{\rm SE} = 2.5$ and $E_{\rm cut} = 0.5$\,keV, and the normalisations of the intrinsic and soft-excess components are tied via the constant \textsc{const}, which is allowed to vary freely.  

\begin{figure}
 
\centering

\includegraphics[trim={1cm 0 0cm 1cm},clip,scale=0.4]{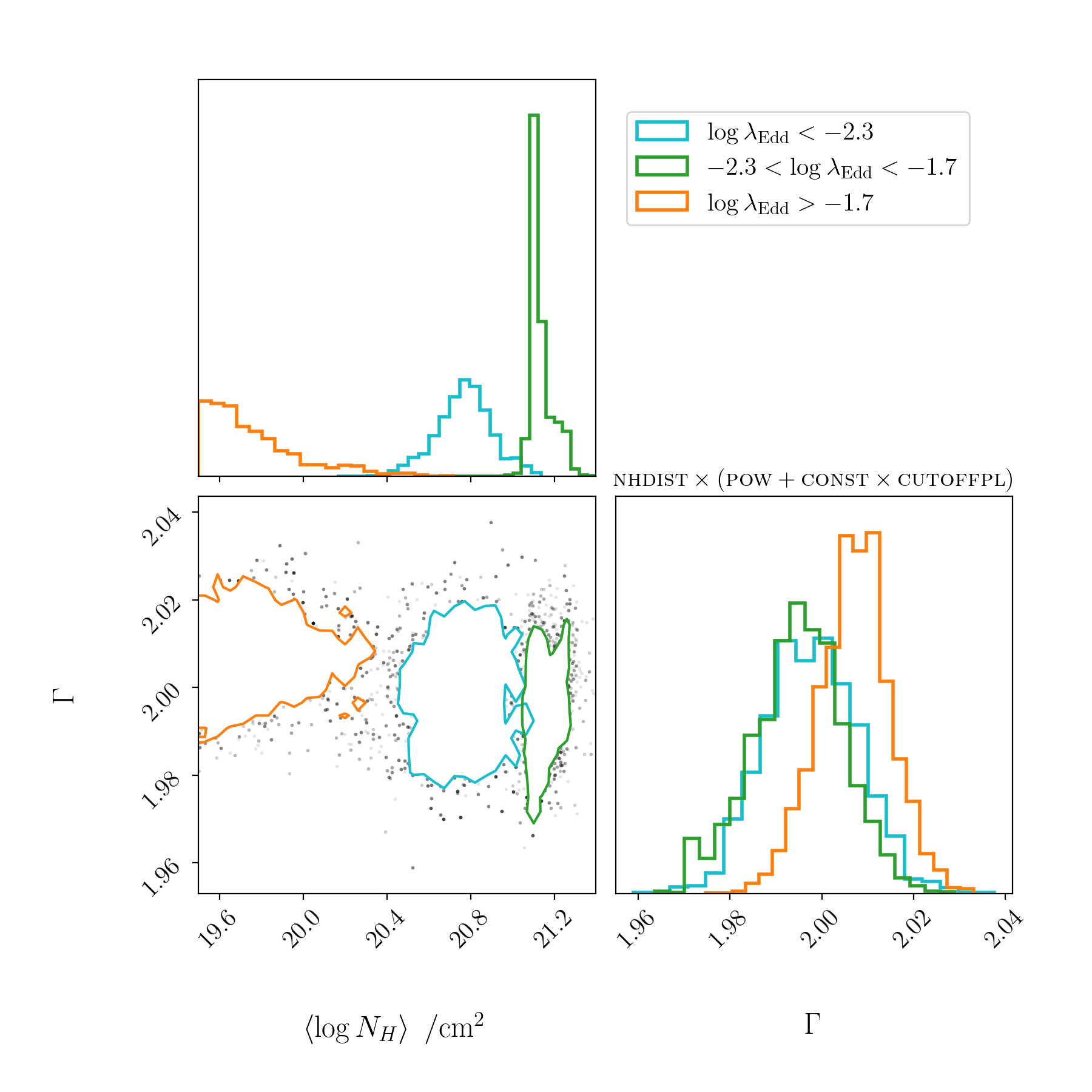}
\includegraphics[trim={1cm 0 0cm 1cm},clip,scale=0.4]{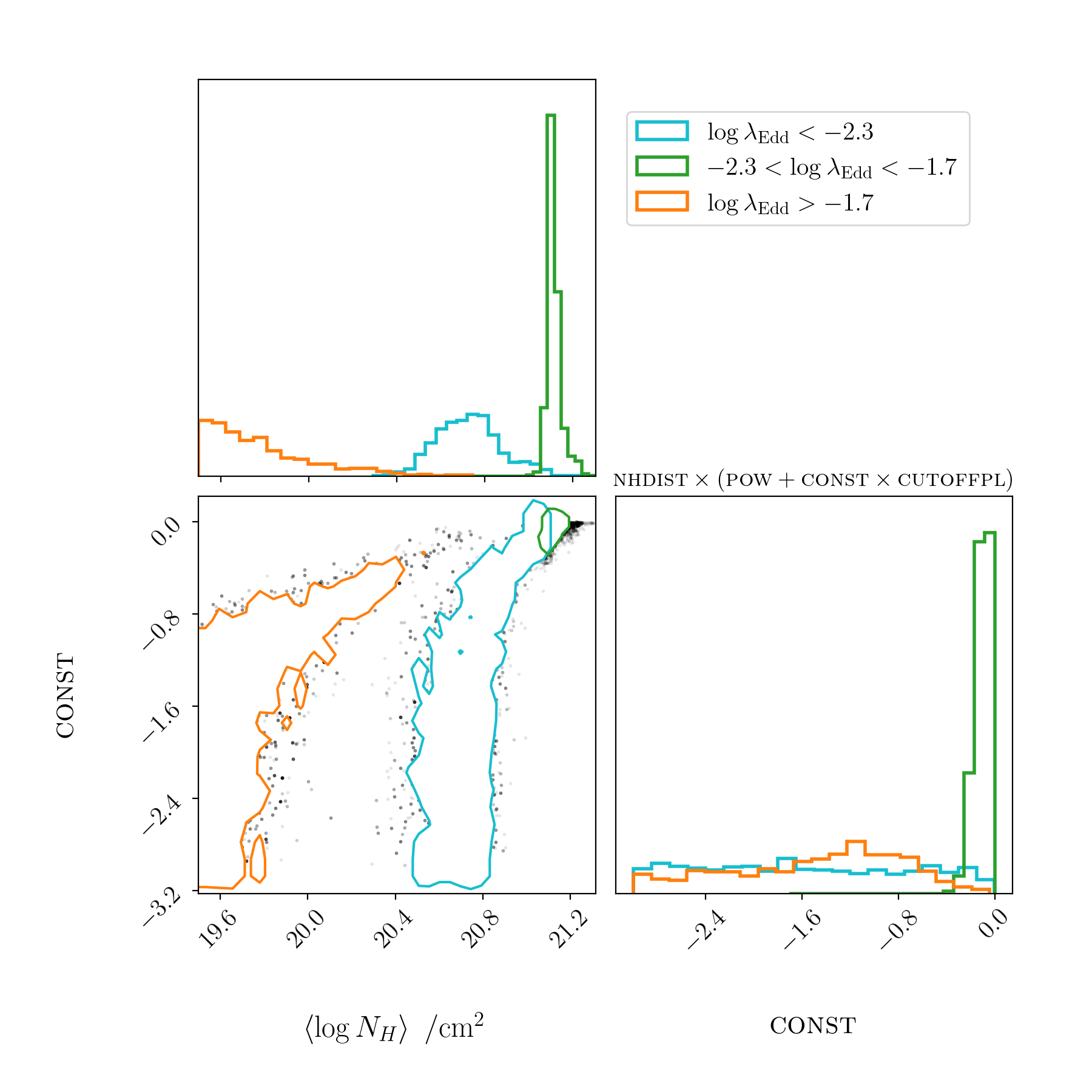}

\caption{Contour plots showing the posterior distributions of $N_{\rm H}$, $\Gamma$, and \textsc{const} (i.e., the normalisation factor between the intrinsic X-ray powerlaw and the soft-excess component). The left figure shows the posterior distributions for the absorbed powerlaw model (\textsc{nhdist}$\times$\textsc{powerlaw}), while the right figure shows the posterior distributions for the model that also includes the soft excess component (\textsc{nhdist} $\times$ (\textsc{powerlaw} + \textsc{constant}$\times$\textsc{cutoffpl}).} 
\label{fig:posteriors_SE}
\end{figure}

Figure\,\ref{fig:posteriors_SE} shows the posterior distributions of $\Gamma$, $N_{\rm H}$, and \textsc{const}. We find that \textsc{const} is well constrained only for the medium–$\lambda_{\rm Edd}$ stack ($-2.3 < \log \lambda_{\rm Edd} < -1.7$), indicating that an additional soft-excess component improves the modelling of its spectrum. For the low and high–$\lambda_{\rm Edd}$ stacks, \textsc{const} is unconstrained, suggesting that no extra component is required to model the soft X-ray emission below 2\,keV. Importantly, the main conclusion remains the same: the high–$\lambda_{\rm Edd}$ stack exhibits significantly lower obscuration, with its $\langle N_{\rm H} \rangle$ posterior shifted toward much smaller values and with no overlap with the $N_{\rm H}$ posteriors of the low and medium–$\lambda_{\rm Edd}$ stacks.

\section{ {\rm [O\,\textsc{iii}]} profile and $N_{\rm H}$ at different effective Eddington ratios } 

\begin{figure}
 
\centering
\includegraphics[trim={0cm 0 0cm 0cm},clip,scale=0.32]{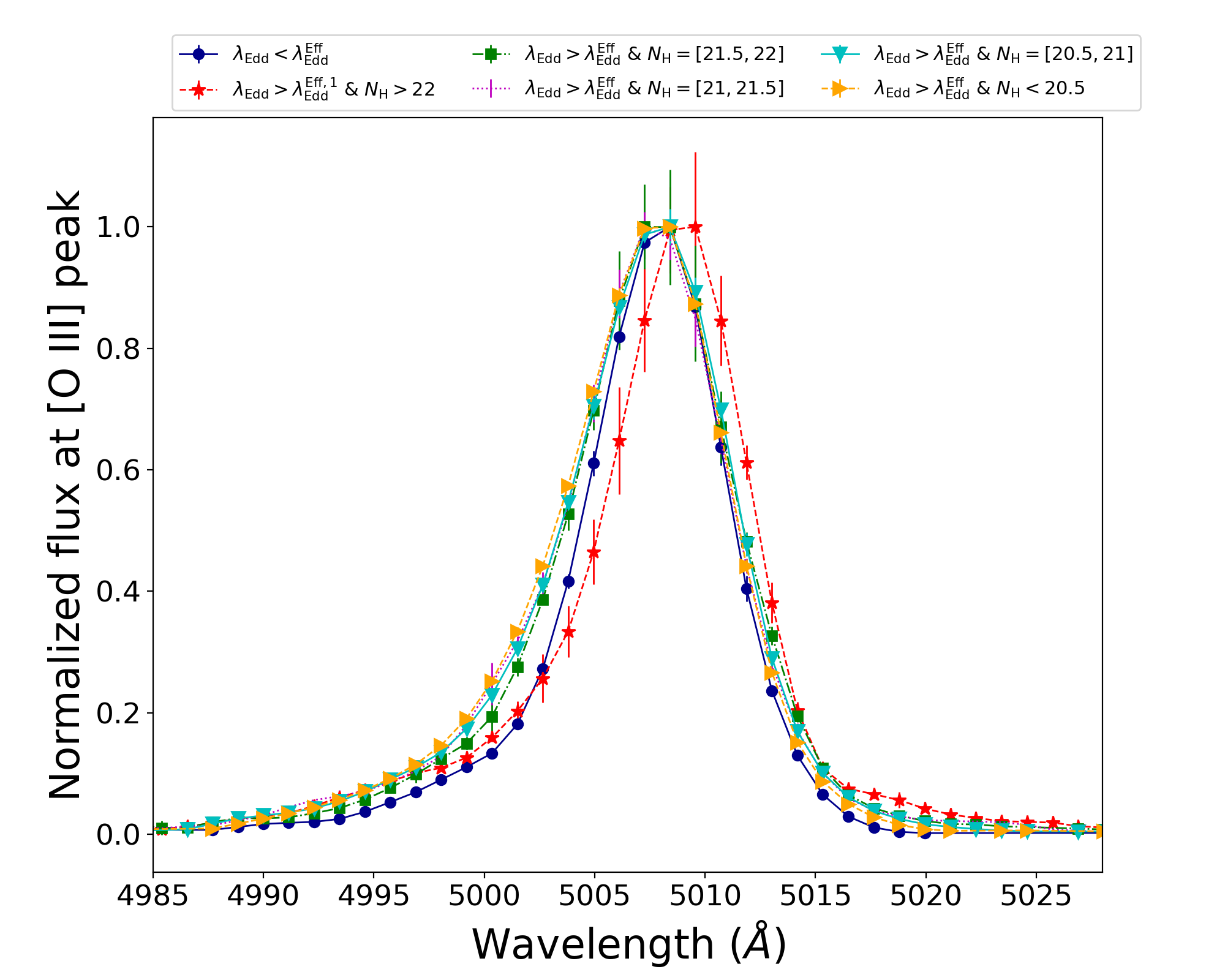}
\caption{Stacked \OIII spectra for the same sections of the $N_{\rm H}-\lambda_{\rm Edd}$ plane of Figure\,\ref{fig:NH_EddRatio}, and using the same marking style and colour-code. The stacks are normalised to their maximum flux to highlight differences in the \OIII line profiles across the Eddington ratio and $N_{\rm H}$ bins.} 
\label{fig:OIIIstacks_EffEddRatio}
\end{figure}

Figure\,\ref{fig:OIIIstacks_EffEddRatio} shows the stacked \OIII profiles of the sources lying on the different regions of the $N_{\rm H}-z$ plane depicted in Figure\,\ref{fig:NH_ER_3ERbins}. Table\,\ref{t:W80_stacks_EffER} reports the $W_{80}$ and $\Delta v$ values derived from the \OIII stacks.

\begin{figure}
 
\centering
\includegraphics[trim={0cm 0 1cm 1cm},clip,scale=0.42]{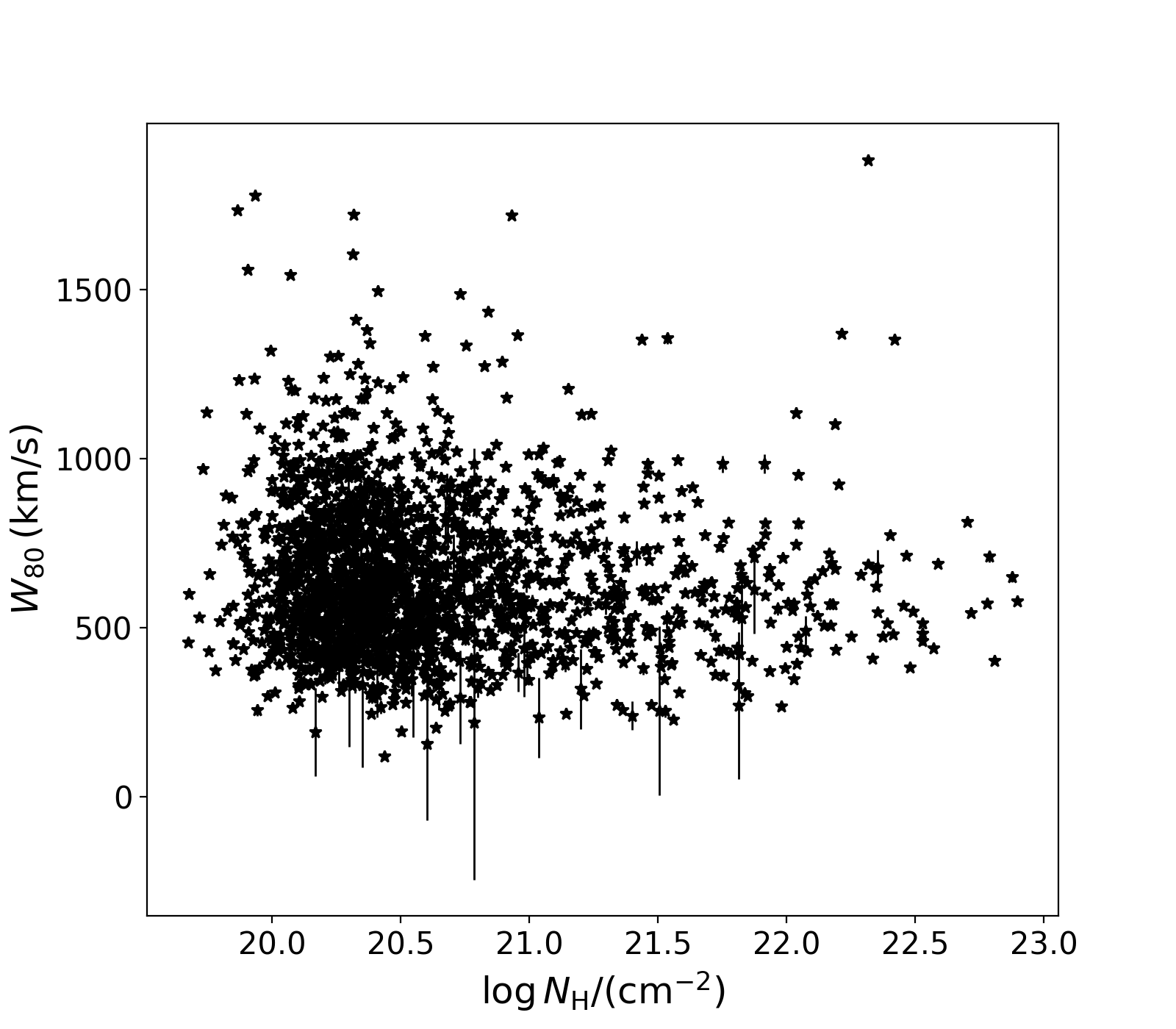}
\caption{\OIII velocity $W_{80}$ against the X-ray column density $N_{\rm H}$, for our sample.} 
\label{fig:W80_NH}
\end{figure}

Figure\,\ref{fig:W80_NH} depicts the \OIII velocity against the X-ray column density for our sample. We find no correlation between those two parameters.

\begin{table*}
\caption{Properties of the \OIII kinematics of the spectral stacks in the location of the $N_{\rm H}-\lambda_{\rm Edd}$ defined in Figure\,\ref{fig:NH_EddRatio}.}
\centering
\begin{tabular}{lccc}
\hline\hline
\noalign{\smallskip}
Sample for stack& Sample size & $W_{80} \rm \, (km \, s^{-1})$ & $\Delta v \rm \, (km \, s^{-1})$\\
\noalign{\smallskip}
\hline
$\lambda_{\rm Edd}<\lambda_{\rm Edd}^{\rm Eff}$ & 239  & $531\pm 4$ & $-142 \pm 2$\\
\noalign{\smallskip}

$\lambda_{\rm Edd}>\lambda_{\rm Edd}^{\rm Eff}$ \& $N_{\rm H}>10^{22} \, \rm cm^{-2} $ (Forbidden region) &  21   & $747\pm 10$ & $-180 \pm 3$\\
\noalign{\smallskip}

$\lambda_{\rm Edd}>\lambda_{\rm Edd}^{\rm Eff}$ \& $N_{\rm H}=10^{21.5}-10^{22} \, \rm cm^{-2} $ &  41   & $695\pm 12$ & $-108 \pm 7 $\\
\noalign{\smallskip}

$\lambda_{\rm Edd}>\lambda_{\rm Edd}^{\rm Eff}$ \& $N_{\rm H}=10^{21}-10^{21.5} \, \rm cm^{-2} $ &  112   & $704\pm 5$ & $-107 \pm 4$\\
\noalign{\smallskip}

$\lambda_{\rm Edd}>\lambda_{\rm Edd}^{\rm Eff}$ \& $N_{\rm H}=10^{20.5}-10^{21} \, \rm cm^{-2} $ &  425   & $689\pm 3$ & $-178 \pm 2 \pm $\\
\noalign{\smallskip}
$\lambda_{\rm Edd}>\lambda_{\rm Edd}^{\rm Eff}$ \& $N_{\rm H}<10^{20.5} \, \rm cm^{-2} $ &  1128   & $674\pm 1$ & $-180 \pm 1$\\
\noalign{\smallskip}
\hline
\end{tabular} \label{t:W80_stacks_EffER}
\end{table*}

\end{appendix}

\end{document}